\documentclass[journal]{IEEEtran}

\usepackage[numbers,sort&compress]{natbib}
\usepackage{url}
\usepackage{bm}
\usepackage{amsfonts}
\usepackage{amsmath}
\usepackage{amssymb}
\usepackage{amsthm}
\usepackage{color}

\usepackage{array,color}
\usepackage{tabularx}
\usepackage{multirow}
\usepackage{booktabs}
\usepackage{amsmath}  
\usepackage{makecell}  

\usepackage{graphicx}  
\usepackage{subfig}
\usepackage{epstopdf}
\usepackage{graphics}
\usepackage{epsfig}
\usepackage{psfrag}
\usepackage{overpic}

 \usepackage{stfloats}
 \usepackage{float}   

 \usepackage{indentfirst} 

\begin{document}

\newcommand {\Data} [1]{\mbox{${#1}$}}  

\newcommand {\DataN} [2]{\Data{\Power{{#1}}{{{#2}}}}}  
\newcommand {\DataIJ} [3]{\Data{\Power{#1}{{{#2}\!\times{}\!{#3}}}}}  

\newcommand {\DatassI} [2]{\!\Data{\Index{#1}{\!\Data 1},\!\Index{#1}{\!\Data 2},\!\cdots,\!\Index{#1}{\!{#2}}}}  
\newcommand {\DatasI} [2]{\Data{\Index{#1}{\Data 1},\Index{#1}{\Data 2},\cdots,\Index{#1}{#2},\cdots}}   
\newcommand {\DatasII} [3]{\Data{\Index{#1}{{\Index{#2}{\Data 1}}},\Index{#1}{{\Index{#2}{\Data 2}}},\cdots,\Index{#1}{{\Index{#2}{#3}}},\cdots}}  

\newcommand {\DatasNTt}[3]{\Data{\Index{#1}{{#2}{\Data 1}},\Index{#1}{{#2}{\Data 2}},\cdots,\Index{#1}{{#2}{#3}}} } 
\newcommand {\DatasNTn}[3]{\Data{\Index{#1}{{\Data 1}{#3}},\Index{#1}{{\Data 2}{#3}},\cdots,\Index{#1}{{#2}{#3}}} } 

\newcommand {\Vector} [1]{\Data {\mathbf {#1}}}
\newcommand {\Rdata} [1]{\Data {\hat {#1}}}
\newcommand {\Tdata} [1]{\Data {\tilde {#1}}} 
\newcommand {\Udata} [1]{\Data {\overline {#1}}} 
\newcommand {\Fdata} [1]{\Data {\mathbb {#1}}} 
\newcommand {\Prod} [2]{\Data {\prod_{\SI {#1}}^{\SI {#2}}}}  
\newcommand {\Sum} [2]{\Data {\sum_{\SI {#1}}^{\SI {#2}}}}   
\newcommand {\Belong} [2]{\Data{ {#1} \in{}{#2}}}  

\newcommand {\Abs} [1]{\Data{ \lvert {#1} \rvert}}  
\newcommand {\Mul} [2]{\Data{ {#1} \times {#2}}}  
\newcommand {\Muls} [2]{\Data{ {#1} \! \times \!{#2}}}  
\newcommand {\Mulsd} [2]{\Data{ {#1} \! \cdot \!{#2}}}  
\newcommand {\Div} [2]{\Data{ \frac{#1}{#2}}}  
\newcommand {\Trend} [2]{\Data{ {#1}\rightarrow{#2}}}  
\newcommand {\Sqrt} [1]{\Data {\sqrt {#1}}} 
\newcommand {\Sqrtn} [2]{\Data {\sqrt[2]{#1}}} 

\newcommand {\Power} [2]{\Data{ {#1}^{\TI {#2}}}}  
\newcommand {\Index} [2]{\Data{ {#1}_{\TI {#2}}}}  

\newcommand {\Equ} [2]{\Data{ {#1} = {#2}}}  
\newcommand {\Equs} [2]{\Data{ {#1}\! =\! {#2}}}  
\newcommand {\Equss} [3]{\Equs {#1}{\Equs {#2}{#3}}}  

\newcommand {\Equu} [2]{\Data{ {#1} \equiv {#2}}}  

\newcommand {\LE}[0] {\leqslant}
\newcommand {\GE}[0] {\geqslant}
\newcommand {\NE}[0] {\neq}
\newcommand {\INF}[0] {\infty}
\newcommand {\MIN}[0] {\min}
\newcommand {\MAX}[0] {\max}

\newcommand {\Funcfx} [2]{\Data{ {#1}({#2})}}  
\newcommand {\Funcfzx} [3]{\Data{ {\Index {#1}{#2}}({#3})}}  
\newcommand {\Funcfnzx} [4]{\Data{ {\Index {\Power{#1}{#2}}{#3}}({#4})}}  
\newcommand {\SI}[1] {\small{#1}}
\newcommand {\TI}[1] {\tiny {#1}}
\newcommand {\Text}[1] {\text {#1}}

\newcommand {\VtS}[0]{\Index {t}{\Text {s}}}
\newcommand {\Vti}[0]{\Index {t}{i}}
\newcommand {\Vt}[0]{\Data {t}}
\newcommand {\VLES}[1]{\Index {\tau} {\SI{\Index {}{ \Text{#1}}}}}
\newcommand {\VLESmin}[1]{\Index {\tau} {\SI{\Index {}{ \Text{min\_\Text{#1}}}}}}
\newcommand {\VAT}[0]{\Index {\Vector A}{\Text{Time}}}
\newcommand {\VPbus}[1]{\Index {P}{\Text{Bus-}{#1}}}
\newcommand {\VPbusmax}[1]{\Index {P}{\Text{max\_Bus-}{#1}}}

\newcommand {\EtS}[2]{\Equs {\Index {t}{\Text {s}}}{#1} {#2}}
\newcommand {\Eti}[2]{\Equs {\Index {t}{i}}{#1} {#2}}
\newcommand {\Et}[2]{\Equs {t}{#1} {#2}}
\newcommand {\EMSR}[2]{\Equs {\Index {\tau} {\SI{\Index {}{ \Text{MSR}}}}}{#1} {#2}}
\newcommand {\EMSRmin}[2]{\Equs {\Index {\tau} {\SI{\Index {}{ \Text{MSR}}}}}{#1} {#2}}

\newcommand {\EAT}[2]{\Equs {\Index {\Vector A}{\Text{Time}}}{#1} {#2}}
\newcommand {\EPbus}[3]{\Equs {\Index {P}{\Text{Bus-}{#1}}}{#2} \Text{ #3}}
\newcommand {\EPbusmax}[3]{\Equs {\Index {P}{\Text{max\_Bus-}{#1}}}{#2} \Text{ #3}}

\newcommand {\Vgam}[1]{\Index {\gamma}{#1}}
\newcommand {\Egam}[2]{\Equs {\Vgam{#1}}{#2}}

\newcommand {\Emu}[2]{\Equs {{\mu}{#1}}{#2}}
\newcommand {\Esigg}[2]{\Equs {{\sigma}^2{#1}}{#2}}

\newcommand {\Vlambda}[1]{\Index {\lambda}{#1}}

\newcommand {\VV}[1]{\Index {\Vector V}{#1}}
\newcommand {\Vv}[1]{\Index {\Vector v}{#1}}
\newcommand {\Vsv}[1]{\Index {v}{#1}}
\newcommand {\VX}[1]{\Index {\Vector X}{#1}}
\newcommand {\VsX}[1]{\Index {X}{\SI{\Index {}{#1}}}}
\newcommand {\Vx}[1]{\Index {\Vector x}{\SI{\Index {}{#1}}}}
\newcommand {\Vsx}[1]{\Index {x}{\SI{\Index {}{#1}}}}
\newcommand {\VZ}[1]{\Index {\Vector Z}{#1}}
\newcommand {\Vz}[1]{\Index {\Vector z}{\SI{\Index {}{#1}}}}
\newcommand {\Vsz}[1]{\Index {z}{\SI{\Index {}{#1}}}}
\newcommand {\VIndex}[2]{\Index {\Vector {#1}}{#2}}
\newcommand {\VY}[1]{\Index {\Vector Y}{#1}}
\newcommand {\Vy}[1]{\Index {\Vector y}{#1}}
\newcommand {\Vsy}[1]{\Index {y}{\SI{\Index {}{#1}}}}

\newcommand {\VRV}[1]{\Index {\Rdata {\Vector V}}{#1}}
\newcommand {\VRsV}[1]{\Index {\Rdata {V}}{#1}}
\newcommand {\VRX}[1]{\Index {\Rdata {\Vector X}}{#1}}
\newcommand {\VRx}[1]{\Index {\Rdata {\Vector x}}{\SI{\Index {}{#1}}}}
\newcommand {\VRsx}[1]{\Index {\Rdata {x}}{\SI{\Index {}{#1}}}}
\newcommand {\VRZ}[1]{\Index {\Rdata {\Vector Z}}{#1}}
\newcommand {\VRz}[1]{\Index {\Rdata {\Vector z}}{\SI{\Index {}{#1}}}}
\newcommand {\VRsz}[1]{\Index {\Rdata {z}}{\SI{\Index {}{#1}}}}
\newcommand {\VTX}[1]{\Index {\Tdata {\Vector X}}{#1}}
\newcommand {\VTx}[1]{\Index {\Tdata {\Vector x}}{\SI{\Index {}{#1}}}}
\newcommand {\VTsx}[1]{\Index {\Tdata {x}}{\SI{\Index {}{#1}}}}
\newcommand {\VTsX}[1]{\Index {\Tdata {X}}{\SI{\Index {}{#1}}}}
\newcommand {\VTZ}[1]{\Index {\Tdata {\Vector Z}}{#1}}
\newcommand {\VTz}[1]{\Index {\Tdata {\Vector z}}{\SI{\Index {}{#1}}}}
\newcommand {\VTsz}[1]{\Index {\Tdata {z}}{\SI{\Index {}{#1}}}}
\newcommand {\VOG}[1]{\Vector{\Omega}{#1}}

\newcommand {\Sigg}[1]{\Data {{\sigma}^2({#1})}}
\newcommand {\Sig}[1]{\Data {{\sigma}({#1})}}

\newcommand {\Mu}[1]{\Data{{\mu} ({#1})}}
\newcommand {\Eig}[1]{\Data {\lambda}({\Vector {#1}}) }
\newcommand {\Her}[1]{\Power {#1}{\!H}}
\newcommand {\Tra}[1]{\Power {#1}{\!T}}

\newcommand {\VF}[3] {\DataIJ {\Fdata {#1}}{#2}{#3}}
\newcommand {\VRr}[2] {\DataN {\Fdata {#1}}{#2}}

\newcommand {\Tcol}[2] {\multicolumn{1}{#1}{#2} }
\newcommand {\Tcols}[3] {\multicolumn{#1}{#2}{#3} }
\newcommand {\Cur}[2] {\mbox {\Data {#1}-\Data {#2}}}

\newcommand {\VDelta}[1] {\Data {\Delta\!{#1}}}

\newcommand {\STE}[1] {\Fdata {E}{\Data{({#1})}}}
\newcommand {\STD}[1] {\Fdata {D}{\Data{({#1})}}}

\newcommand {\TestF}[1] {\Data {\varphi(#1)}}
\newcommand {\ROMAN}[1] {\uppercase\expandafter{\romannumeral#1}}

\def \FuncC #1#2{
\begin{equation}
{#2}
\label {#1}
\end{equation}
}

\def \FuncCC #1#2#3#4#5#6{
\begin{equation}
#2=
\begin{cases}
    #3 & #4 \\
    #5 & #6
\end{cases}
\label{#1}
\end{equation}
}

\def \Figff #1#2#3#4#5#6#7{   
\begin{figure}[#7]
\centering
\subfloat[#2]{
\label{#1a}
\includegraphics[width=0.23\textwidth]{#4}
}
\subfloat[#3]{
\label{#1b}
\includegraphics[width=0.23\textwidth]{#5}
}
\caption{\small #6}
\label{#1}
\end{figure}
}

\def \Figffb #1#2#3#4#5#6#7#8#9{   
\begin{figure}[#9]
\centering
\subfloat[#2]{
\label{#1a}
\includegraphics[width=0.23\textwidth]{#5}
}
\subfloat[#3]{
\label{#1b}
\includegraphics[width=0.23\textwidth]{#6}
}

\subfloat[{#4}]{
\label{#1c}
\includegraphics[width=0.48\textwidth]{#7}
}
\caption{\small #8}
\label{#1}
\end{figure}
}

\def \Figffp #1#2#3#4#5#6#7{   
\begin{figure*}[#7]
\centering
\subfloat[#2]{
\label{#1a}
\begin{minipage}[t]{0.24\textwidth}
\centering
\includegraphics[width=1\textwidth]{#4}
\end{minipage}
}
\subfloat[#3]{
\label{#1b}
\begin{minipage}[t]{0.24\textwidth}
\centering
\includegraphics[width=1\textwidth]{#5}
\end{minipage}
}
\caption{\small #6}
\label{#1}
\end{figure*}
}

\def \Figf #1#2#3#4{   
\begin{figure}[#4]
\centering
\includegraphics[width=0.48\textwidth]{#2}

\caption{\small #3}
\label{#1}
\end{figure}
}

\definecolor{Orange}{RGB}{249,106,027}
\definecolor{sOrange}{RGB}{251,166,118}
\definecolor{ssOrange}{RGB}{254,213,190}

\definecolor{Blue}{RGB}{008,161,217}
\definecolor{sBlue}{RGB}{090,206,249}
\definecolor{ssBlue}{RGB}{200,239,253}

\title{A Random Matrix Theoretical Approach to Early Event Detection in Smart Grids}

\author{Xing~He,  Robert~C. Qiu,~\IEEEmembership{Fellow,~IEEE}, Qian~Ai, ~\IEEEmembership{Member,~IEEE}, Yinshuang~Cao, Jie~Gu, Zhijian~Jin
\thanks{This work was partly supported by National Natural Science Foundation of China (No. 51577115).}
\thanks{Xing~He, Robert~C. Qiu, Yinshuang~Cao, Qian~Ai, Jie~Gu, and Zhijian~Jin are with the Department of Electrical Engineering, Research Center for Big Data Engineering Technology, State Energy Smart Grid R$\&$D Center, Shanghai Jiaotong University, Shanghai 200240, China. (e-mail: {hexing\_hx@126.com)}}
\thanks{Robert~C.~Qiu is also with the Department of Electrical and Computer Engineering,
Tennessee Technological University, Cookeville, TN 38505, USA. (e-mail: {rqiu@tntech.edu})}
}

\maketitle

\begin{abstract}
Power systems are developing very fast nowadays, both in size and in complexity; this situation is a challenge for Early Event Detection (EED). This paper proposes a data-driven unsupervised learning method to handle this challenge.
Specifically, the random matrix theories (RMTs) are introduced as the statistical foundations for random matrix models (RMMs); based on the RMMs, linear eigenvalue statistics (LESs) are defined via the test functions as the system indicators. By comparing the values of the LES between the experimental and the theoretical ones, the anomaly detection is conducted. Furthermore, we develop 3D power-map to visualize the LES; it provides a robust auxiliary decision-making mechanism to the operators.
In this sense, the proposed method conducts EED with a pure statistical procedure, requiring no knowledge of system topologies, unit operation/control models, etc.
The LES, as a key ingredient during this procedure, is a high dimensional indictor derived directly from raw data. As an unsupervised learning indicator, the LES is much more sensitive than the low dimensional indictors obtained from supervised learning.
With the statistical procedure, the proposed method is universal and fast; moreover, it is robust against traditional EED challenges (such as error accumulations, spurious correlations, and even bad data in core area).
Case studies, with both simulated data and real ones, validate the proposed method.
To manage large-scale distributed systems, data fusion is mentioned as another data processing ingredient.
\end{abstract}

\begin{IEEEkeywords}
big~data, early~event~detection, data-driven, unsupervised~learning, smart~grid,  random~matrix~model, linear eigenvalue statistics, 3D~power-map, data~fusion
\end{IEEEkeywords}

%
\IEEEpeerreviewmaketitle

\section{Introduction}

%
\IEEEPARstart{D}{ata} have become a strategic resource for power systems. Data are readily accessible caused by developments of various technologies and devices, such as Information Communication Technology (ICT), Advanced Metering Infrastructure (AMI), Phasor Measurement Units (PMUs), Intelligent Electronic Devices (IEDs), Supervisory Control and Data Acquisition (SCADA) \cite{xu2013power}. As a result, data with features of volume, velocity, variety, and veracity (i.e. 4Vs data) \cite{IBM2014fourv}, as well as curses of dimensionality \cite{moulin2004support}, are inevitably generated and daily aggregated in power systems.

Early event detection (EED), by continuously monitoring and processing 4Vs data, detects and identifies emerging patterns of anomalies in power systems to make corresponding emergency responses \cite{neill2010multivariate}.  In other words, EED aims to tell signals from noises---we treat receivable sample errors and irregular fluctuations (from distributed generators and small loads) as noises; whereas we treat system faults, network reconfigurations, and dramatic changes from loads/generators (often unintended)  as signals. A smart grid, especially the one large in scale, is a complex big data system essentially \cite{zhou2013review, he2015arch}. For such a system, it is a big challenge to conduct EED within a tolerable elapsed time and hardware resources.

The integration of \emph{big data analytics} and \emph{unsupervised learning mechanism} is an effective approach to this challenge.
For the  \emph{former}, big data analytics is a scientific trend dealing with complex data \cite{nature2008bigd, science2011bigd}.
It is a data-driven tool and aims to work out statistical characteristics (especially correlations) indicated by linear eigenvalue statistics (LESs) \cite{qiu2015smart}. That means, it conducts data processing in high dimensions, rather than builds and analyzes individual models based on assumptions and simplifications, to help understand and gain insight into the systems \cite{qiu2013bookcogsen}.
Big data analytics has already been successfully applied in numerous phenomena, such as quantum systems \cite{brody1981random}, financial systems  \cite{laloux2000random}, biological systems \cite{howe2008big}, wireless communication networks \cite{qiu2013bookcogsen,qiu2014Intial70N,qiu2014MIMO}; we believe that it will also have a wide applied scope in power systems  \cite{qiu2015smart,IBM2009Manag, kezunovic2013role}.
For the \emph{latter}, the supervised learning methods are prevailing in data processing. The key parts are the inferred functions and empirical models; these functions/models, produced via a subjective training procedure (often in low dimensions), lead to a determinate parameter as the system indicator \cite{mohri2012foundations}. However, for a complex system, it is hard to find a convincing training way to ensure the validity of the last indicator; besides, it is impossible to train all the scenarios to infer an event identification framework which is robust enough to manage all the 4Vs data.
In contrast, the unsupervised method proposed in this paper utilizes the data in the form of random matrix models (RMMs), which are derived from the raw data in a statistical manner. Hence, the unsupervised methods is more suitable for EED in smart grids.

\subsection{Contribution}
This paper proposes a novel data-driven unsupervised learning method to conduct EED in smart grids, and a comparison is made with supervised ones (e.g. a dimensionality reduction method based on Principal Component Analysis (PCA) \cite{xie2014dimensionality}). First, random matrix theories (RMTs) are briefly introduced as the solid mathematic foundations for RMMs. Built on RMMs, two major data processing ingredients---LES designs and data fusion---are systematically studied.
1) LESs are high dimensional statistical indicators; with different test functions, LESs gain insight into the systems from different perspectives. Moreover, some theoretical values related to LESs are predictable as the reference points via the latest theorems.
Additionally, 3D power-map is developed to visualize the LES for the decision-making auxiliary functions. This visualization is sensitive to events---it is able to detect serious system faults, as well as some small fluctuations; moreover, the visualization is robust against bad data---even with data loss in the core area, we can still achieve the proper judgements.
2) Besides, data fusion, by putting together diverse data sources, provides us a comprehensive view towards systems. It is a deep research; we just give a brief mention.
In general, based on RMTs and the big data applying architecture \cite{he2015arch}, this paper presents a series of work associated with EED: the theorems of LESs, a briefly discussion about data fusion, the indicators derived from LESs, and the visualization of the results. Case studies, with both simulated data and real ones, validate the proposed method, and related theories and theorems.

\subsection{Related Work}
\subsubsection{Our Previous Work}
Paper \cite{he2015arch} is the first one; it is also the first attempt to apply big data analytics systematically to power systems. It provided a feasible architecture as the approach with two independent procedures---engineering procedure and mathematical one.
Random matrix theories (RMTs), such as Ring Law and M-P Law, were elaborated as the solid foundations; Mean spectral radius (MSR) was proposed to indicate the statistical correlations (now we know that MSR is a specific LES).
Specifically, moving split-window (MSW) technology was introduced to deal with data in temporal dimensions; in spatial dimensions, block calculation was presented. Simulated data in various fields were studied as cases.
Then we moved forward to the second stage. Paper \cite{he2015corr}, based on the first one (specially, Ring Law, M-P Law, the architecture, MSR, MSW, the simulated 118-buses system), talked about the correlation analysis approaches.
An augmented matrix was studied as the key ingredient. Actually, augmented matrix is a form of data fusion---it consists of two data sources with totally different sizes and meanings: the status one for the basic part, and the factor one for the augmented part.
As the same stage work, paper \cite{qiu2015MIMO}, using the 70 nodes network testbed, tested some data fusion ingredients based on RMTs: the product of non-hermitian random matrices, the geometric mean, the arithmetic mean, and the product of random Ginibre matrices. With the experimental data, the effectiveness of these data fusion ingredients, is validated for signal detection in Massive MIMO systems.

\subsubsection{Others' Work}
Data are a core resource and data management is the keypoint for EED in smart grids. There have been numerous discussions about utilizing PMUs to improve wide-area monitoring, protection and control (WAMPAC)  \cite {phadke2008wide,terzija2011wide,xie2012distributed}. Xu initiated power disturbance data analytics, and showed a wide scope in the future \cite {xu2013power}. The mathematical foundations, system frameworks, and applying architectures are missing yet. Recently, Xie proposed an early event detection algorithm based on principal component analysis (PCA) \cite {xie2014dimensionality}. However, it is a supervised learning method; the bad subspace (i.e., linear combinations of the labelled data from several empirical chosen PMUs named as pilot PMUs), caused by improper training procedure, will lead to bad results. Although many researches (especially those methods based on specific physical models) were done in various fields, little attention has been paid to the data-driven unsupervised learning methods, which are based on solid mathematical foundations and with universal statistical procedures. Additionally, related mathematical work is introduced in \textit{Sec \ROMAN 2}.

\section{Random Matrix Theories}

The nomenclature is given as Table \ref{tab:Nomenclature}.

\begin{table}[htbp]
\caption{Some Frequently Used Notations in the Theories}
\label{tab:Nomenclature}
\centering

\begin{tabularx}{0.48\textwidth} { l !{\color{black}\vrule width1pt} p{6.8cm} } 

\toprule[1.5pt]
\hline
\textit{Notations} & \textit{Meanings}\\

\Xhline{1pt}

\VX{},\Vx{},\Vsx,\Vsx{i\!,j} & a matrix, a vector, a single value, an entry of a matrix \\

\VRX{},\VRx{},\VRsx{} & hat: raw data\\
\VTX{},\VTx{},\VTsx{},\VTZ{} & tilde: intermediate variables, formed by normalization\\

\Data{N,T,c} & the numbers of rows and columns; \Equs{c}{N/T}\\
\VF {C}{N}{T} & \Muls{N}{T} dimensional complex space\\
\VX{u} &  the singular value equivalent of \VTX{}\\

\Vector S & covariance matrix of \VX{}: \Belong {\Equs {\Vector S}{\Div 1 N\VX{}\Her{\VX{}}}}{\VF CNN}\\
\Vector M & another form of covariance matrix of \VX{}: \Equs {\Vector M}{c{\Vector S}}\\

\VZ{},\Data{L} & \Data{L} independent matrices product: \Data {\VZ{}=\Prod{i=1}{L} \VX {u,i}}\\

\Vlambda{\Vector S}, \Vlambda{\VTZ{}}, \Vlambda{\Vector M} & the eigenvalue of matrix \Vector S, \VTZ{}, \Vector M\\

\Vlambda{\Vector S,i} & the \Data i-th eigenvalue of matrix \Vector S\\
\Data{r} & the circle radius on the complex plane of eigenvalues\\

\VLES{}{} &  linear eigenvalue statistics\\

\VLES{MSR}{} &  mean value of radius for all eigenvalues of \VTZ{}: \Mu{\Vector r_{\Vlambda{\VTZ{}}}}\\
$\varphi,\widehat{\varphi}$&the test function and its Fourier transformation\\

\VsX{} & random variable\\
\Fdata {E}(\VsX{}),\Fdata {D}(\VsX{})& expectation, variance for \VsX{}\\
\Mu{\Vx{}},\Data{\Sigg{\Vx{}}} & mean, variance for \Vx{}\\
$X^\circ$ & $ X-$\STE{X} \\

$\kappa_i$ & $i$-th cumulant of a random variable \VsX{}\\
${\left[ {\zeta \left( \theta  \right)} \right]_{\theta  = {\theta _2}}^{\theta  = {\theta _1}}}$ &  $\zeta \left( {{\theta _1}} \right) - \zeta \left( {{\theta _2}} \right)$\\

\toprule[1pt]

\end{tabularx}

\end{table}

\subsection{From Physical System to Random Matrix}
For a system, we assume \Data {t} times observation for \Data {n}-dimensional vectors \DatassI {\VRx{}}{t} (\Belong{\VRx{j}}{\VF{C}{n}{1}}, \Equs{j}{1,\!\cdots\!,t}), and a data source, denoted as \VOG{} (in size of $n \times t$), is obtained.
\VOG{} is in a high-dimensional space but not an infinite one (Or more explicitly, we are interested in the practical regime in which \Equs{\Data {n}}{}{100--10000}, and \Data {t} is sufficient large); this disables most classical tools. In contrast, RMTs enable us to select arbitrary data---both in temporal dimensions (e.g. \Data T from \Data t) and in spatial dimensions (e.g. \Data N from \Data n)---to form  \Belong {\VRX {}} {\VF CNT} naturally; {\VRX {}} is a random matrix due to the presence of ubiquitous noises. Furthermore, we can convert \VRX {} into a normalized matrix {\VTX{}}  row-by-row (see (\ref{eq:StdMatrix})); thus, the random matrix model (RMM) is built to map the system.

\normalsize{}In our previous work \cite{he2015arch, he2015corr, qiu2015MIMO}, we have already elaborated RMTs, Ring Law, M-P Law, and MSR; here we just give some key results\footnote{Although the asymptotic convergence in RMTs is considered under infinite dimensions, the asymptotic results are remarkably accurate for relatively moderate matrix sizes such as tens. This is the very reason why RMTs can handle practical massive systems.}. Whereas, the data processing ingredients, including linear eigenvalue statistic (LES) designs and data fusion, are the major parts; they are elaborated in \textit{Sec \ROMAN 3}.

\subsection{Random Matrix Theories (RMTs)}
RMTs are effective to power systems, which has already been validated \cite{he2015corr}. First, we assume that:
for the rectangular \Muls {N}{T} random matrix \VX{}, the entries are independent identically distributed (i.i.d.) variables, satisfying the conditions:
 \begin{equation}
 \label{eq:randomX}
 \Equs {\STE{\VsX{i,j}}}{0}, \quad  \Equs {\STE{{\VsX{i,j}}{\VsX{m,n}}}}{\delta_{i,m}\delta_{j,n}{\sigma^2}}
 \end{equation}
 \small{where $\sigma$ is the variance, and ${\delta_{\alpha,\beta}}$ is the Kronecker Delta Function:}
 \normalsize{}
\[
\delta_{\alpha,\beta}=
\begin{cases}
    1 & \alpha=\beta \\
    0 & \alpha\neq\beta
\end{cases}
\]

\subsubsection{Ring Law}
{\Text{\\}}

Consider a $L$ independent matrices product \Data {\VZ{}=\Prod{i=1}{L} \VX {u,i}}, where \Belong{\VX u}{\VF CNN} is the singular value equivalent \cite{ipsen2014weak} of \VTX{} (see (\ref {eq:Xu})); \VTX{} is obtained directly from raw data \VRX{} (see (\ref{eq:StdMatrix})). Furthermore, the matrices product \VZ{} is converted into \VTZ {} (see (\ref{eq:StdZ})). Thus, the empirical spectrum density (ESD) of \VTZ {} converges almost surely to the same limit given by

\FuncCC {eq:RingLaw}
{\Funcfzx {\rho}{\Data{r\!i\!n\!g}}{\Vlambda {}}}
{ \Div{1}{\pi{}c\Data {L}}{\Power{\Abs{\lambda}}{(2/\Data {L}-2)}}}  {{\Text {, }} \Power {(1-c)}{\Data {L}/2} \LE \Abs{\lambda} \LE \Data 1   }
{0}   {\Text {, otherwise}}
as \Trend {N,T}{\INF} with the ratio \Equ {c}{\Belong {N/T} {(0,1]} }.

\subsubsection{Marchenko-Pastur Law (M-P Law)}
{\Text{\\}}

\normalsize{}M-P Law describes the asymptotic behavior of the the covariance matrix:
\begin{equation}
\label {eq:Sc}
\Belong {\Equs {\Vector S}{\Div 1 T\VX{}\Her{\VX{}}}}{\VF CNN}
\end{equation}
\small{where \VX{}  is the rectangular \Muls {N}{T}  non-Hermitian random matrix satisfying condition (\ref{eq:randomX}).}\normalsize{}

Then, the ESD of \Vector S converges to the distribution of M-P Law \cite{qiu2012bookcogpp, marvcenko1967distribution} with density function:
\FuncCC {eq:MPLaw}
{\Funcfzx {\rho}{\Data{\!m\!p}}{\Vlambda{}}}
{ \Div{1}{2\pi{}\lambda c{\sigma^2}}\Sqrt{ (a_+-\lambda)(\lambda-a_-)} }{{\Text {, }} a_-\LE \lambda \LE a_+ }
{0}  {\Text {, otherwise}}
\small{where \Equs {a_\pm}{\sigma^2(1\pm\Sqrt c)^2}.}\normalsize{}

\subsubsection{Mean Spectral Radius (MSR)}
{\Text{\\}}

MSR is a high-dimensional indicator. For a specific matrix \VTZ{} (as described in \textit{Ring Law} part above), we can calculate the eigenvalues \Vlambda{\VTZ{}} on the complex plane. The mean value of all these eigenvalues' radii length is denoted as  {\VLES{MSR}:
\FuncC {eq:MSR}{
\Equs {\VLES{MSR}{}}{ \Sum {i=1}{N}{\Div{1}{N}\Abs {\Vlambda{\VTZ{},i}} }}
}

\section{Data Processing Ingredients}
\subsection{Linear Eigenvalue Statistics and Related Research}
{\Text{\\}}
\subsubsection{Definition}
{\Text{\\}}

 The linear eigenvalue statistic (LES) of a matrix \Belong{\VX{}}{\VF CNN} is defined via the continuous test function  \Data{\varphi: \Fdata R \rightarrow \Fdata C}  \cite{lytova2009clrforles,shcherbina2011central}

\begin{equation}
\mathcal{N}_N[\varphi]=\Sum{i=1}{N}{\varphi({\lambda_i})}
\end{equation}

\subsubsection{Law of Large Numbers}
{\Text{\\}}

The law of Large Numbers is the first step in studies of eigenvalue distributions for a certain random matrix ensemble. The result, for the Wigner ensemble, obtained initially in \cite{wigner1958distribution}, was improved in \cite{pastur1972spectrum}, where the Stieltjes transformation was introduced and the famous semicircle law was shown under the minimal conditions on the distribution of \Vector W (the Lindeberg type conditions) \cite{shcherbina2011central}.
The law of Large Numbers tells us that $N^{-1}\mathcal{N}_N[\varphi]$ converges in probability to the limit
\begin{equation}
\label{LES1}
\lim_{N \to \INF}\Div 1N\mathcal{N}_N[\varphi]\!=\!\int\varphi(\lambda)\rho(\lambda)\,d\lambda
\end{equation}
\small{where $\rho(\lambda)$ is the probability density function (PDF) of the eigenvalues.}\normalsize{}
In particular, for the Gaussian orthogonal ensemble (GOE) \cite{lytova2009clrforles} (see  (\ref{eq:GOE1}),(\ref{eq:GOE2}), and (\ref{eq:GOE3}) in the appendix), $\rho(\lambda)$ is according to the semicircle law:
\FuncCC {eq:semicirlce density}
{\rho_{sc}(\lambda)}
{\Div{1}{2\pi\omega^2}\sqrt{4\omega^2-\lambda^2}}{\lambda^2<4\omega^2}
{0}  {\lambda^2{\GE}4\omega^2}

 The covariance matrix ensemble is another classical type ; M-P Law, as describe in \textit{Sec \ROMAN 2}, is adapted to this ensemble.
 The covariance matrix is widely used in engineering due to the rectangular form---we can also study  the RMM \Belong{\VX {}}{\VF CNT} with $N\!\ne{\!T}$. We will further discuss this ensemble below.

\subsubsection{Central Limit Theorems (CLTs)}
{\Text{\\}}

Central Limit Theorems (CLTs), as the natural second step, aim to study the LES fluctuations. Lots of papers devote to proofs of CLTs for different random matrix ensembles  (see \cite{johansson1998fluctuations,guionnet2002large,bai2004clt,anderson2006clt,lytova2009clrforles,shcherbina2011central,pan2011universality}). CLTs for LESs with polynomial test functions of some generalizations for the Wigner and covariances matrices were proved in \cite{anderson2006clt} via moment methods.
In contrast, CLTs for LES with real analytic test functions of the Wigner and covariances matrices were established in \cite{bai2004clt} under additional assumptions that
\[
\left|
\begin{aligned}
    & \Equs {\STE{{W_{i,i}}^2}}{2}, \Equs {\STE{{W_{i,j}}^4}}{\Equs {3\Fdata{E}^2({W_{i,j}}^2)}{3}} & \text{Wigner}\\
   & \Equs {\STE{{X_{i,j}}^4}}{3\Fdata{E}^2({X_{i,j}}^2)} & \text{Covariance}
\end{aligned}
\right.
\]

 In the recent paper \cite{lytova2009clrforles}, CLTs for LESs of the Wigner and covariances matrices were proved under assumptions that \Equs {\STE{{W_{i,i}}^2}}{2}, the third and the forth moments of all entries are the same, but $ {\Fdata{E}({W_{i,j}}^4)}$ is not necessary $3$. Moreover, the test functions are not supposed to be real analytic. It was assumed that the Fourier transformation $\widehat{\varphi}$  satisfies the inequality
\[
\int ({1+|k|^5})|\widehat{\varphi}(k)|\,dk<\INF
\]
which means that $\varphi$ has more than 5 bounded derivatives.

\subsection{CLT for Covariance Matrices}
\Text{\\}

Parallel to (\ref{eq:Sc}), we study another typical covariance matrix:
\begin{equation}
\label {eq:Sc2}
\Belong {\Equs {\Vector M}{\Div 1 N\VX{}\Her{\VX{}}}=\Div {1}{c}\Vector S}{\VF CNN}
\end{equation}
its PDF is also according to M-P Law:
\FuncCC {eq:MPLaw2}
{\Funcfzx {\rho}{\Data{\!m\!p}_2}{\Vlambda{}}}
{ \Div{1}{2\pi{}\lambda {\sigma^2}}\Sqrt{ (a_+-\lambda)(\lambda-a_-)} }{{\Text {, }} a\LE \lambda \LE b }
{0}  {\Text {, otherwise}}
\small{where \Equs {a_\pm}{\sigma^2(1\pm1/\Sqrt c)^2}, and  $\sigma$ is the variance.}\normalsize{}

The CLT for {\Vector M} is given as follows \cite{shcherbina2011central}:
\normalsize{}
\newtheorem{thm1}{Theorem}[section]
\begin{thm1}[M. Sheherbina, 2009]
Consider a rectangular \Muls {N}{T}  non-Hermitian random matrix \VX{}, with entries \VsX{i,j} satisfying the condition (\ref{eq:randomX}); \Vector M is the covariance matrix  (see (\ref{eq:Sc2})). Let the real valued test function $\varphi$ satisfy condition ${{\left\| \varphi  \right\|}_{3/2+\varepsilon }}<\infty   \left( \varepsilon >0 \right)$. Then ${\mathcal{N}_N}^\circ[\varphi]$, in the limit $N,T\to \infty , \Equs {c}{N/T}\le 1$, converges in the distribution to the Gaussian random variable with zero mean and the variance:
\small{}
\begin{equation}
\label {eq:CLTforLes}
\begin{aligned}
     {{V}_{SC}}\left[ \varphi  \right]=    &    \frac{2}{c\pi^2 }\iint\limits_{-\frac{\pi }{2}<{{\theta }_{1}},{{\theta }_{2}}<\frac{\pi }{2}}{{{\psi }^{2}}\left( {{\theta }_{1}},{{\theta }_{2}} \right)}\left( 1-\sin {{\theta }_{1}}\sin {{\theta }_{2}} \right) d{{\theta }_{1}}d{{\theta }_{2}} \\
 &                +\frac{{{\kappa}_{4}}}{{\pi }^{2}}\left( \int_{-\frac{\pi }{2}}^{\frac{\pi }{2}}{\varphi \left( \zeta \left( \theta  \right) \right)\sin \theta  d{{\theta }}} \right)^2 \\
\end{aligned}
\end{equation}
\small{where $\psi \left( {{\theta }_{1}},{{\theta }_{2}} \right)=\frac{\left[ \varphi \left( \zeta \left( \theta  \right) \right) \right]_{\theta ={{\theta }_{2}}}^{\theta ={{\theta }_{1}}}}{\left[ \zeta \left( \theta  \right) \right]_{\theta ={{\theta }_{2}}}^{\theta ={{\theta }_{1}}}}$, and $\zeta \left( \theta  \right) = 1 + 1/c + {2}/{\sqrt c} \sin \theta$; $\kappa_4=\mathbb{E}\left( {{X}^{4}} \right) -3$ is the $4$-th cumulant of entries of \VX{}.}\normalsize{}
\end{thm1}
\normalsize{}

\subsection{LES Designs and Theoretical Values}

For Gauss variable $X$, \Equs {\STE{X}}{0}, \Equs {\STE{X^2}}{1}, and \Equs {\STE{X^4}}{3} (see (\ref{eq:E124})). A typical scenario is assumed:
\Equs {N}{118} and \Equs {T}{240}, thus $c=N/T=0.4917$;

\subsubsection{LES for Ring Law}
\Text{\\}

MSR (see \textit{Sec \ROMAN 2}) is a special LES\footnote{Since $\lambda_{\widetilde{\mathbf Z},i}$ are highly correlated random variables (each one is a complicated function of the random matrices $\widetilde{\mathbf X}_i~(i\!=\!1,2,\!\ldots\!,L)$), $\tau_{\text{MSR}}$ is a random variable.}; it is defined as follows:
\begin{equation}
\tau_{\text{MSR}}=\sum_{i=1}^N \frac{1}{N}\vert \lambda_{\widetilde{\mathbf Z},i}\vert
\end{equation}
\small{where $\lambda_{\widetilde{\mathbf Z},i}~(i\!=\!1,2,\!\ldots\!,N)$ are the eigenvalues of $\widetilde{\mathbf Z}$, and $\vert\lambda_{\widetilde{\mathbf Z},i}\vert$ is the radius of $\lambda_{\widetilde{\mathbf Z},i}$ on the complex plane.}\normalsize{}

According to (\ref{LES1}), the theoretical expectation of \Data{r} when $N \to \INF$ (\STE{\VLES{MSR}{}}), are calculated as follows:
\begin{equation}\label{eq:ExpMSR}
\begin{aligned}
\STE{\VLES{MSR}{}}={\STE{r}}
&\Data{= \iint_{\text {Area}}\Fdata {P}(r){\times}r{\cdot}r\,{\rm d}r   {\rm d}\theta}\\
&\Data{= \int^{2\pi}_0\!\int^1_{\sqrt{1-c}}   \Div{1}{c\pi}r{\cdot}r  {\rm d}r \;  {\rm d}\theta}=0.8645\\
\end{aligned}
\end{equation}
\small{where \Data{\Fdata {P}(r)} is given in formula (\ref{eq:RingLaw}), and \Equs{c}{0.4917} for this scenario.}\normalsize{}

Also, we can calculate the theoretical variances (\STD{\VLES{MSR}{}}):
\begin{equation}\label{eq:VarR}
\begin{aligned}
\Data{\Fdata {E}(r^2)}&\Data{= \iint_{\text {Area}}\Fdata {P}(r){\times}r^2{\cdot}r\,{\rm d}r   {\rm d}\theta}=0.7542\\
\STD{\VLES{MSR}{}}&\Data{=\Fdata {D}(r)} \Data{= \Fdata {E}(r^2)-\Fdata {E}(r)^2=0.0068}\\
\end{aligned}
\end{equation}

\subsubsection{LESs for Covariance Matrices}
\Text{\\}

1. Chebyshev Polynomials ${\varphi(\lambda)=T_2=2x^2-1}$
\begin{equation}\label{eq:LEST2}
\tau_{T_2}=\sum_{i=1}^N(2{\lambda_i}^2-1)
\end{equation}

For the scenario, according to (\ref{LES1}) and (\ref{eq:MPLaw2}), we get \STE{\tau_{T_2}}:
\begin{equation}
\begin{aligned}
\STE{\tau_{T_2}}  &=N\!\int{\varphi(\lambda)}\rho_{mp2}(\lambda)\,d\lambda=6600\\
\end{aligned}
\end{equation}
and according to (\ref{eq:CLTforLes}), we get \STD{\tau_{T_2}}:
\begin{equation}
\STD{\tau_{T_2}}=1080
\end{equation}

Similarly, we can design other test functions to obtain diverse LESs as the indicators, and some theoretical values; here we list some classical test functions:

2. Chebyshev Polynomials: $T_3=4x^3-3x$

3. Chebyshev Polynomials: $T_4=8x^4-8x^2+1$

4. Determinant: $\text{DET}=\ln(x)$

5. Likelihood-ratio function: $\text{LRF}=x-\ln(x)-1$

In \textit{Sec \ROMAN 5}, we will show their applications in smart grids.

\subsection{Universality Principle}
Akin to CLTs, universality \cite{qiu2013bookcogsen} refers to the phenomenon that the asymptotic distributions of various covariance matrices (such of eigenvalues and eigenvectors) are identical to those of Gaussian covariance matrices. These results let us calculate the exact asymptotic distributions of various test statistics without restrictive distributional assumptions of matrix entries. The presence of the universality property suggests that high-dimensional phenomenon is robust to the precise details of the model ingredients \cite{van570probability}. For example, one can perform various hypothesis tests under the assumption that the matrix entries not Gaussian distributed but use the same test statistic as in the Gaussian case.

The data of real systems can be viewed as a spatial and temporal sampling of the random graph. Randomness is introduced by the uncertainty of spatial locations and the system uncertainty. Under real-life applications, we cannot expect the matrix entries follow i.i.d. distribution. Numerous studies based on both simulations \cite{he2015arch} and experiments, however, demonstrate that the Ring Law and M-P Law are universally followed. In such cases, universality properties provide a crucial tool to reduce the proofs of general results to those in a tractable special case---the i.i.d. case in our paper.

\subsection{Data Fusion}
\Text{\\}

Data fusion (including the augmentation, the blocking, the sum, and the product of matrices) is another data processing ingredient.
Comparing to the LES designs, which aim to define the LES $\tau$ via the test functions $\varphi(\lambda)$ for a determinate $\VX{}$, data fusion manages to handle multiple data sources (i.e. $\VX{1},\VX{2},\cdots$), even with diverse features (e.g. in totally different size).
The theories about data fusion are deep and novel: {G{\"o}tze}, {K{\"o}sters}, and et al, in \cite{Got2014Datafusion, Got2014Datafusion2, Koster2015Datafusion}, have already studied the performance of the matrices in the form of :
\[
\Vector F_n^{(1)}+\cdots+\Vector F_n^{(m)}
\]
\small{where $\Vector F_n^{(i)}=\Vector X_n^{(i,1)}\cdots \Vector X_n^{(i,k)}$,
and $\Vector X_n^{(0,0)},\cdots,\Vector X_n^{(i,j)},\cdots,\Vector X_n^{(m,k)}$ are independent $\Muls nn$ matrices with independent entries.}\normalsize{}

We will not go so far and just show some typical applications; data fusion is very common and meaningful in engineering.
Our previous work  \cite{he2015corr} conducted data fusion as follows:
the status data source (a matrix \Vector B in size of $\Muls NT$), and the factor data source (a vector \Vector c in size of $\Muls 1T$), in a certain manner, are put together  to form the augmented matrix \Vector A (\Equs {\Vector A}{[\Vector B { ; } \Vector C]}); this \Vector A, as a new random matrix model, is used to conduct correlation analysis. Besides, Zhang \cite{qiu2015MIMO}, using the data from a 70 nodes network testbed, validated the data fusion in the field of signal detection for Massive MIMO systems.

\section{Unsupervised Learning Method}
This section makes a comparison between the unsupervised learning methods (e.g. Ring Law Analysis) and the supervised ones (e.g. PCA) from diverse perspectives.
\subsection{Procedure of Ring Law Analysis}

The Ring Law Analysis conducts EED with following steps:

\begin{table}[H]
\centering

\begin{tabular*}{8.8cm} {p{8.6cm}}
\toprule[1.5pt]
\textbf {Steps of Ring Law Analysis} \\
\toprule[0.5pt]
1) Select arbitrary raw data (or all available data) as the data source \Vector \Omega\\
2) Forming random matrix model \VRX{} at a certain time \Index {t}{i}; \\
3) Obtain \VTZ{} by variable transformations (\Data{\VRX{}\rightarrow\VTX{}\rightarrow\VX{u}\rightarrow\VZ{}\rightarrow\VTZ{}});\\
4) Calculate eigenvalues \Vlambda{\VTZ{}} and plot the Ring on the complex plane;\\
5) Conduct high-dimensional analysis;\\
\quad 4a) Observe the experimental ring and compare it with the reference one;\\
\quad 4b) Calculate \VLES{MSR}{} as the experimental value;\\
\quad 4c) Compare \VLES{MSR}{} with the theoretical value \Data{N}{\STE{r}};\\
6) Repeat 2)-5) at the next time point (\Equs {\Index {t}{i}}{\Index {t}{i}+1});\\
7) Visualize \VLES{MSR}{} on the time series;\\
8) Make engineering explanations.\\

\toprule[1pt]
\end{tabular*}
\end{table}

\textit{Steps 2)--7)} conduct high dimensional analysis only using raw data, and then visualize the indicator \VLES{}{}; they are unsupervised statistical proceedings without assumptions and simplifications.
In \textit{step 2)},  for different purposes, arbitrary raw data, even ones from distributed nodes or intermittent time series, are able to be focused on to form the RMM \VRX{}. It is also an online data-driven method requiring no knowledge of the physical models/topologies. In addition, the size of \VRX{} is controllable during \textit{step 2)}; it relieves the curse of dimensionality in some ways.

\subsection{Procedure of Principal Component Analysis (PCA)}
PCA is a prevailing data-driven method \cite{xie2012distributed}.
Xie proposed a dimensionality reduction method based on PCA as a approach to EED in smart grids \cite {xie2014dimensionality}; the steps are listed as follows:

\begin{table}[H]
\centering

\begin{tabular*}{8.8cm} {p{8.6cm}}
\toprule[1.5pt]
\textbf {Steps of PCA} \\
\toprule[0.5pt]

\hangafter1\hangindent1.2em\noindent 1) Select data \Equs {\VY{}}{[\, {\DatassI{\Vy{}}{N}}]} \!\Belong \! {}{\VF{C}{n}{N}}, \Equs {\Vy{i}}{[\, {\!\Vsy{1,i},\!{\Vsy{2,i}},\!\cdots,\!\Vsy{n,i}}]^T};   \\

\hangafter1\hangindent1.2em\noindent 2) \Equs{\VIndex{C}{\VY{}}}{\VY{}^{\!H}\VY{}}\Belong {}{\VF{C}{N}{N}}, calculate \Eig{\VIndex{C}{\VY{}}}; \\

\hangafter1\hangindent1.2em\noindent 3) Rearrange and select the top \Data {m} eigenvalues: \Data{\lambda_1{\rightarrow}\VIndex p1,\cdots,\lambda_m{\rightarrow}\VIndex pm};    \\

\hangafter1\hangindent1.2em\noindent 4) Form \Data m dimensional principal component subspace \Fdata L\Data {(\,\DatassI{\Vector p}{m})} and project the original \Data N variables onto it:
 Select \Data {m' \LE m } vector-based variables as the pilot PMUs from \Data N PMUs to form the linear basis matrix \Equs{\VY{\!B}}{[\, {\DatassI{\Vy{b}}{m'}}\,]}\Belong {}{\VF{C}{n}{m'}}. The selected \Data{m'} variables  should be as orthogonal to each other as possible, which means \Data{\min\Equs{(\cos {\theta})}{(\Vy{bi}{\cdot}\Vy{bj})/(\Abs{\Vy{bi}}\Abs{\Vy{bi}})} \quad  (\Equs{i,j}{1\!,2\!,\!\cdots\!,m'; \  i\!\ne\!j}) }\\

\hangafter1\hangindent1.2em\noindent 5) Represent non-pilot PMUs \Vy{ci}  \ (\Equs{i}{1\!,2\!,\!\cdots\!,N\!-\!m'}) for training :  Let \Equs{\Vv{ci}}{[\, {\!\Vsv{1,ci},\!{\Vsv{2,ci}},\!\cdots,\!\Vsv{m',ci}}\,]^T} be the vector of regression coefficients:
 \[ \Data{\Vy{ci} = \langle  (\underbrace{{\DatassI{\Vy{b}}{m'}}}_{\Text{Basis}})\cdot({\!\Vsv{1,ci},\!{\Vsv{2,ci}},\!\cdots,\!\Vsv{m',ci}} )           \rangle \Equs{}{\VY {\!B}\Vv {ci}}  } \]
\Data{\Rightarrow   \Vv{ci}\Equs{}{(\VY{\!B}^{\!H}\!\VY{\!B})^{\!-\!1}\VY{\!B}^{\!H}\Vy{ci}}         }; \\
\hangafter1\hangindent1.2em\noindent 6) Train at \EtS{\!s\!-\!1}{}:
\[ \Data{   \Vv{ci}^{(\!s\!-\!1)}\Leftarrow f(\VY{\!B}^{(\!s\!-\!1)}, \Vy{ci}^{(\!s\!-\!1)} ) }; \]\\
\hangafter1\hangindent1.2em\noindent 7) Judge at \Et{s}{}:
\[\Data{\lVert \Vy{ci}^{\!s},\VY{\!B}^{\!s}\Vv{ci}^{(\!s\!-\!1)} \rVert     }.\]\\

\toprule[1pt]
\end{tabular*}
\end{table}

\textit{Step 6)} and \textit{step 7)} are the executive parts based on data processing procedure---\textit{steps 1)-5)}. The key steps are \textit{step 4)} and  \textit{step 5)}; they constitute the training procedure. By choosing  \Data{m'} PMUs (as pilot PMUs) from total \Data{N}  PMUs, the procedure tags the system in a reduced subspace (\Data{N{\rightarrow}m'}) ; in this way, the function \Vv{ci} is inferred.

\subsection{Comparison of Supervised and Unsupervised Methods}

The supervised learning methods are prevailing among massive systems. The key idea for these methods is to tag the systems---by labelled parameters, principal eigenvectors, inferred functions, etc; the detailed steps, by dealing with prior models and data in a low dimensional space, make up an empirical training procedure. In a word, supervised learning methods are of 'labelling' steps based on assumptions and simplifications. For supervised learning methods, there are some problems, essentially, hard to be solve:\\\small{}
a). The error accumulations, spurious correlations, incidental correlations are unavoidable as systems grow large; there does not exist a solid mechanism to eliminate or to relieve them.\\
b). The training procedure is not strict. Taking paper \cite {xie2014dimensionality} for an example, it chooses the pilot PMUs in an unconvinced way: the most unrelated ones, rather than the best suitable ones, are chosen. Some additional devices, especially the empirical ones, are essential as mentioned in the paper---e.g., the PMUs of some topologically and physically significant buses should be pilot ones, while the PMUs which are historically eventful should not.
Additionally, the improper setting of the pilot PMUs total number \Data {m'}, or the pre-specified variance threshold \Data {\varsigma}, will make the result worse.\\
c). The bad data in the core area (e.g. the incomplete, the inaccurate, and the unavailable data), or the worst situation---loss of all the data of the event area, will almost disable the methods.\\
d). Moreover, it is impossible to train over all the events or scenarios. There must be something unexpected in the large scale systems, even of which we can not give a proper description in low dimensions. Just a simple intuitive example: we can easily deduce \Equs{\!-\Vy{ci}}{\!-\VY{B}\Vv{ci}} from \Equs{\Vy{ci}}{\VY{B}\Vv{ci}}. Whereas, it is hard to make a verdict that the simultaneous reverse of the pilot PMUs data \VY{B} and the non-pilot PMUs data \Vy{ci} is an anomaly or not.
\normalsize{}

In general, traditional model-based methods or data-driven supervised learning methods are in low dimensions; they highly depend on only a few parameters related to physical models, subjective hypotheses, empirical causality, training procedure, etc.

On the other hand, data-driven unsupervised ones are usually based on high-dimensional statistical models which are built on solid mathematical foundations.
In other words, the related theories, algorithms, and statistics (e.g. RMTs and LESs) are always based on probability and designed in a multi-dimensional space.
Thus, the unsupervised methods tend to analyze the interrelations and interactions (seen as correlations) directly using the raw data without any label;
they are fast, sensitive, and universal.
Moreover, they are pure statistical data processing in high-dimensions which will not bring in systematical errors.
It comes to the conclusion that the unsupervised learning methods, with advantages described as above, perform better than the supervised ones.

\section{Case Studies}
\normalsize{}In this section, we use both simulated data and real ones to validate the proposed approach. For the simulated case, we adopt the standard IEEE 118-bus system \cite{ni2007new} (shown as Figure \ref{fig:IEEE118network}). Detailed information about the simulation is referred to the \emph {case118.m} in \textit{Matpower package} and \textit{Matpower 4.1 User’s Manual} \cite{MATPOWER2011matpower}.
There are generally two scenarios in the system: 1) only white noises, e.g., small random fluctuations of loads and Gaussian sample errors; 2) Signals plus noises, it means that there are also sudden changes, or even serious faults.
For the real case, we use a 48-hours database of some power grid in China.
Above all, the EED may be modeled as binary hypothesis testing: normal hypothesis $ {\cal H}_0 $ (no signal present) and abnormal hypothesis $ {\cal H}_1 $ (signal  present).

\subsection{Simulated 118-bus System}
\normalsize{}According to \textit{From Physical System to Random Matrix} in \textit{Sec \ROMAN{2}}, the data source $\Omega_{\Vector V}={\Rdata v_{i,j}}\Belong{}{\VF R{118}{1500}}$ (\Equs{n}{118}, \Equs{t}{1500}) is obtained to map the simulated system.

\subsubsection{Conduct EED Based on Ring Law, M-P Law, MSR}
\Text{\\}

\normalsize{}
For details about this work, see our previous work \cite{he2015arch}. Here, we assume the events as Table \ref{Tab: Event Series}; Figure \ref{fig:Case 118} shows the results.

\begin{figure}[htbp]
\centering
\subfloat[Ring Law at \EtS{600}{s}]{
\label{RMa}
\includegraphics[width=0.23\textwidth]{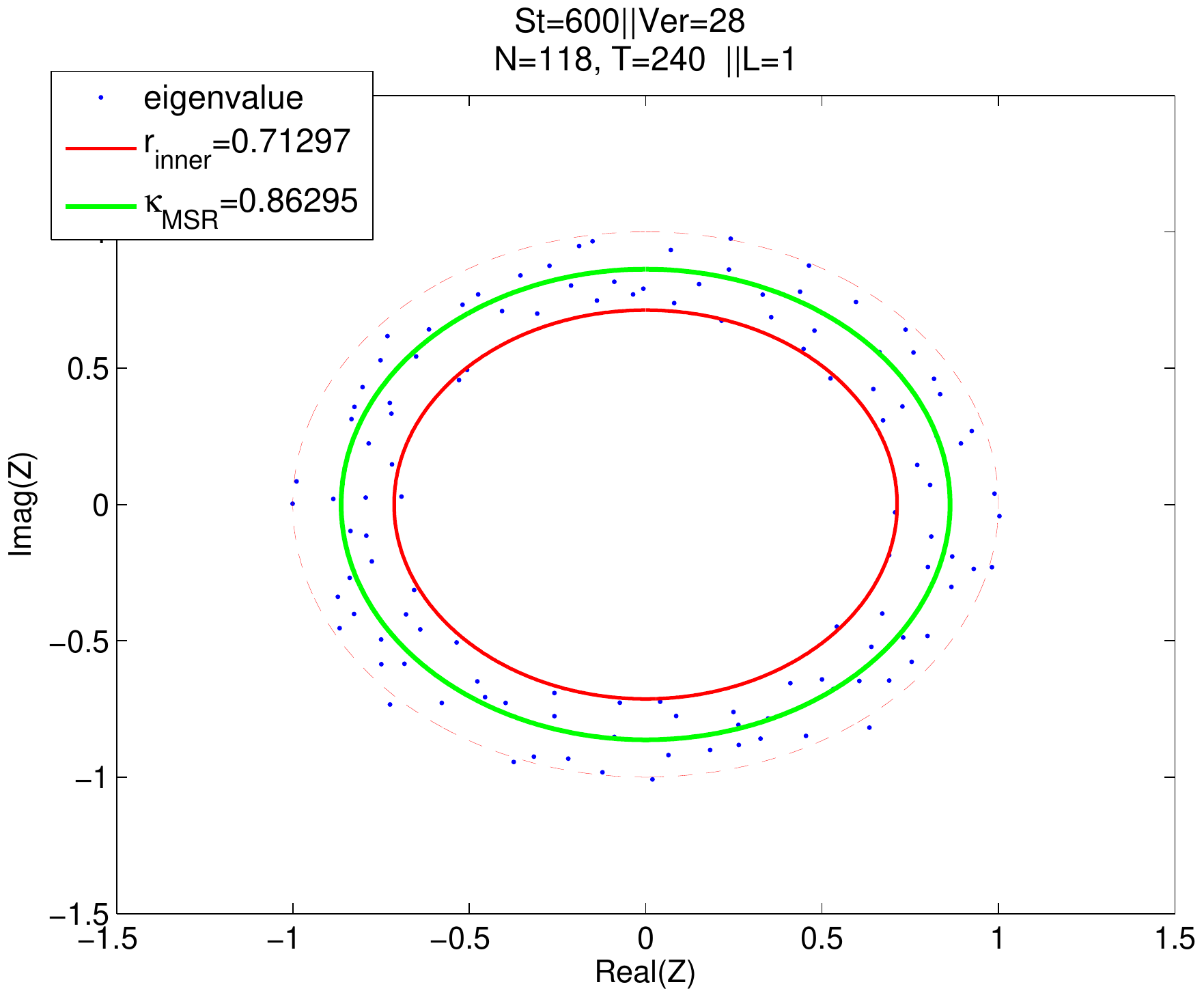}
}
\subfloat[Ring Law at \EtS{601}{s}]{
\label{RMb}
\includegraphics[width=0.23\textwidth]{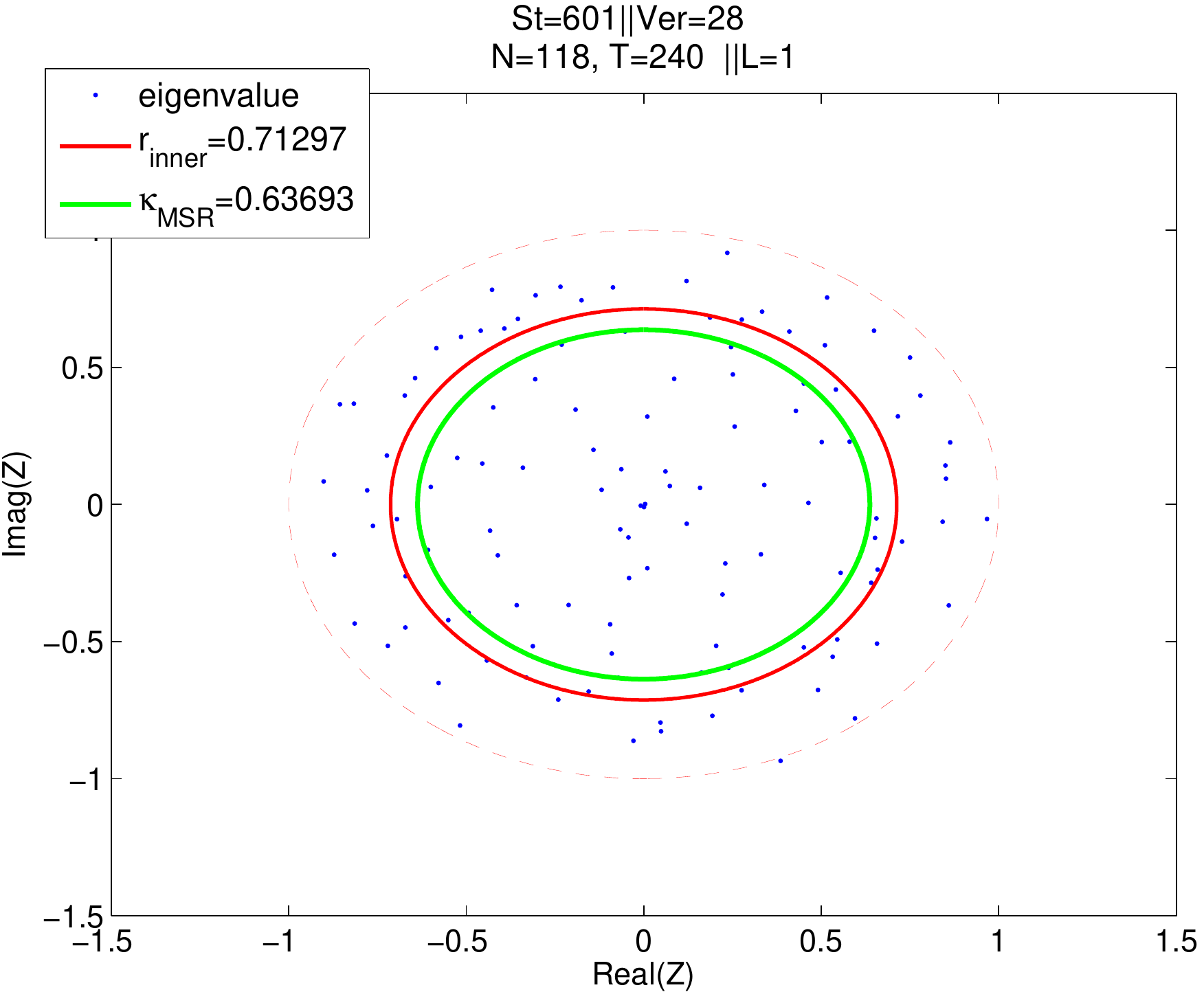}
}

\subfloat[M-P Law at \EtS{600}{s}]{
\label{RMc}
\includegraphics[width=0.23\textwidth]{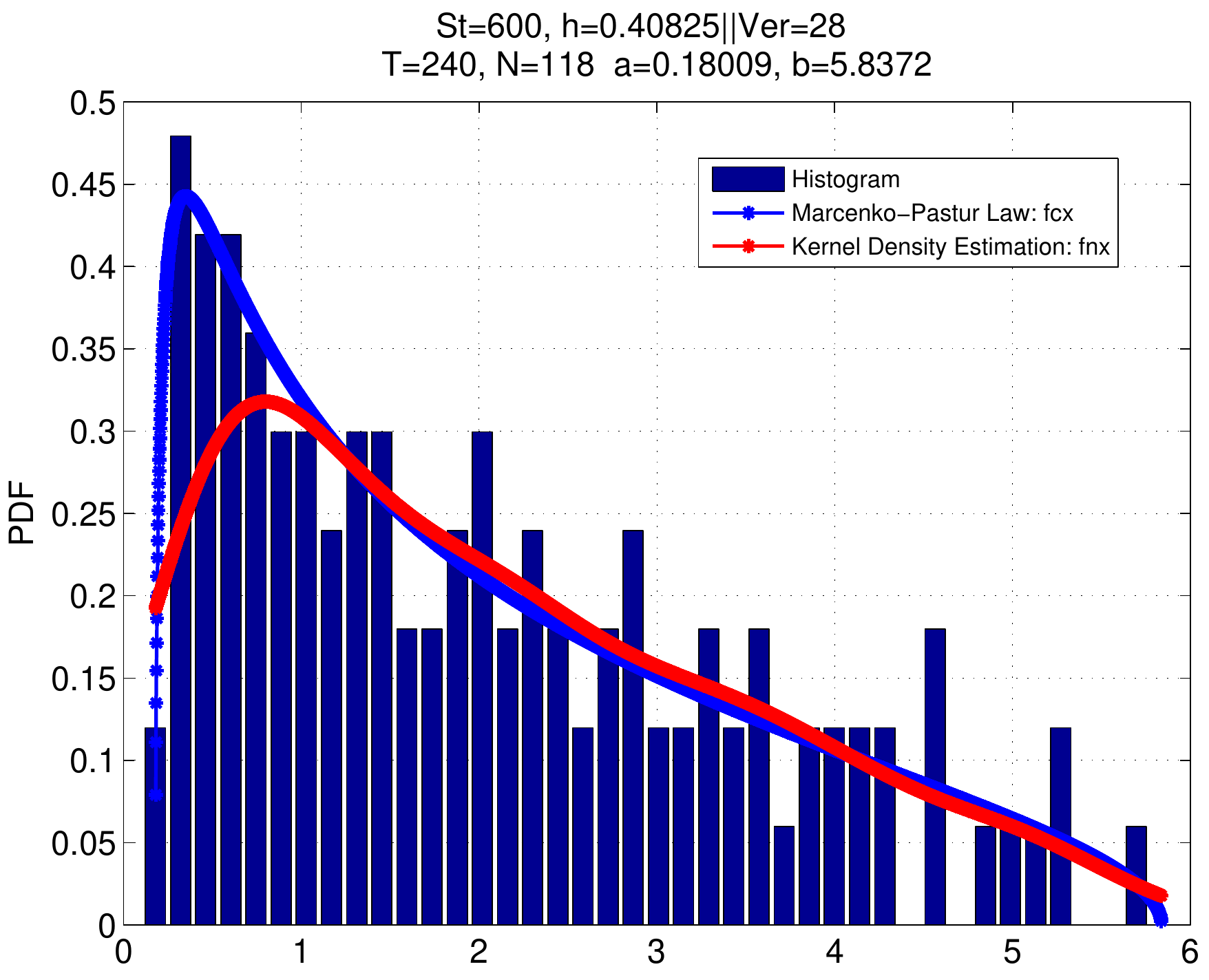}
}
\subfloat[M-P Law at \EtS{601}{s}]{
\label{RMd}
\includegraphics[width=0.23\textwidth]{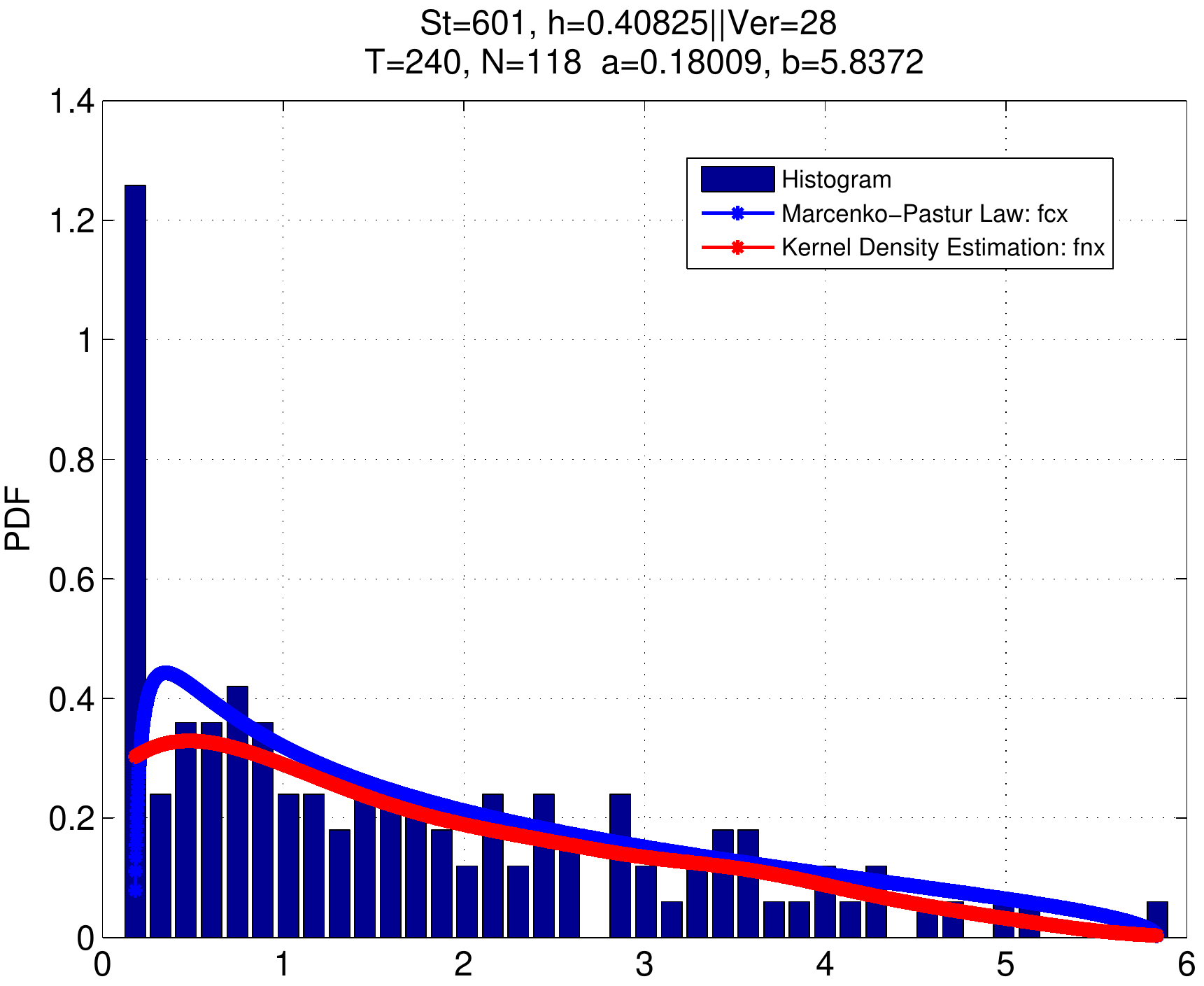}
}

\subfloat[MSR on Time Series]{
\label{RMe}
\includegraphics[width=0.46\textwidth]{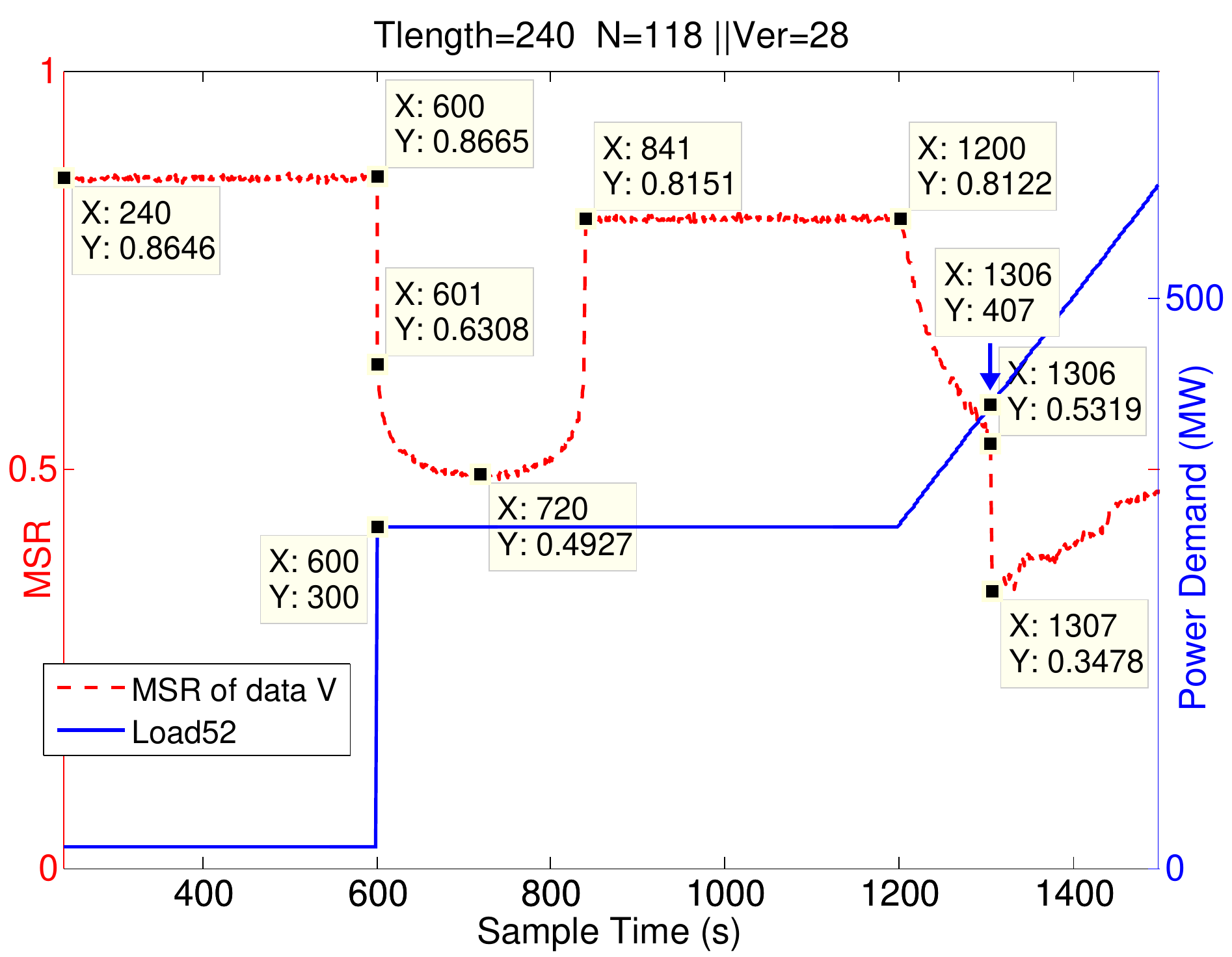}
}
\caption{Ring Law, M-P Law and MSR for Simulated Data \Vector V}
\label{fig:Case 118}
\end{figure}
}

\normalsize{}At sampling time \EtS{600}{s}, the RMM \VRX{} includes a time period \EAT {361\!:\!600}{s}; the noises play a dominant part during the period.
Figure \ref{RMa} shows that the distribution of \Vlambda{\VTZ{}} is more closely to the reference ring (\Data {L=1}); and Figure \ref{RMc} shows that the distribution of \Vlambda{\Vector M} (in blue bars) fits the M-P distribution (in blue line) quite well. Whereas, at sampling time \EtS{601}{s}, Figure \ref{RMb} shows that the eigenvalue points collapse to the center point of the circle, which means that the correlations of the data have been enhanced somehow; and Figure \ref{RMd} also shows the deviation between the experimental distribution and the theoretical one.

From \Cur {\VLES{MSR}}{t} curve in Figure \ref{RMe}, it is observed that \VLES{MSR}{} starts to decrease  {\Data {\small{(0.8665, 0.6308,\cdots,0.4927)}}} at \Et{600}{s}, just when the event (\emph{sudden change of \VPbus{52}}) occurs as the signal. The influence lasts for full time length \mbox{(\Equs{T}{240} s)} and the decreasing lasts for half (\mbox{120 s}); thus, a "U"-shaped curve is observed. In this way, we conduct anomaly detection; the time for the beginning point of "U" (\Et {600}{s} for this case) is right the anomaly start time.

\subsubsection{LES Designs}
\Text{\\}

Designing LESs is a major target in this paper; here, we study diverse LESs with different test functions. Keeping \Equs{T}{240} s, we can divide the temporal space into 5 subspaces (stages) according to the status of \VPbus{52} as follows:

\begin{table}[H]
\centering
\begin{tabularx}{0.48\textwidth} { l|r>{$}c<{$} r l|l   }  
& \multicolumn{4}{|c|}{Time Areas (time length)} & Descripiton\\
\hline
\Vector {S1}& 241 s& --  &600 s&(360 s)&fluctuations around 0 M\\
\Vector {S2}& 601 s& --  &840 s&(240 s)&a step signal\\
\Vector {S3}& 841 s& -- &1200 s&(360 s)&fluctuations around 300 M\\
\Vector {S4}& 1201 s& -- &1306 s&(106 s)&a ramp signal\\
\Vector {S5}& 1307 s& -- &1500 s&(194 s)&static voltage collapse\\
\hline

\end{tabularx}
\end{table}


The results of the LESs, designed according to \textit{LES Designs and the Theoretical Values} in  \textit{Sec \ROMAN 3}, are shown as Table \ref{Tab: LES}:

\begin{table}[H]
\caption {LESs and their Values}
\label{Tab: LES}
\centering

\begin{minipage}[!h]{0.49\textwidth}
\centering

\footnotesize
\begin{tabularx}{\textwidth} { >{\scshape}l !{\color{black}\vrule width1pt}        >{$}l<{$}  >{$}l<{$}  >{$}l<{$}   >{$}l<{$}  >{$}l<{$}  >{$}l<{$}  >{$}l<{$}  >{$}l<{$} }  
\toprule[1.5pt]
\hline
 & {\text{MSR}} &   {T_2} & {T_3} & {T_4} & {\text{DET}}  & {\text{LRF}}\\
\hline
\hline
\multicolumn{7}{l} {\Vector {E0}: Theoretical Value}\\
\hline
\STE{\tau}&0.8645&1.34\text{E}3&1.01\text{E}4&8.35\text{E}4&48.3&73.68\\
\STD{\tau}&0.0068&6.65\text{E}2&9.35\text{E}4&1.30\text{E}7&1.32&1.42\\
$c_v$ &0.0954&0.0193&0.0304&0.0432&0.0238&0.0162\\
\toprule[1pt]
\multicolumn{7}{l} {\Vector {S1}: Only small fluctuations around 0 MW}\\
\hline
\Mu{\tau}&0.8648&1.33\text{E}3&9.93\text{E}3&8.19\text{E}4&73.68&73.3\\
\Sigg{\tau}&0.0080&6.53\text{E}1&2.20\text{E}4&4.67\text{E}6&0.406&0.322\\
\toprule[1pt]
\multicolumn{7}{l} {\Vector {S2}: A step signal (\VPbus{52}: 0 MW $\rightarrow$ 300 MW) is included}\\
\hline
\Mu{\tau}&0.5149&1.29\text{E}4&1.92\text{E}6&3.04\text{E}8&-174&295\\
\Sigg{\tau}&0.0788&3.30\text{E}6&1.51\text{E}11&5.66\text{E}15&890&893\\
\toprule[1pt]
\multicolumn{7}{l} {\Vector {S3}: Only small fluctuations around 300 MW}\\
\hline
\Mu{\tau}&0.8141&1.54\text{E}3&1.73\text{E}4&2.89\text{E}5&27.9&93.1\\
\Sigg{\tau}&0.0250&1.81\text{E}2&3.30\text{E}5&4.20\text{E}8&0.507&0.419\\
\toprule[1pt]
\multicolumn{7}{l} {\Vector {S4}: A Ramp signal as the system incoming}\\
\hline
\Mu{\tau}&0.6448&6.43\text{E}3&6.40\text{E}5&7.54\text{E}7&-61.4&182\\
\Sigg{\tau}&0.0571&7.20\text{E}6&1.68\text{E}11&3.29\text{E}15&1.74\text{E}3&1.73\text{E}3\\
\toprule[1pt]
\multicolumn{7}{l} {\Vector {S5}: Static voltage collapse for the system}\\
\hline
\Mu{\tau}&0.4136&7.49\text{E}3&6.48\text{E}5&7.25\text{E}7&-598&719\\
\Sigg{\tau}&0.1076&7.47\text{E}6&2.99\text{E}11&1.02\text{E}15&4.16\text{E}4&4.16\text{E}4\\

\hline
\toprule[1pt]
\end{tabularx}
\raggedright
\small {*$\Equs{c_v}{\sqrt{\STD{\tau}}/\STE{\tau}}$ is the coefficient of variation}
\end{minipage}
\end{table}

The above results validate that it is feasible to analyse the power system using the random matrix model (RMM); the upper layer should be ontology and we do not go that far. Here, we just make some engineering statements/analyses:

1. Independent of the system models/topologies, we can design the LES $\tau$. For a RMM \VX{} with determinate size \Mul NT, if the test function $\varphi({\lambda})$ is given, the LES $\tau$ is obtained; some related theoretical values (the expectation \STE{\tau}, the variance \STD{\tau}, and the coefficient of variation ${c_v}$) are able to be calculated as well.

2. Among $\varphi({\lambda})$ given above, LRF performs best in the view of ${c_v}$ (Low $c_v$ means high precision and repeatability of the assay \cite{nutter1995improving}; here, we regard 12\% as the upper bound).

3. We can make a summary about the performance of the experimental values (mean \Mu{\tau} and variance \Sigg{\tau}) comparing to the theoretical ones (expectation \STE{\tau} and variance \STD{\tau}):\\
\small{}
 a). During \Vector {S1}, \Mu{\tau} is close to \STE{\tau}; \Sigg{\tau} is much less than \STD{\tau}; \\
 b). During \Vector {S3}, \Mu{\tau} has a little bias (i.e. \Abs{\Mu{\tau}-\STD{\tau}}), but more than \Vector{S1}; \Sigg{\tau} is acceptable ($\sqrt{\Sigg{\tau}}/\Mu{\tau}<12\%$);\\
 c). For \Vector {S2}, \Vector {S4}, and \Vector {S5}, \Mu{\tau} has much more bias; \Sigg{\tau} is always too big to be accepted;\\
 d). variance \Sigg{\tau} is much more sensitive than mean \Mu{\tau}.\\
\normalsize{}
Then, we can conjecture that:\\
\small{}
a). The more stable the system is, the more effective the theoretical values become (\Mu{\tau} is close to \STE{\tau}; \Sigg{\tau} is less than \STD{\tau});\\
b). Different test functions  $\varphi({\lambda})$ have different characteristics and functions. In this sense, we can balance the reliability and sensitivity for anomaly detection in a special system;\\
c). In addition, a test function is akin to a filter in some sense; beyond event detection (distinguish signals from noises), it has the potential to trace a specific anomaly (distinguish the signal from others);\\
d). For a special purpose, e.g. the lowest  ${c_v}$  or the lowest bias, there should exist an optimal combination of the Chebyshev Polynomials as the test function.
\normalsize{}

\subsection{Advantages of LES and Visualization Using 3D Power-Map}
\normalsize{}A LES is an indicator in high dimensions; its value depends on a wide sample data in the form of the entries of the RMM \VRX{}  (\textit{Sec \ROMAN 4}). As a result, LES $\tau{}$, as a statistical indicator, is sensitive and robust against bad data; these advantages will be validated via the visualization of \VLES{} using 3D power-map later.

\begin{figure*}[t]
\centering
\begin{overpic}[scale=0.26
]{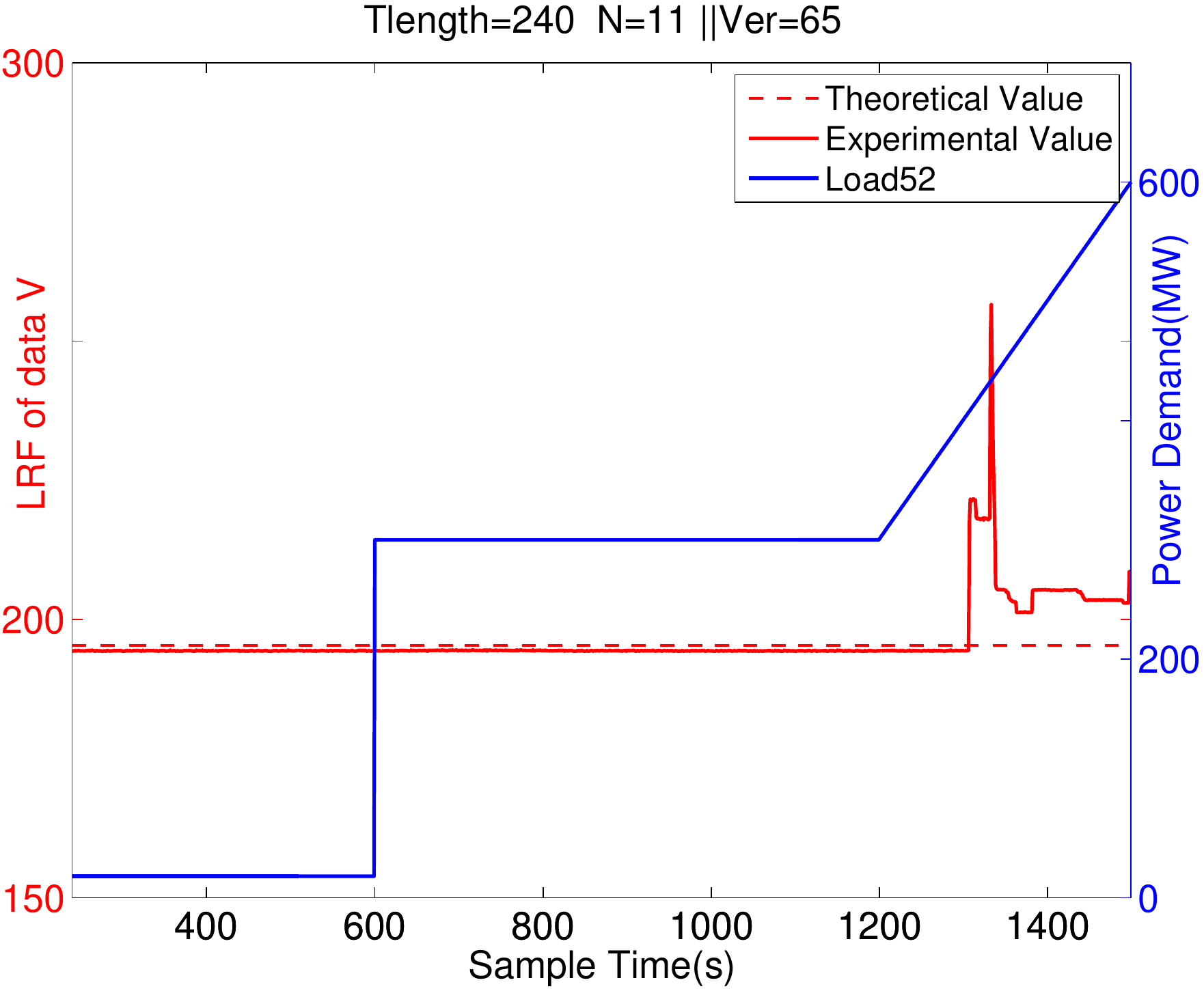}
    \setlength {\fboxsep}{1pt}
   \put(10,70) {\fbox{\small \color{blue}{\textbf A$1$}}}   
\end{overpic}
\begin{overpic}[scale=0.26
]{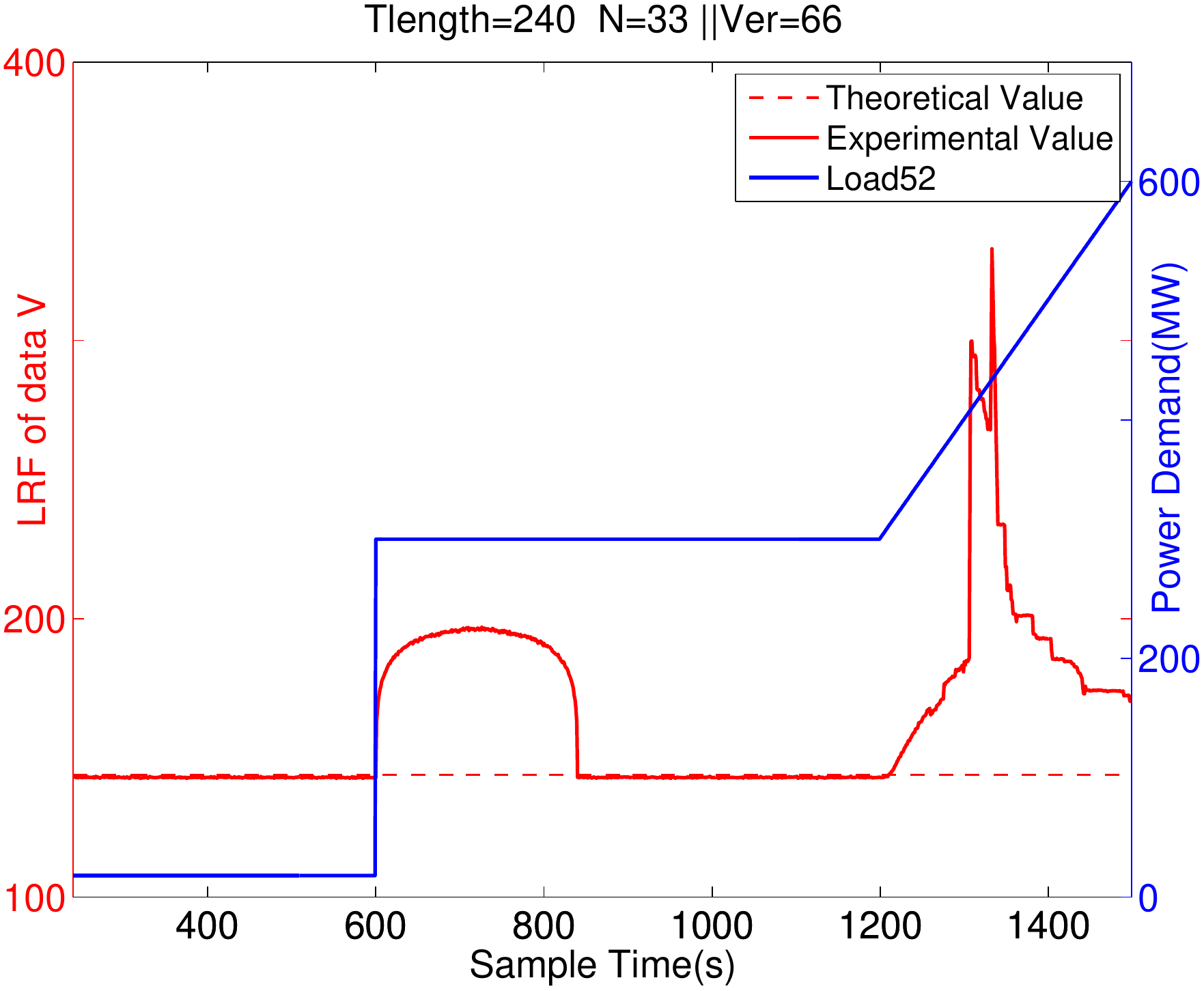}
    \setlength {\fboxsep}{1pt}
    \put(10,70) {\fbox{\small \color{blue}{\textbf A$2$}}}   
\end{overpic}
\begin{overpic}[scale=0.26
]{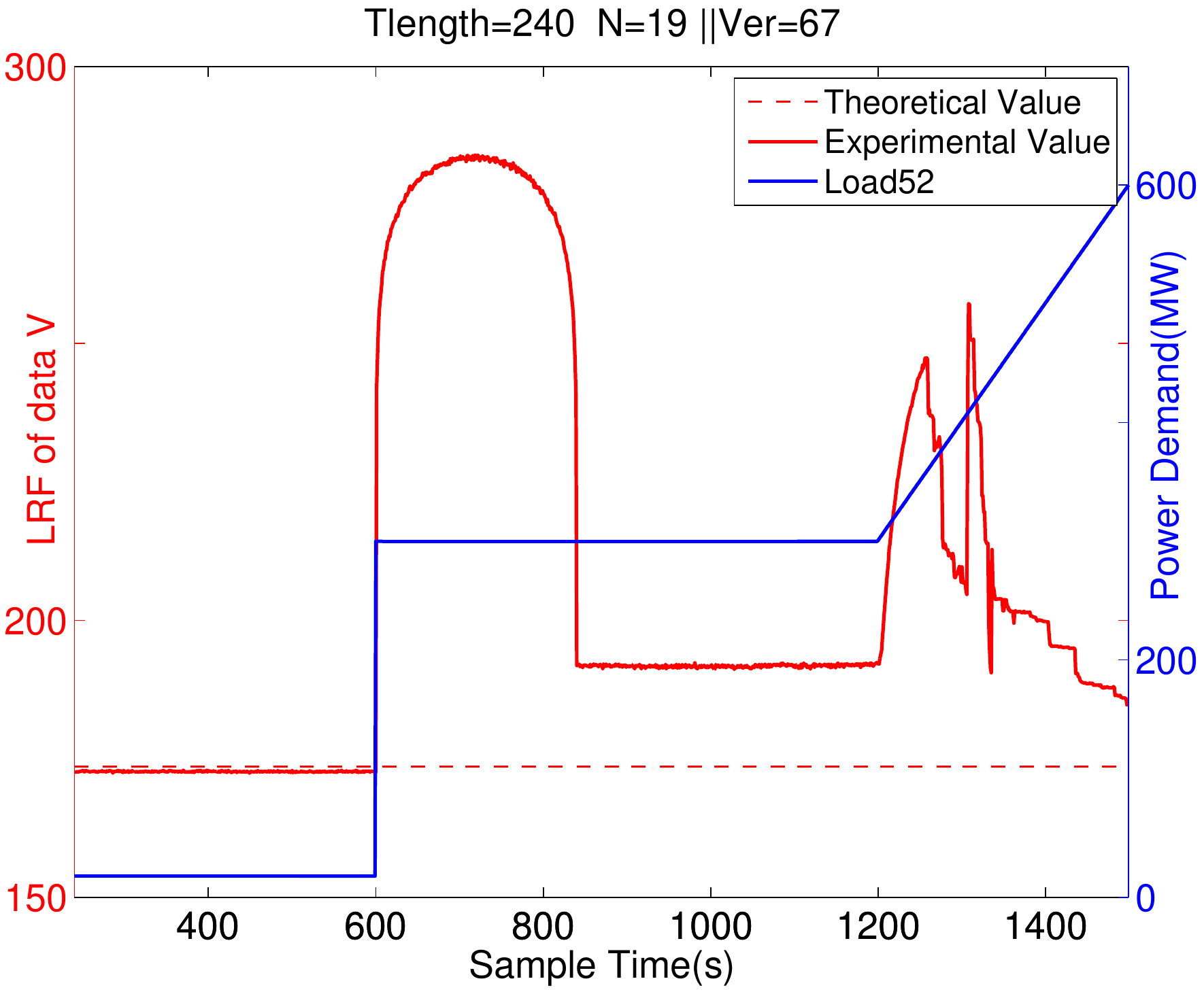}
    \setlength {\fboxsep}{1pt}
    \put(10,70) {\fbox{\small \color{blue}{\textbf A$3$}}}   
\end{overpic}

\begin{overpic}[scale=0.26
]{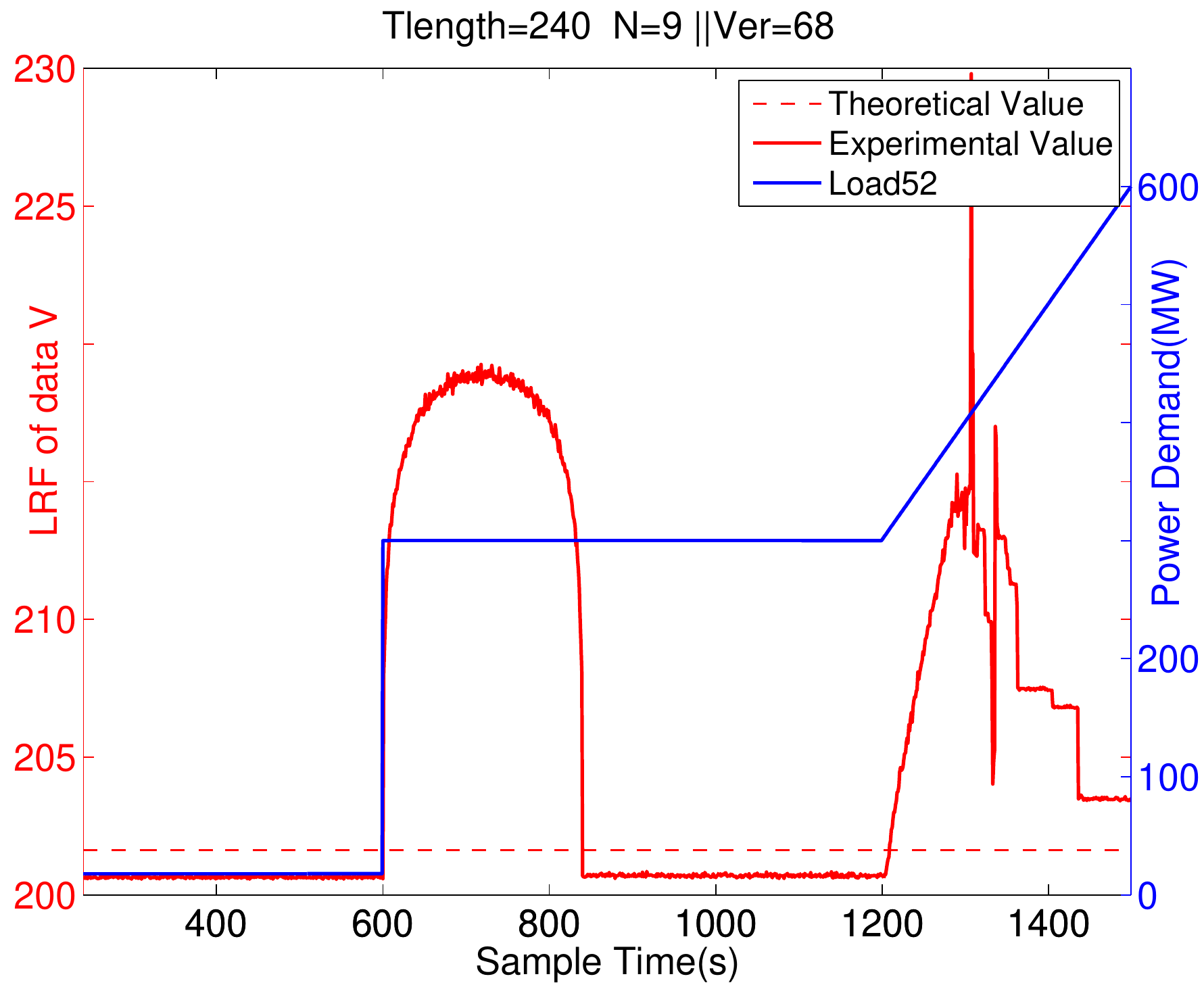}
    \setlength {\fboxsep}{1pt}
    \put(10,70) {\fbox{\small \color{blue}{\textbf A$4$}}}   
\end{overpic}
\begin{overpic}[scale=0.26
]{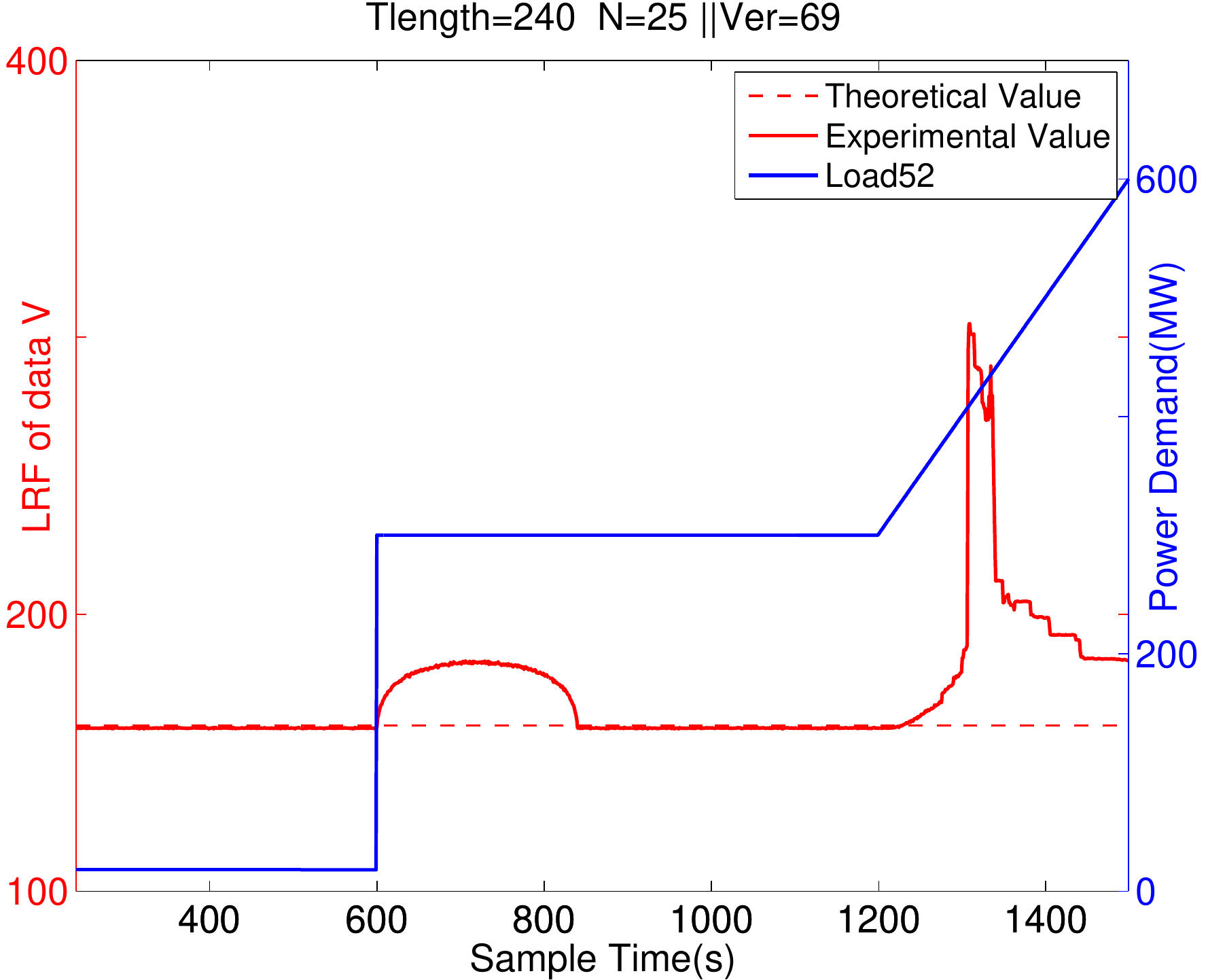}
    \setlength {\fboxsep}{1pt}
    \put(10,70) {\fbox{\small \color{blue}{\textbf A$5$}}}   
\end{overpic}
\begin{overpic}[scale=0.26
]{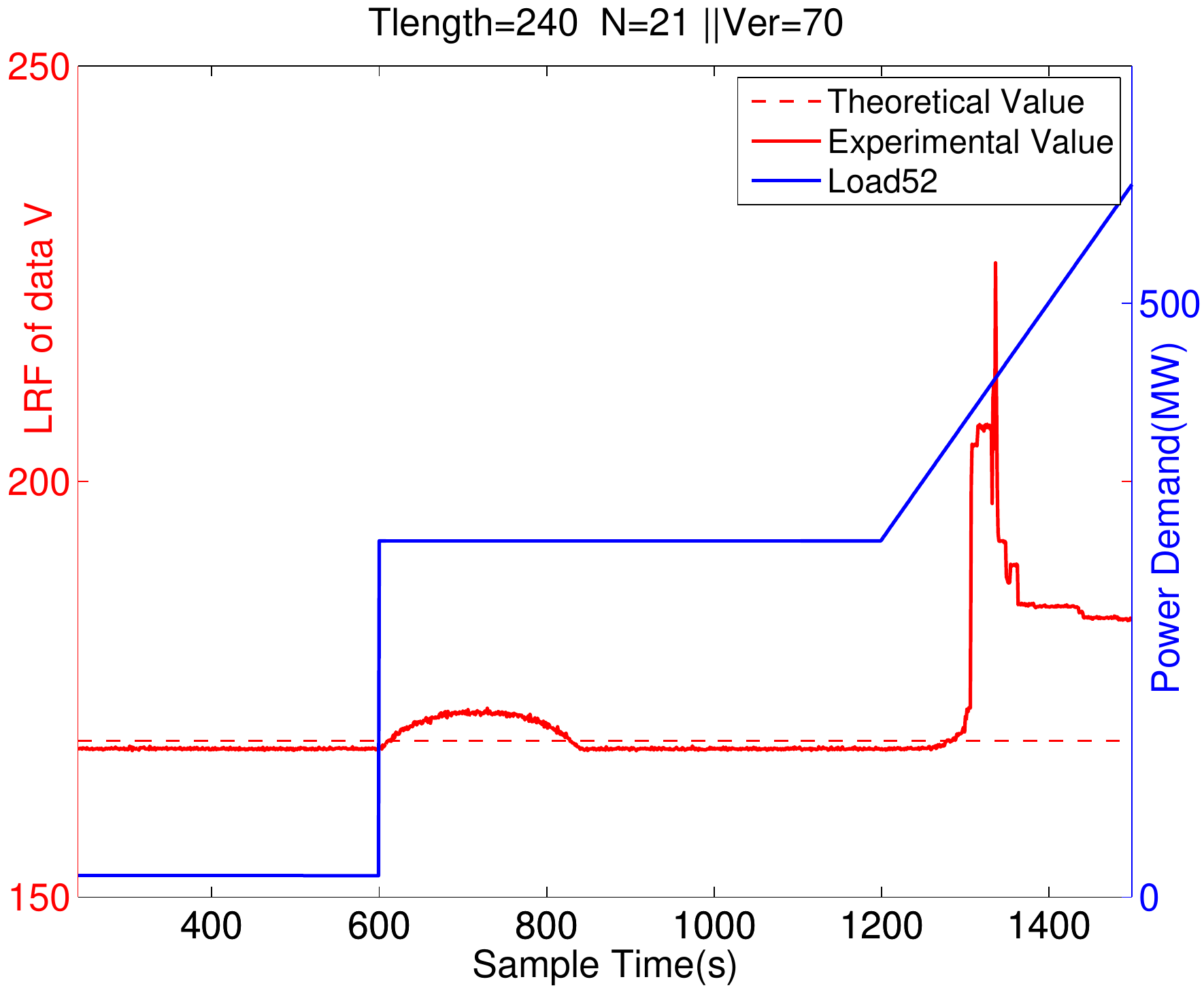}
    \setlength {\fboxsep}{1pt}
    \put(10,70) {\fbox{\small \color{blue}{\textbf A$6$}}}   
\end{overpic}
\caption{\Cur {\VLES{LRF}}{t} Curve for Single Partitioning: A1--A6}
\label{fig:Ples}
\end{figure*}

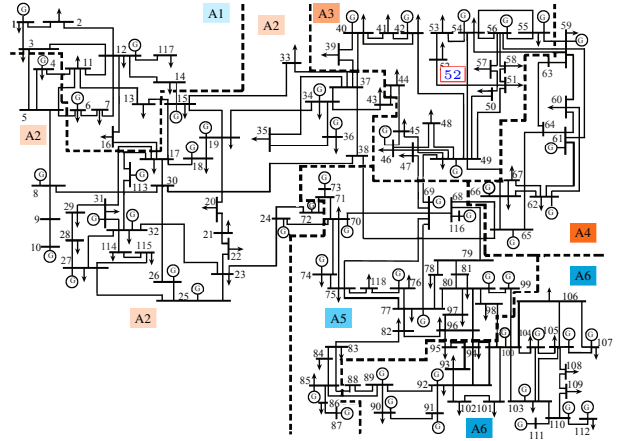
\begin{figure}[H]
\centering
\begin{overpic}[scale=0.58
]{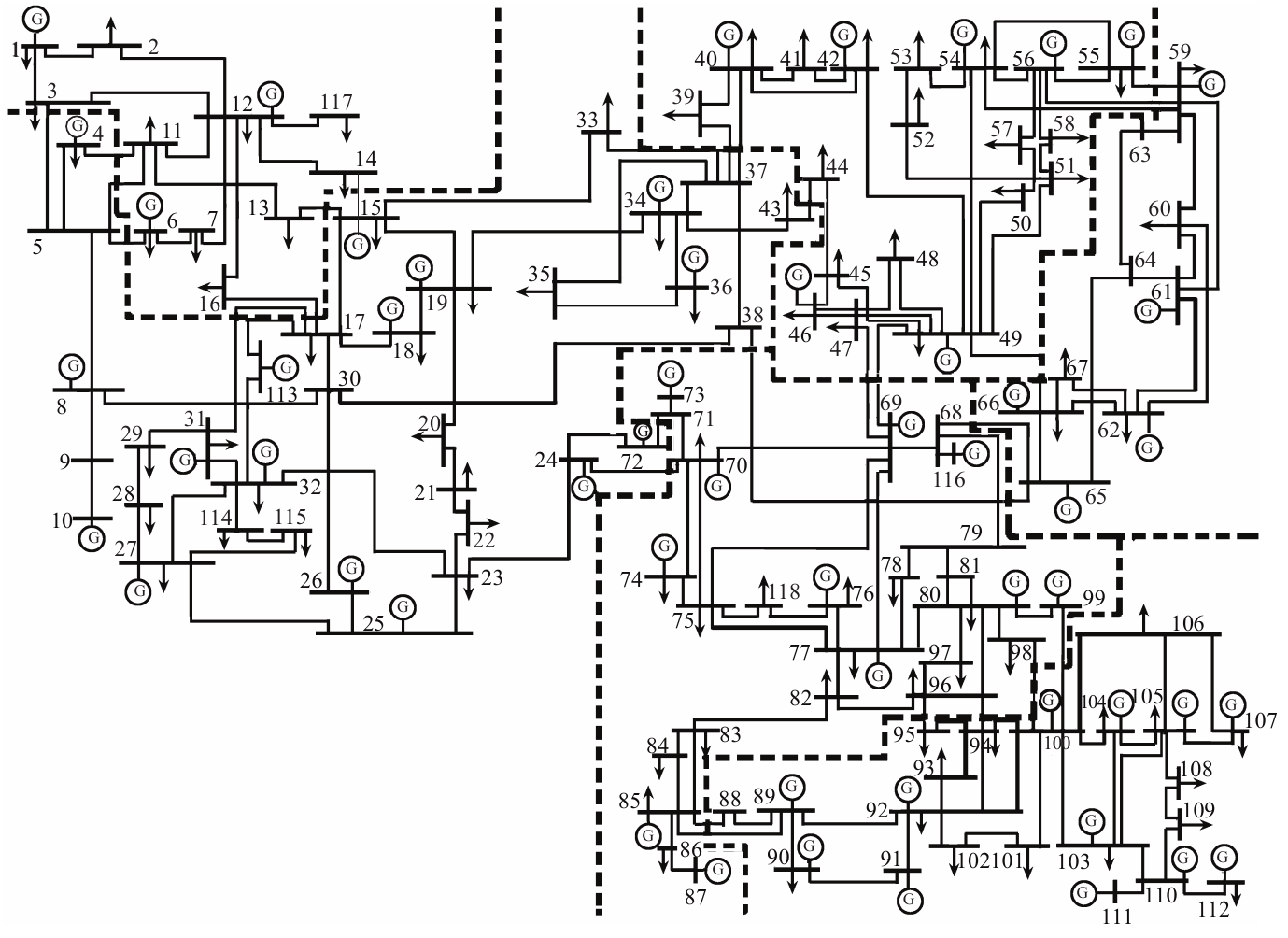}
    \setlength {\fboxsep}{1pt}
    \put(71,60) {\fcolorbox{red}{white}{\tiny \color{blue}{$52$}}} 

      \setlength {\fboxsep}{2pt}
      \put(32,70) {\fcolorbox{white}{ssBlue}{\tiny \color{black}{A1}}}
      \put(20,20) {\fcolorbox{white}{ssOrange}{\tiny \color{black}{A2}}}
      \put(41,68) {\fcolorbox{white}{ssOrange}{\tiny \color{black}{A2}}}
      \put(02,50) {\fcolorbox{white}{ssOrange}{\tiny \color{black}{A2}}}
      \put(50,70) {\fcolorbox{white}{sOrange}{\tiny \color{black}{A3}}}
      \put(92,34) {\fcolorbox{white}{Orange}{\tiny \color{black}{A4}}}
      \put(52,20) {\fcolorbox{white}{sBlue}{\tiny \color{black}{A5}}}
      \put(92,27) {\fcolorbox{white}{Blue}{\tiny \color{black}{A6}}}
      \put(75,02) {\fcolorbox{white}{Blue}{\tiny \color{black}{A6}}}

\end{overpic}
\caption{Partitioning network for the IEEE 118-bus system. There are six partitions, i.e. A1, A2, A3, A4, A5, and A6.}
\label{fig:IEEE118network}
\end{figure}

For this case, we keep the hypothetical scenario as Table \ref{Tab: Event Series}, and take \VLES{LRF} as the indicator.
 We obtain the regional \VLES{LRF} for each region\footnote{this division depends on the specific network structure; however, a potential problem lies here: how small the partition  (see Figure \ref{fig:IEEE118network}) can be that keeps the theories still work.
To find the answer we should study how to turn big data into tiny data \cite {feldman2013turning}; it is another topic and we do not expand here. Theoretically, the bias of \VLES{} is closely related to the size of RMMs \VRX{} (more specifically, $N$)} in a similar way to that for the whole system (118 nodes); thus, the \Cur {\VLES{LRF}}{t} curves are plotted as Figure \ref{fig:Ples}.

\begin{figure*}[!t]
\centering
\subfloat[\EtS{600}{s}]{
\begin{overpic}[width=0.16\textwidth]{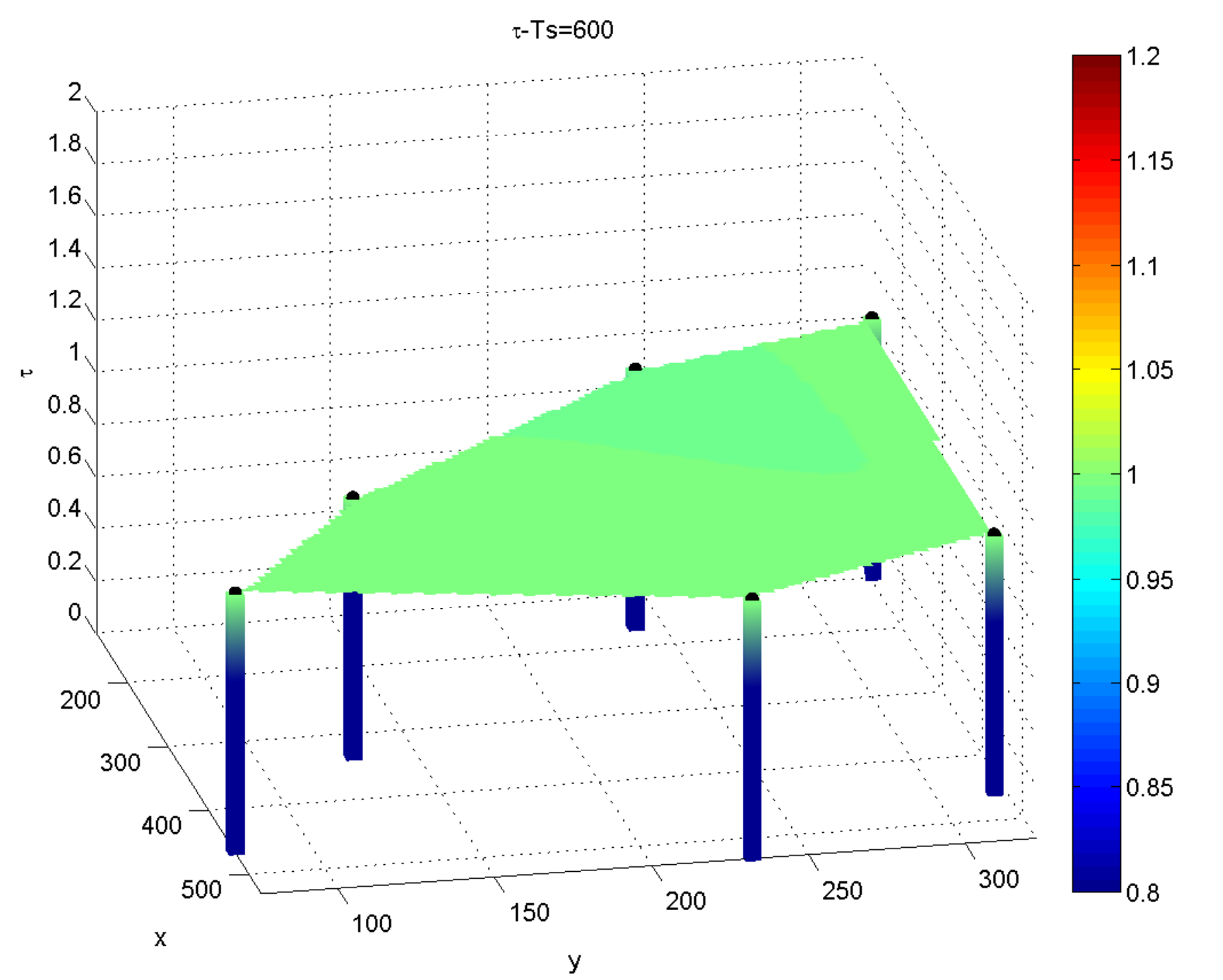}
    \setlength {\fboxsep}{1pt}
   \put(65,58) {\fbox{\tiny \color{blue}{\textbf A$1$}}}   
   \put(45,54) {\fbox{\tiny \color{blue}{\textbf A$2$}}}   
   \put(80,40) {\fbox{\tiny \color{blue}{\textbf A$3$}}}   
   \put(65,24) {\fbox{\tiny \color{blue}{\textbf A$4$}}}   
   \put(22,44) {\fbox{\tiny \color{blue}{\textbf A$5$}}}   
   \put(12,36) {\fbox{\tiny \color{blue}{\textbf A$6$}}}   
\end{overpic}
}
\subfloat[\EtS{601}{s}]{
\includegraphics[width=0.16\textwidth]{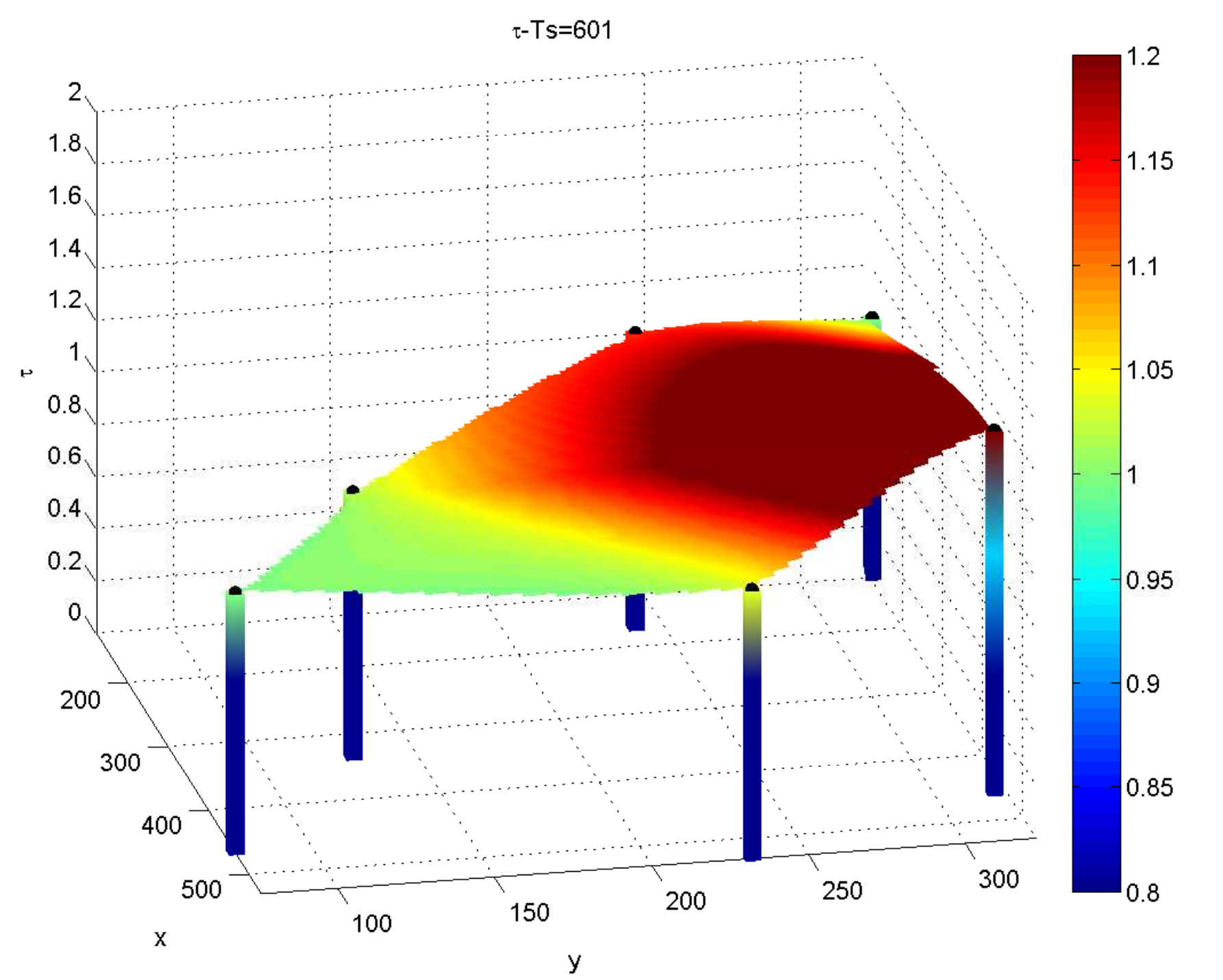}
}
\subfloat[\EtS{720}{s}]{
\includegraphics[width=0.16\textwidth]{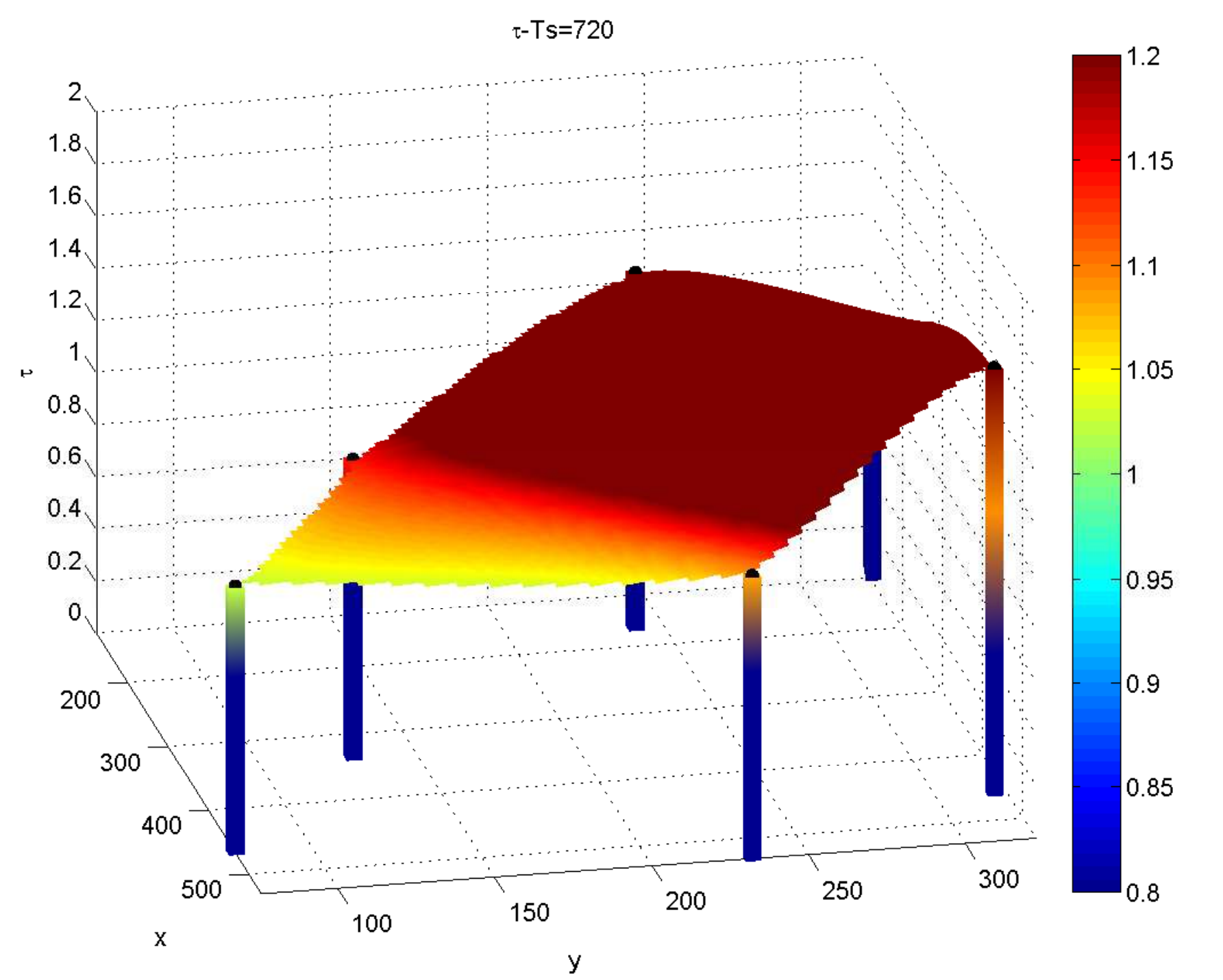}
}
\subfloat[\EtS{1198}{s}]{
\includegraphics[width=0.16\textwidth]{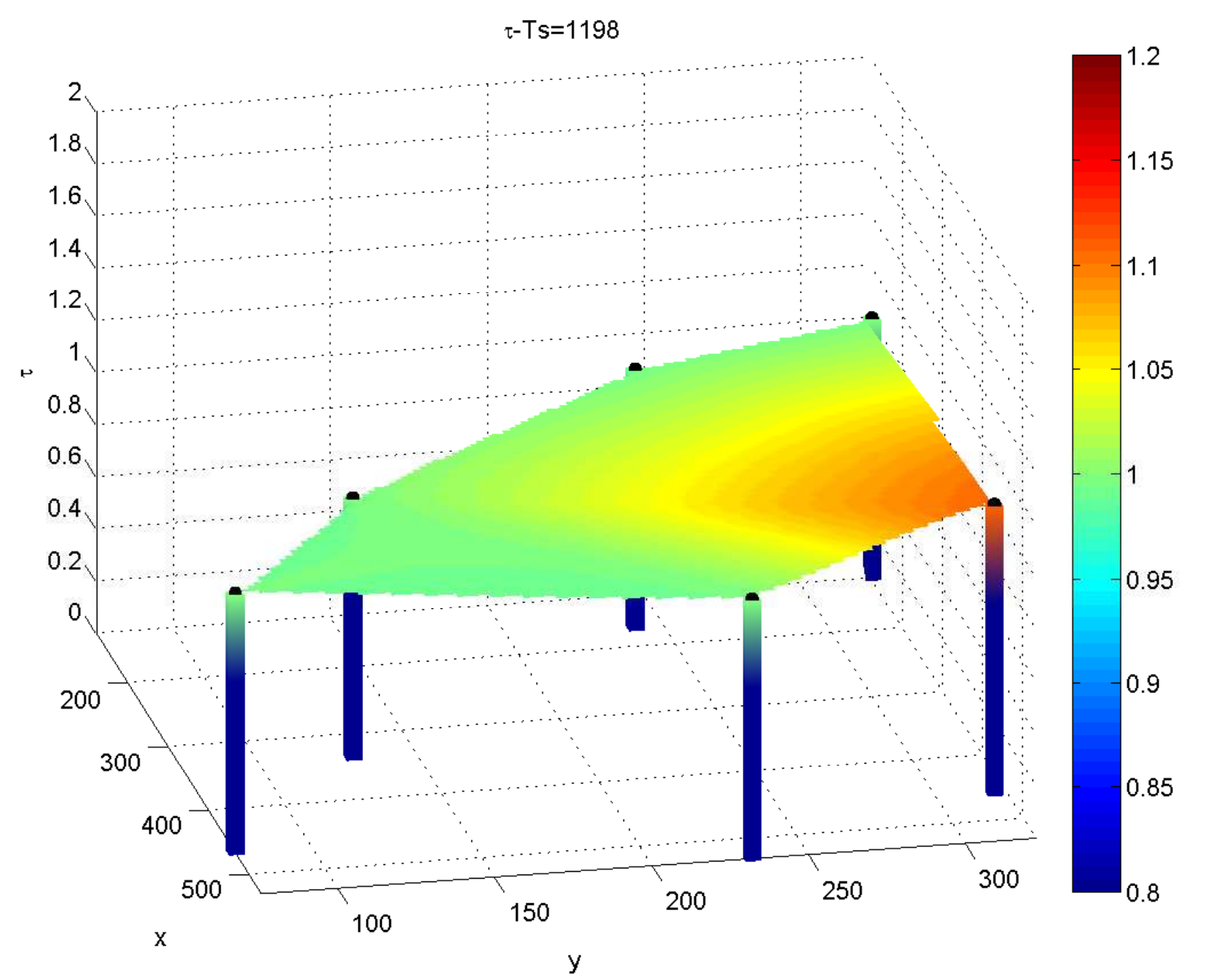}
}
\subfloat[\EtS{1250}{s}]{
\includegraphics[width=0.16\textwidth]{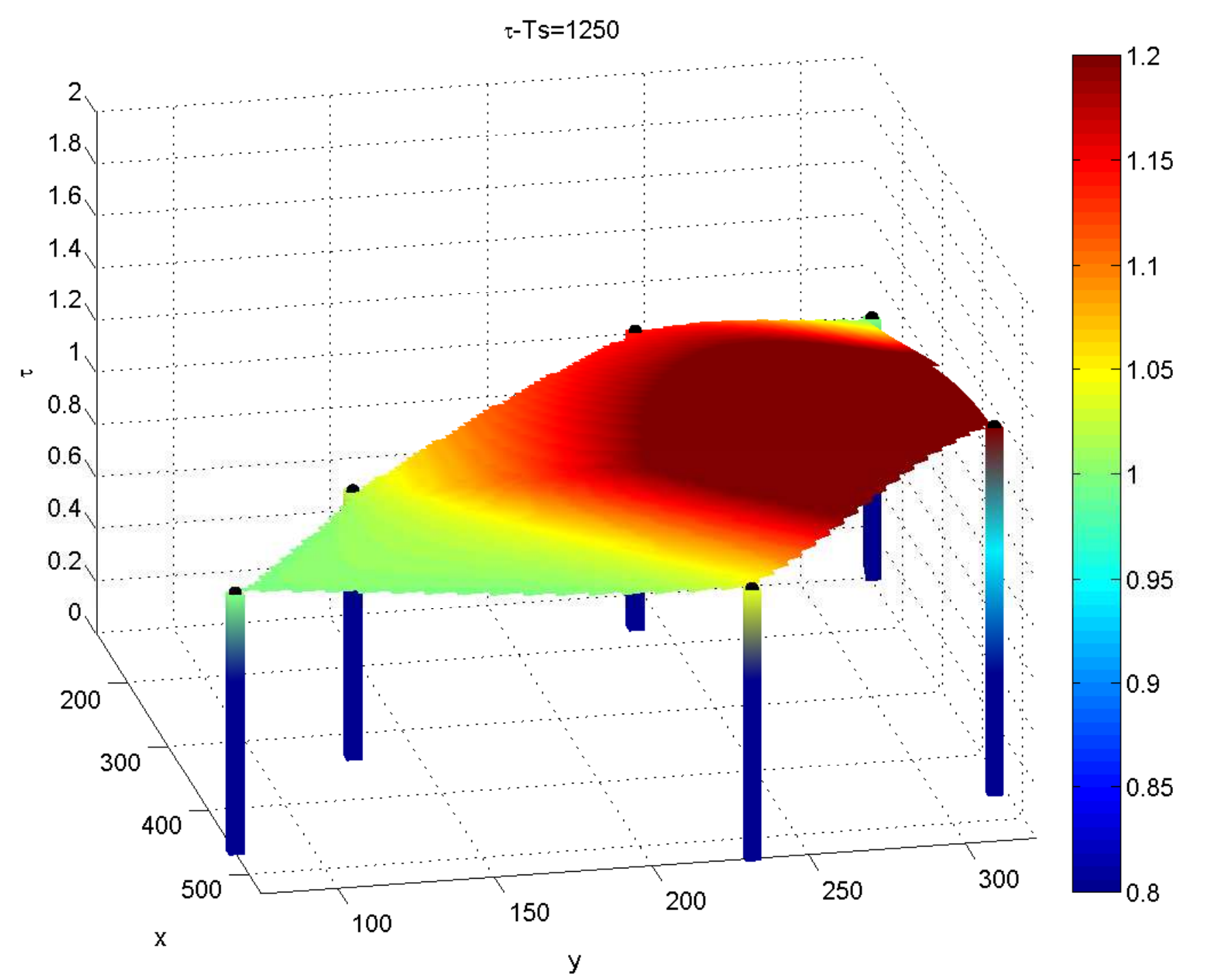}
}
\subfloat[\EtS{1350}{s}]{
\includegraphics[width=0.16\textwidth]{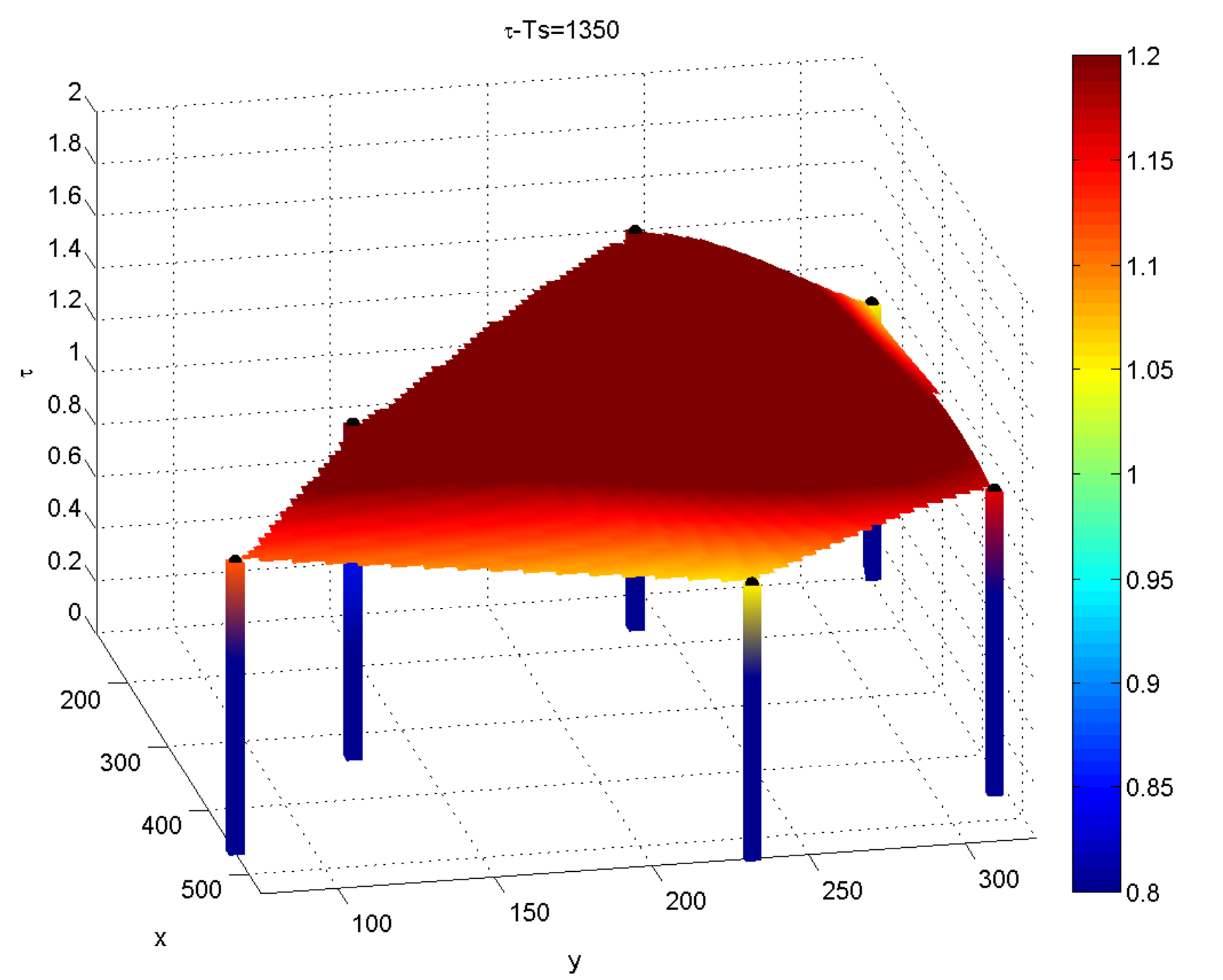}
}
\caption{Visualization of the high-dimensional indictor $\eta$ with Full Data Sets}
\label{fig:FullDataMSR}

\centering
\subfloat[\EtS{600}{s}]{
\includegraphics[width=0.16\textwidth]{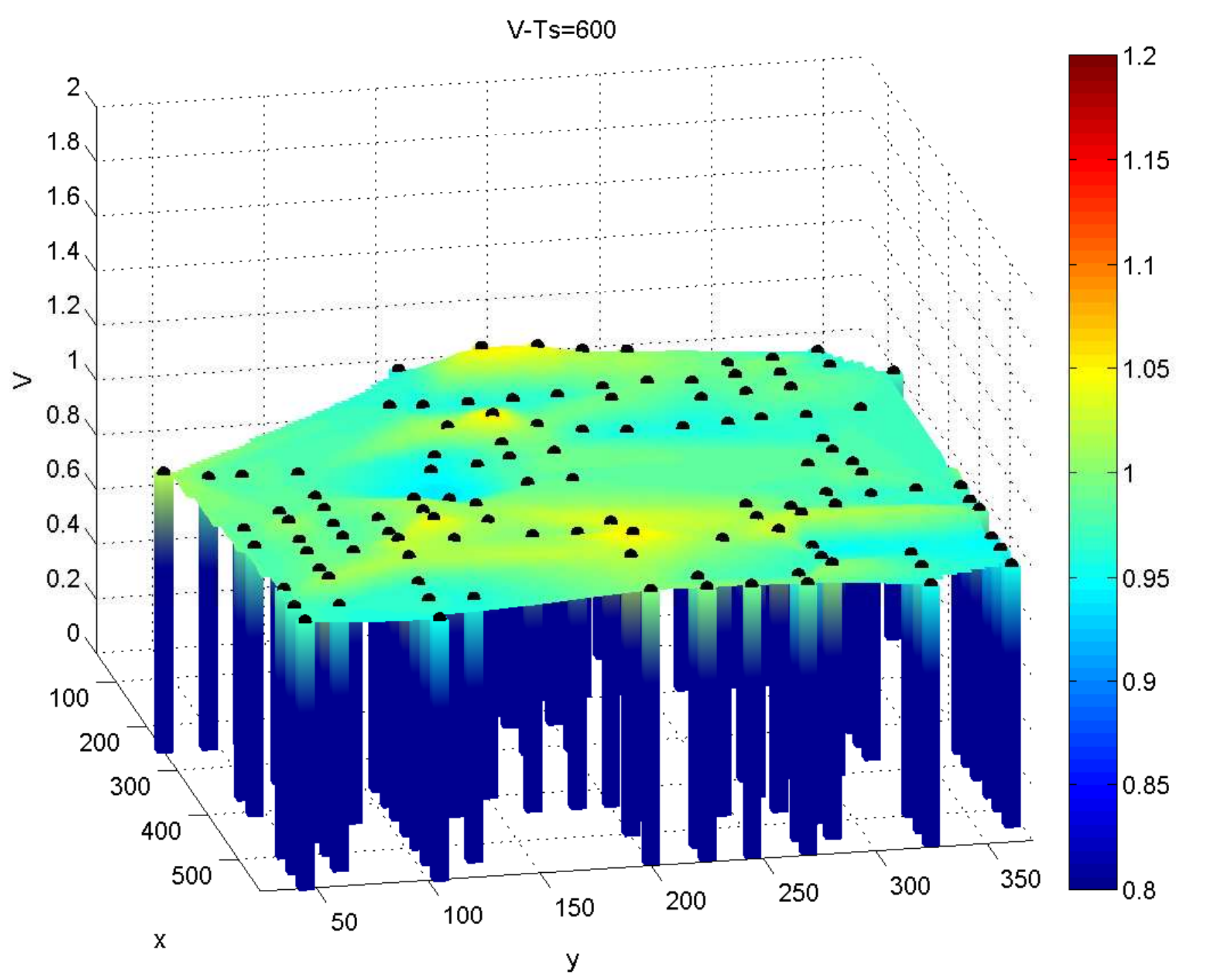}
}
\subfloat[\EtS{601}{s}]{
\includegraphics[width=0.16\textwidth]{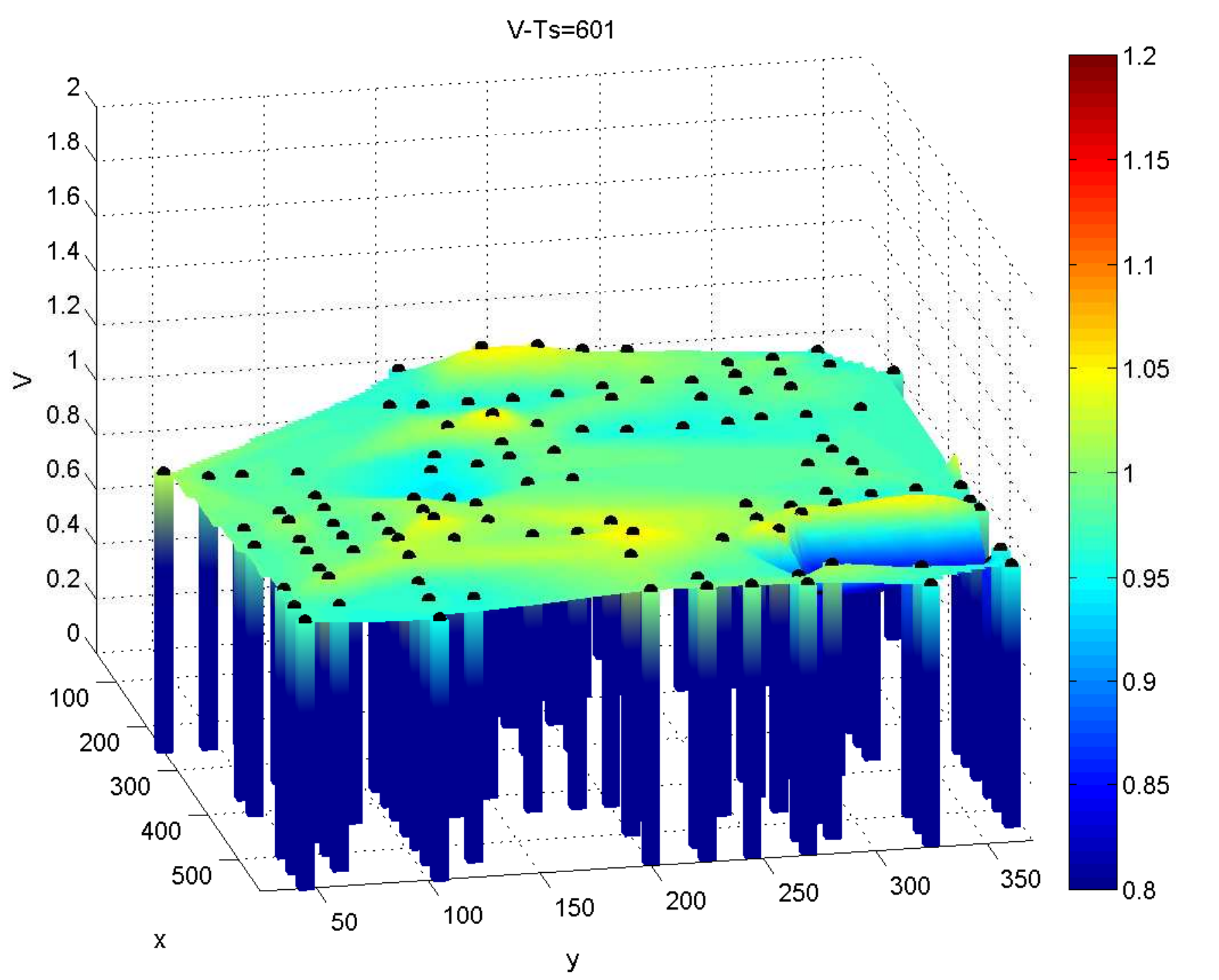}
}
\subfloat[\EtS{720}{s}]{
\includegraphics[width=0.16\textwidth]{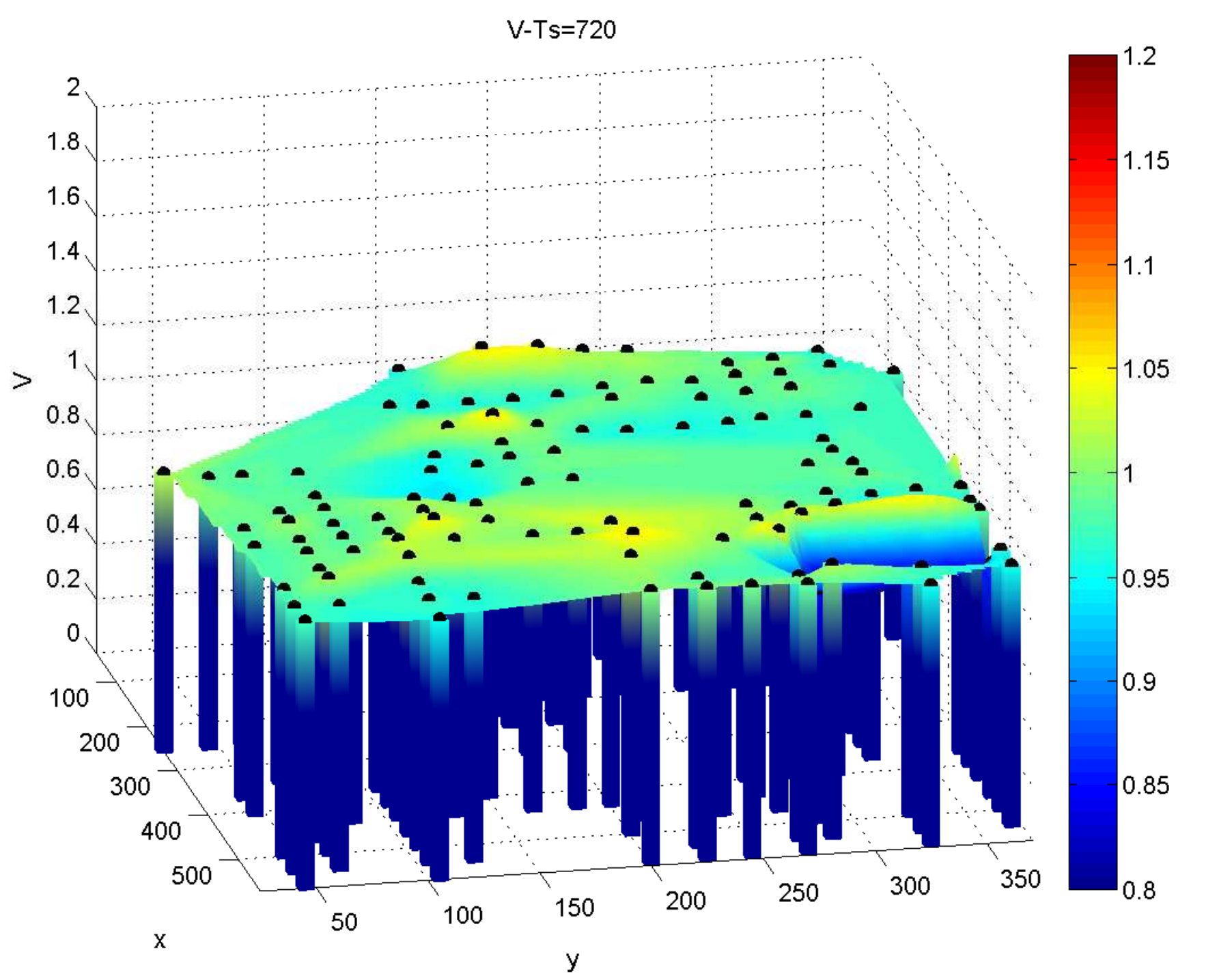}
}
\subfloat[\EtS{1198}{s}]{
\includegraphics[width=0.16\textwidth]{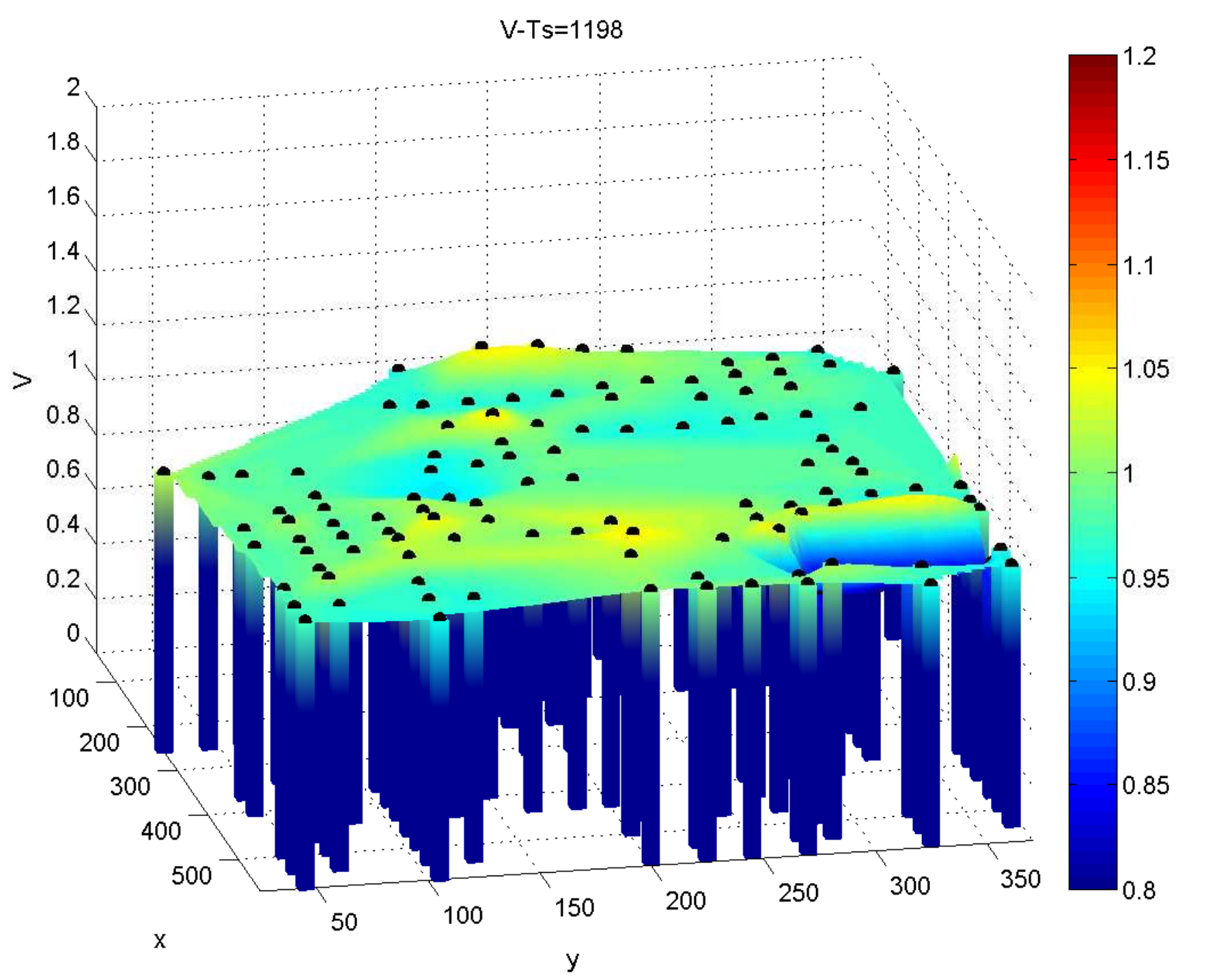}
}
\subfloat[\EtS{1250}{s}]{
\includegraphics[width=0.16\textwidth]{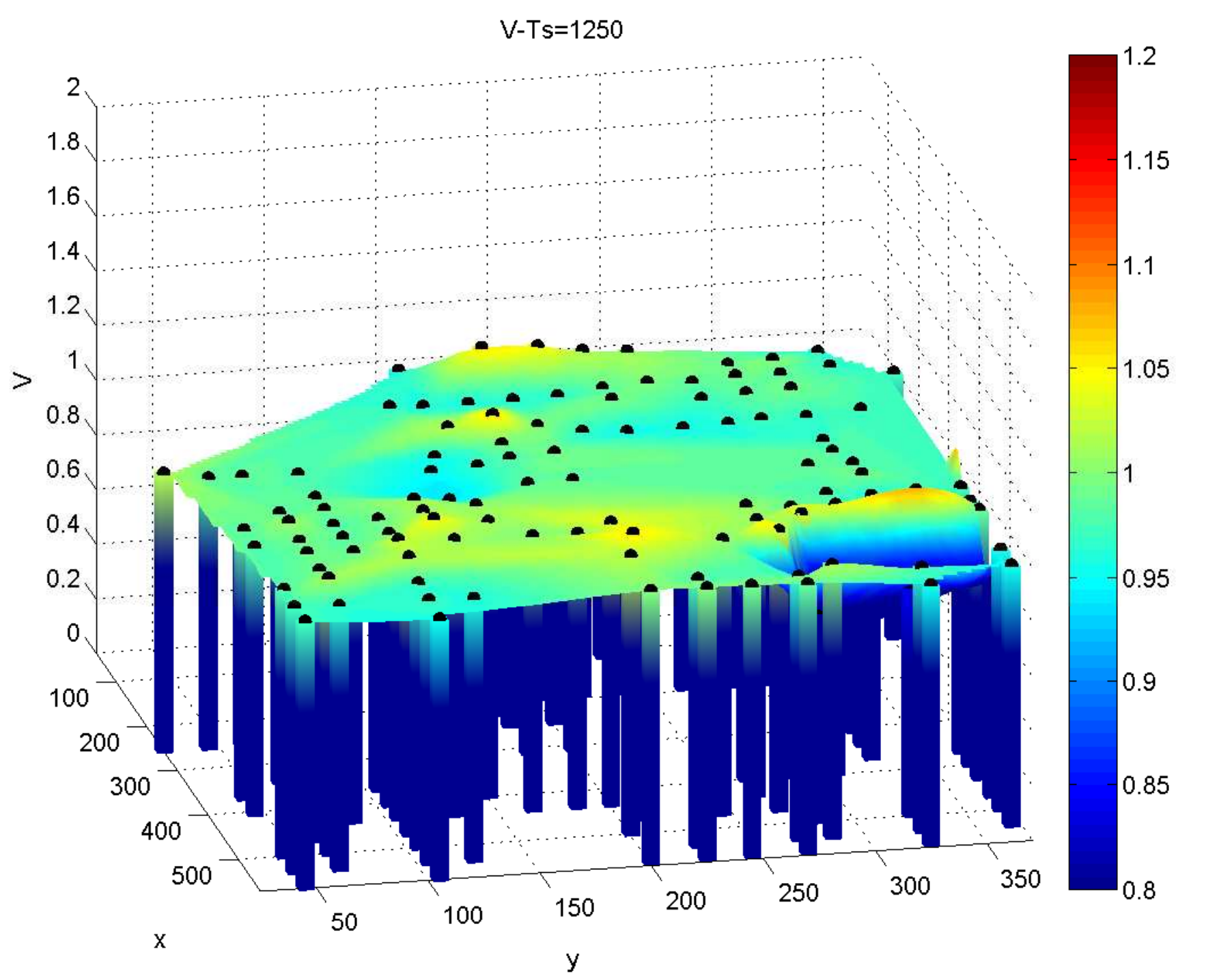}
}
\subfloat[\EtS{1350}{s}]{
\includegraphics[width=0.16\textwidth]{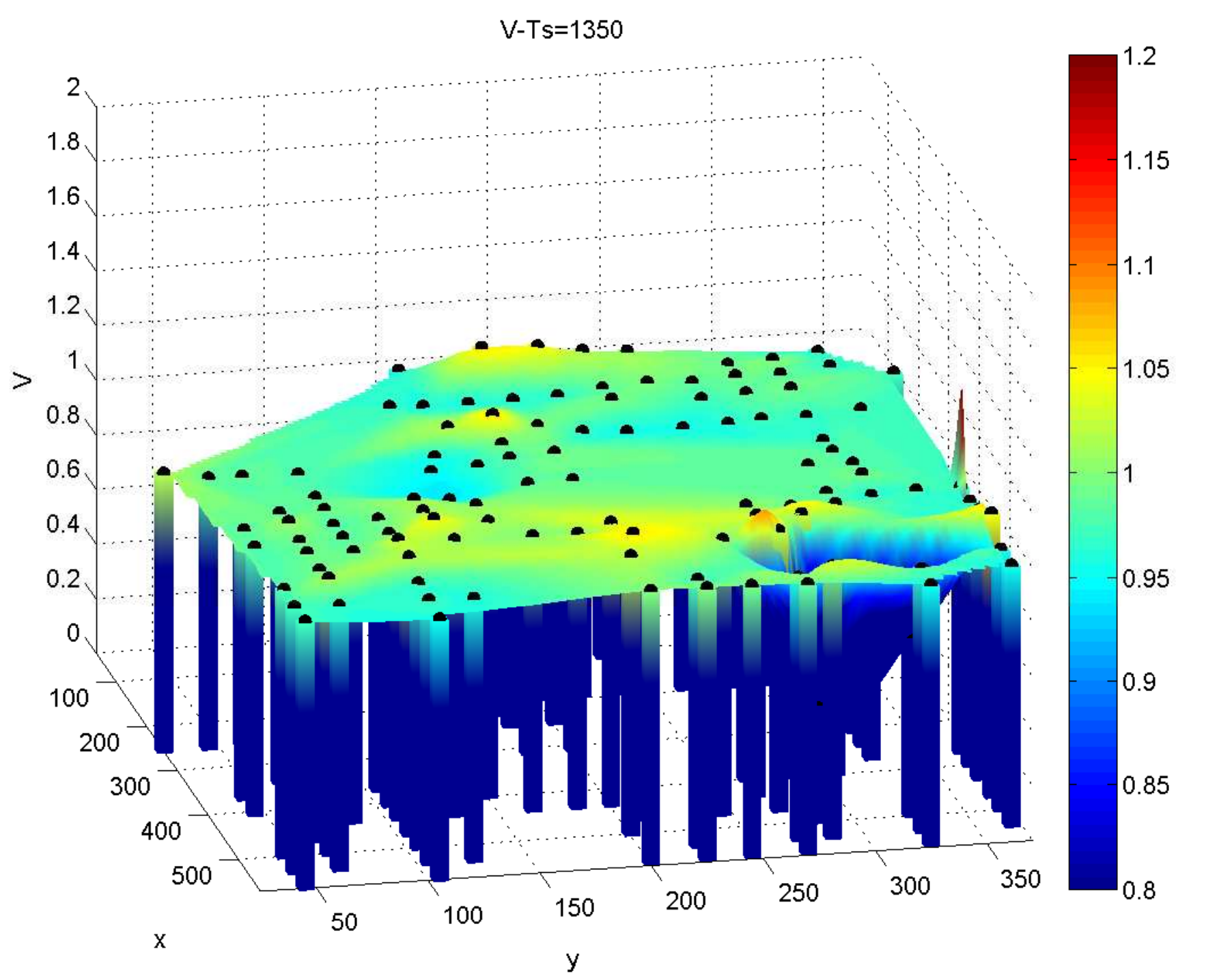}
}
\caption{Visualization of the Voltage \VV{} with Full Data Sets}
\label{fig:FullDataV}

\centering
\subfloat[\EtS{600}{s}]{
\begin{overpic}[width=0.16\textwidth]{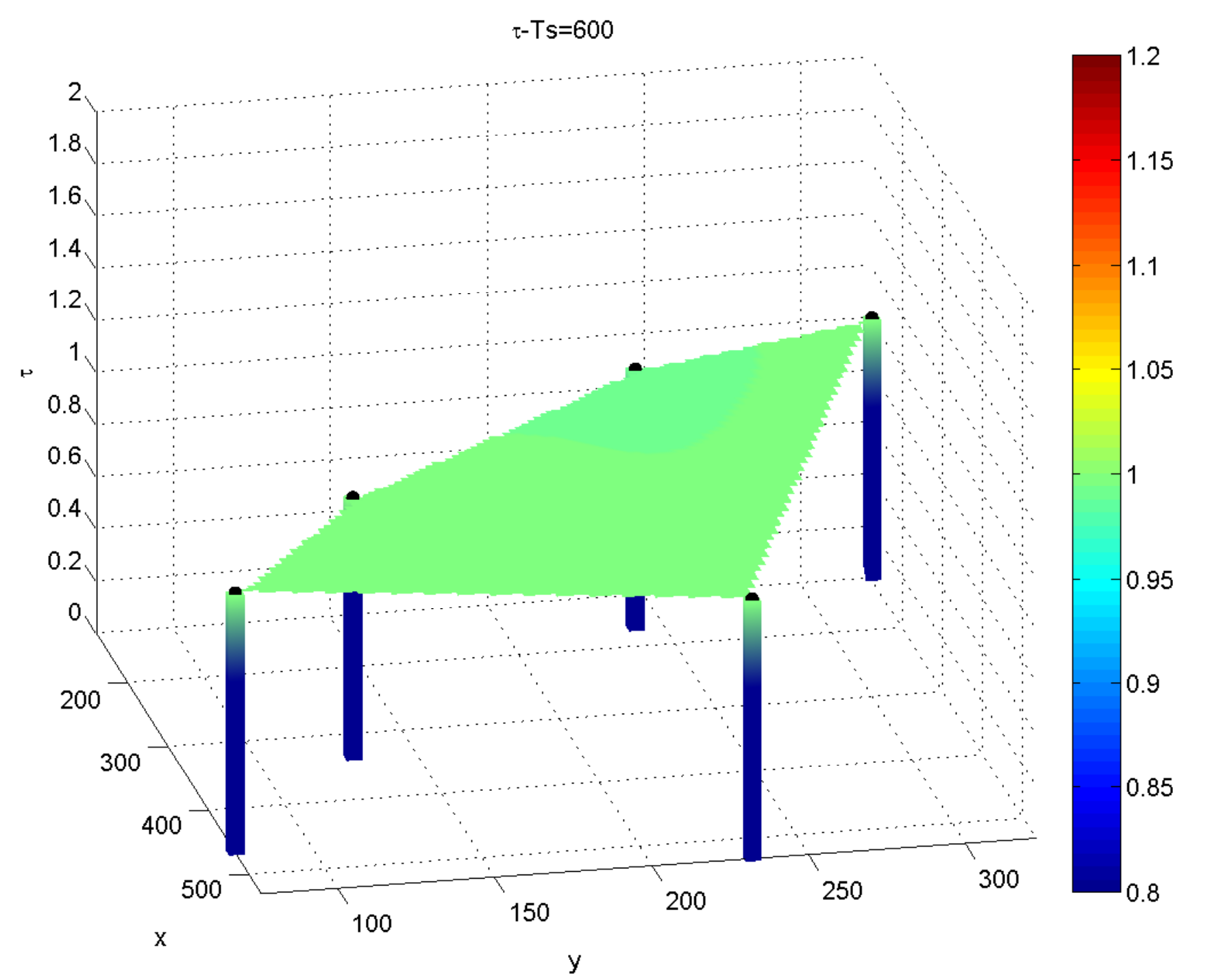}
    \setlength {\fboxsep}{1pt}
   \put(65,58) {\fbox{\tiny \color{blue}{\textbf A$1$}}}   
   \put(45,54) {\fbox{\tiny \color{blue}{\textbf A$2$}}}   
   \put(65,24) {\fbox{\tiny \color{blue}{\textbf A$4$}}}   
   \put(22,44) {\fbox{\tiny \color{blue}{\textbf A$5$}}}   
   \put(12,36) {\fbox{\tiny \color{blue}{\textbf A$6$}}}   
\end{overpic}
}
\subfloat[\EtS{601}{s}]{
\includegraphics[width=0.16\textwidth]{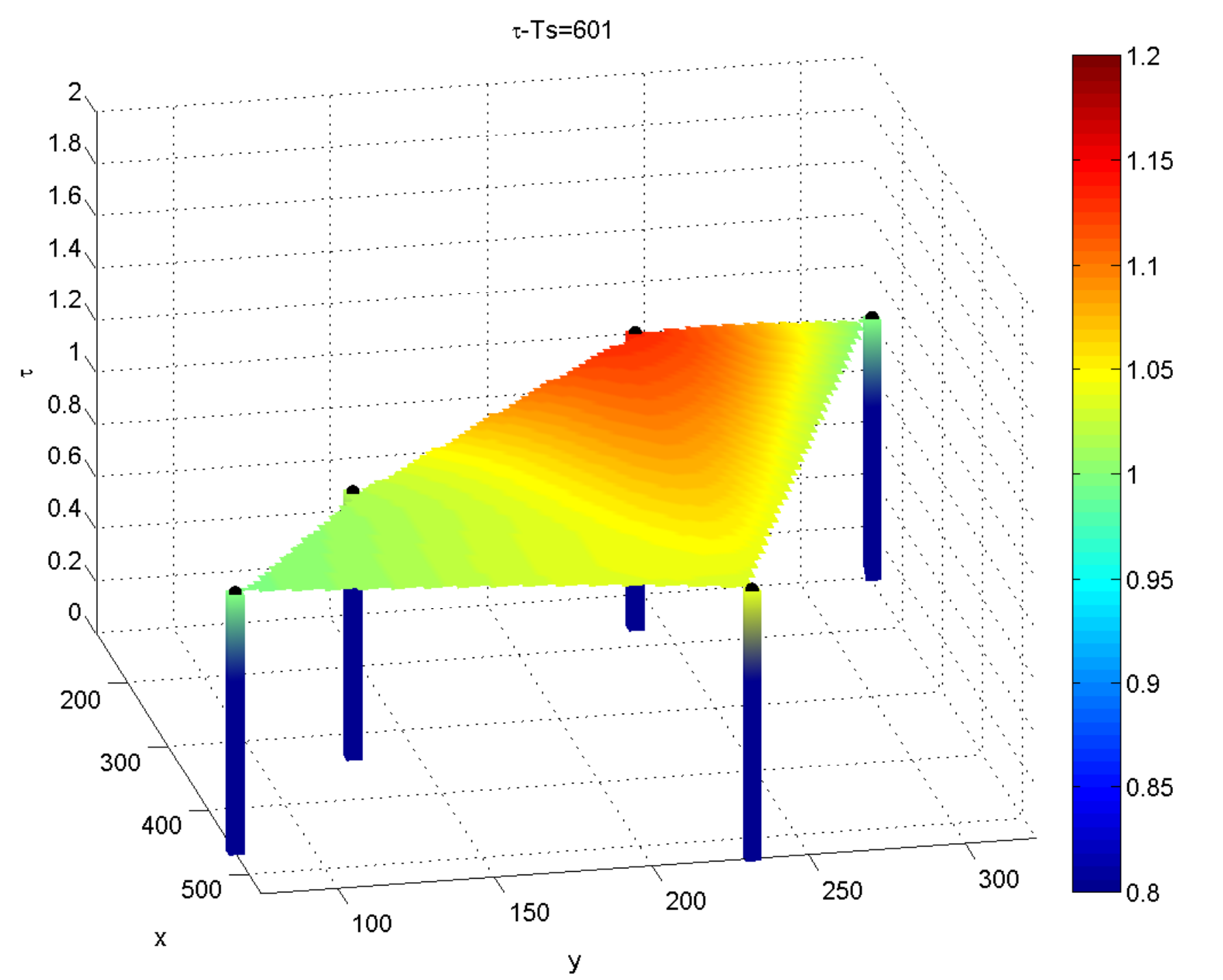}
}
\subfloat[\EtS{720}{s}]{
\includegraphics[width=0.16\textwidth]{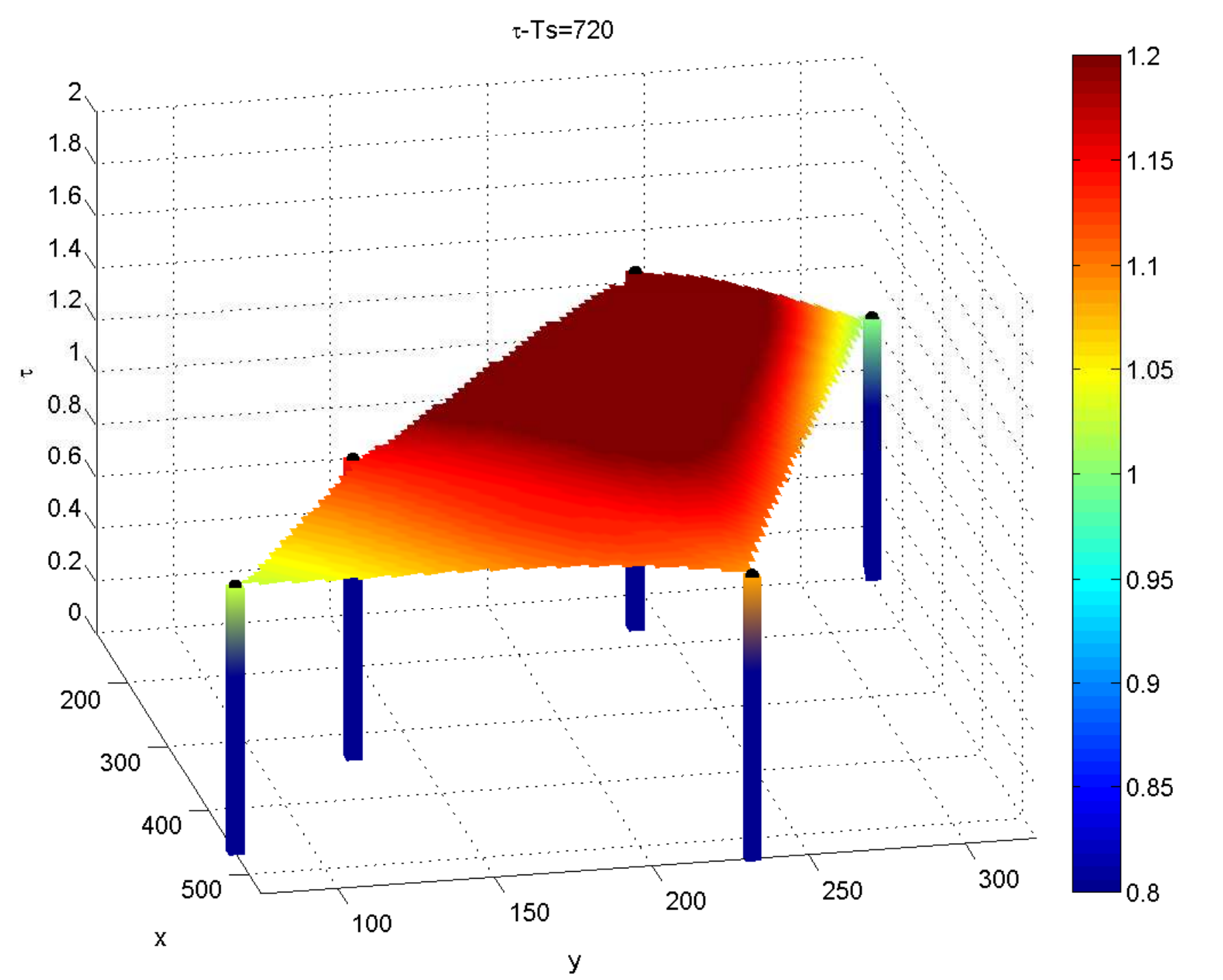}
}
\subfloat[\EtS{1198}{s}]{
\includegraphics[width=0.16\textwidth]{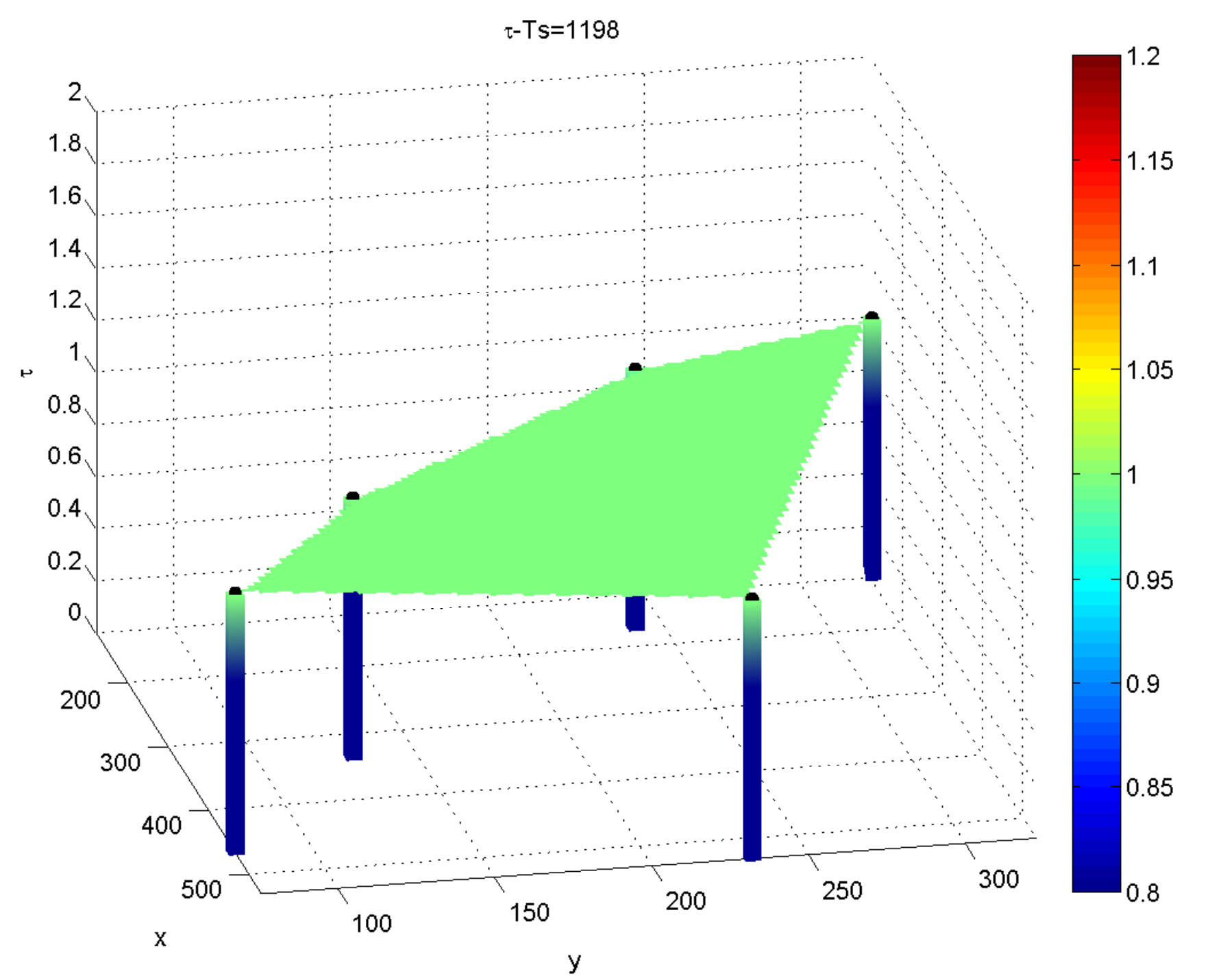}
}
\subfloat[\EtS{1250}{s}]{
\includegraphics[width=0.16\textwidth]{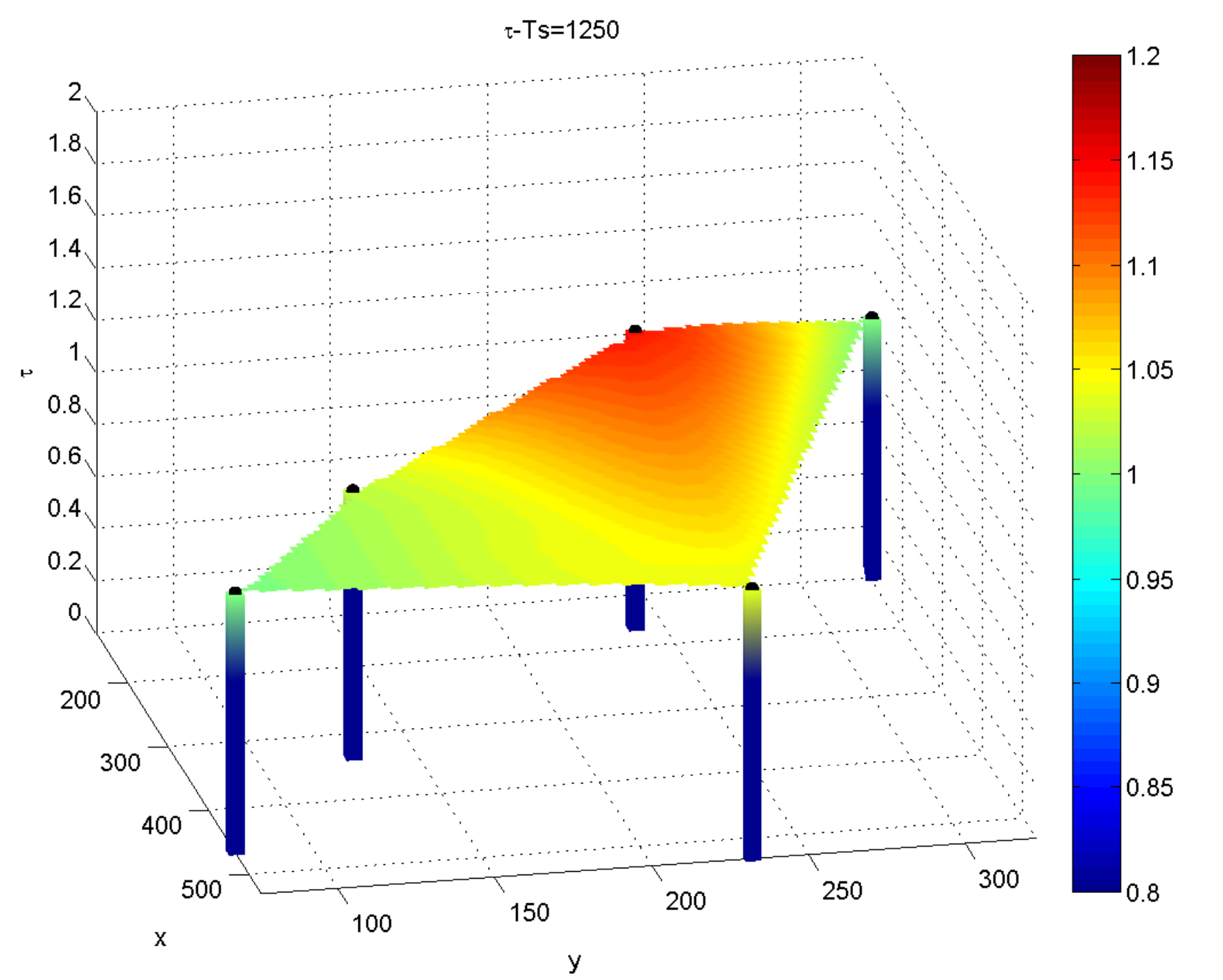}
}
\subfloat[\EtS{1350}{s}]{
\includegraphics[width=0.16\textwidth]{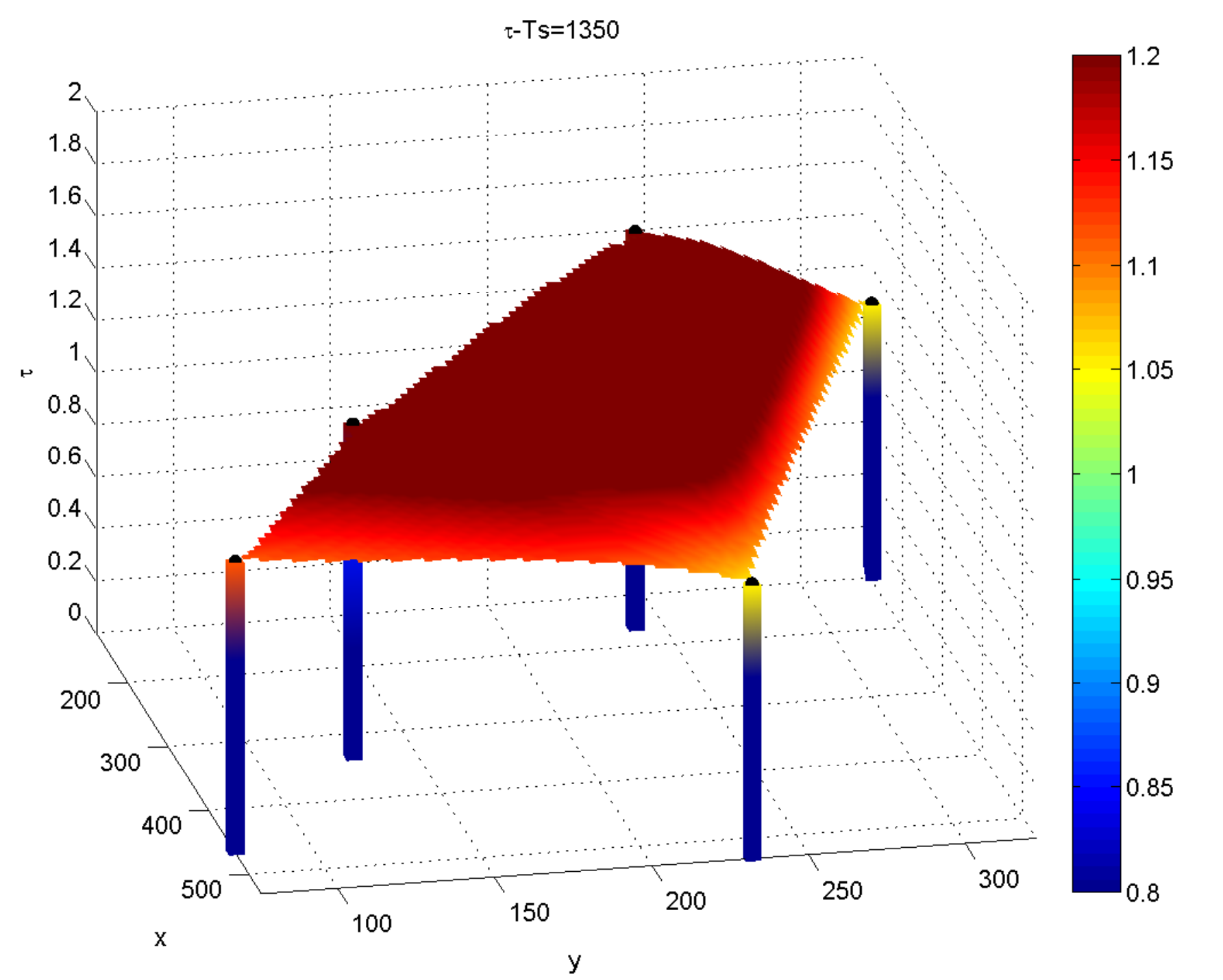}
}
\caption{Visualization of the high-dimensional indictor $\eta$ without Data Set of A3}
\label{fig:DataWithoutA3MSR}

\centering
\subfloat[\EtS{600}{s}]{
\includegraphics[width=0.16\textwidth]{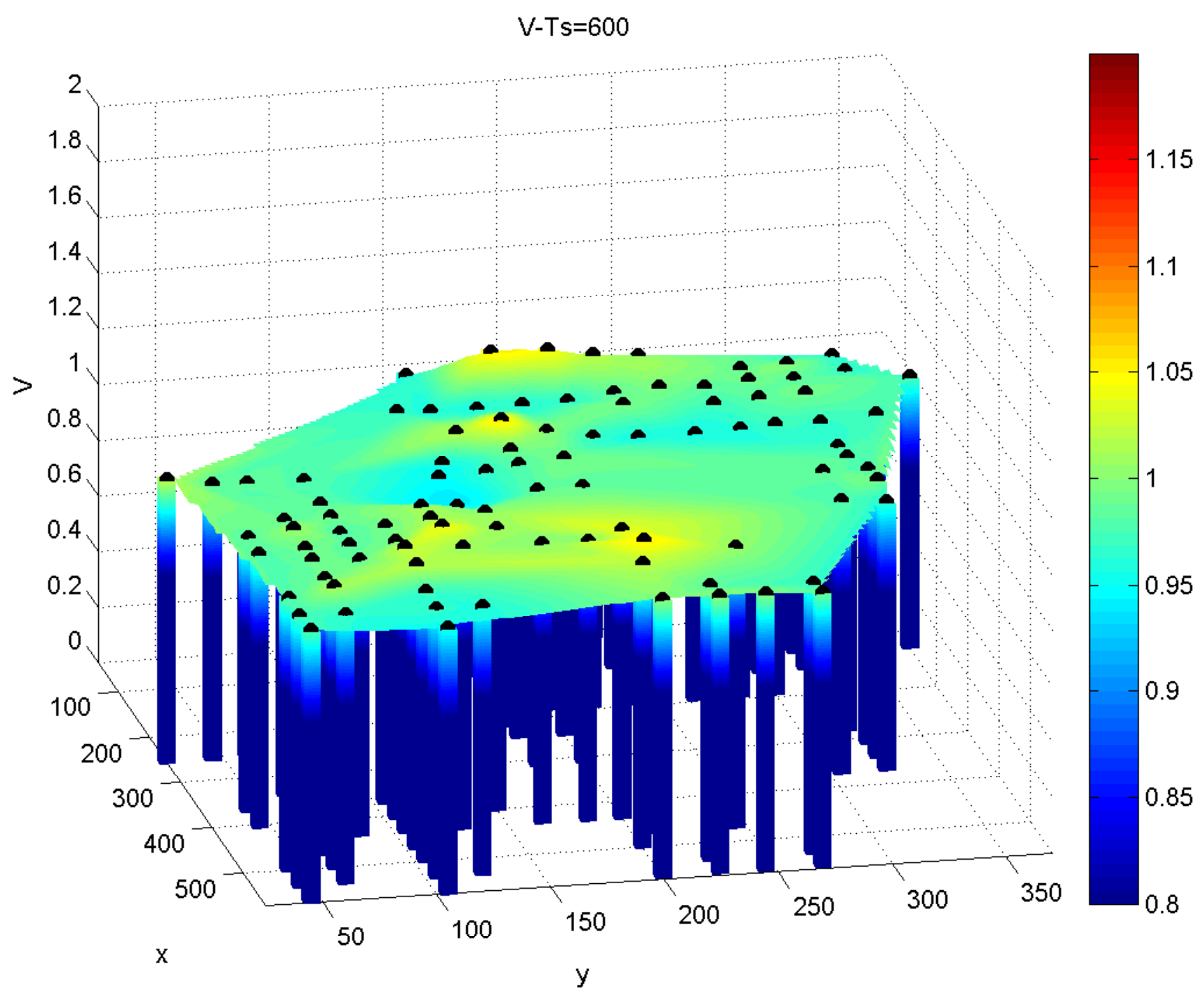}
}
\subfloat[\EtS{601}{s}]{
\includegraphics[width=0.16\textwidth]{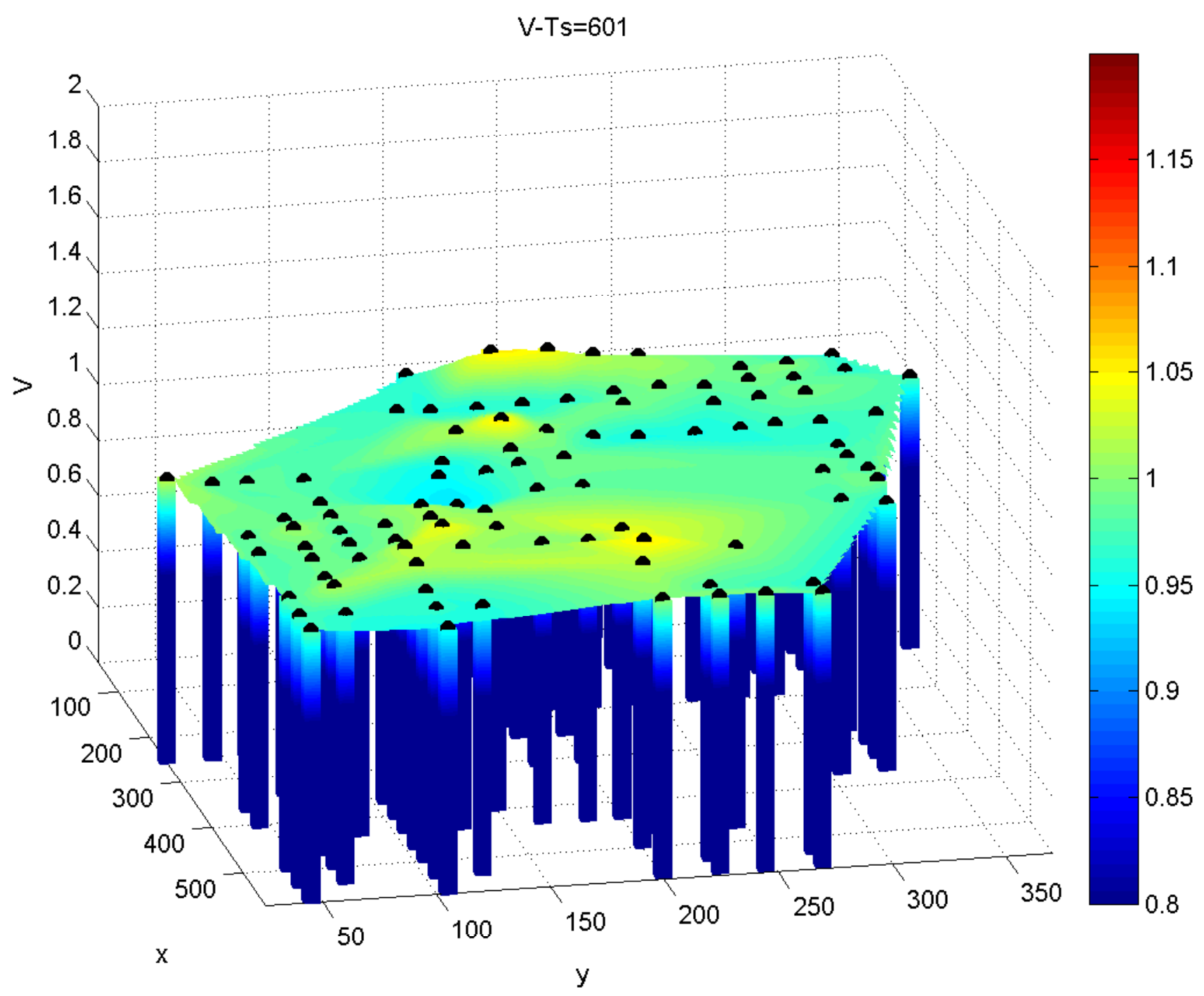}
}
\subfloat[\EtS{720}{s}]{
\includegraphics[width=0.16\textwidth]{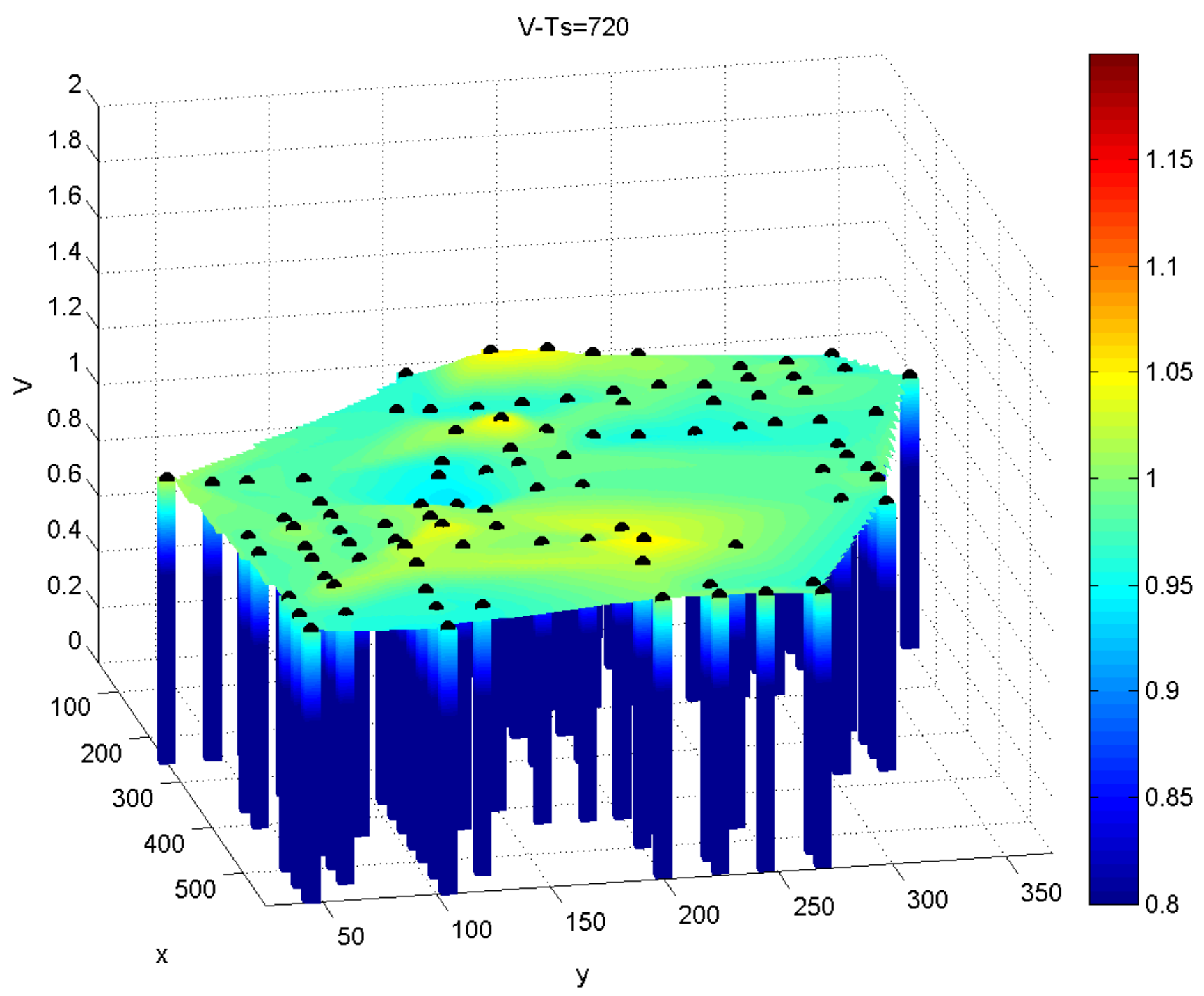}
}
\subfloat[\EtS{1198}{s}]{
\includegraphics[width=0.16\textwidth]{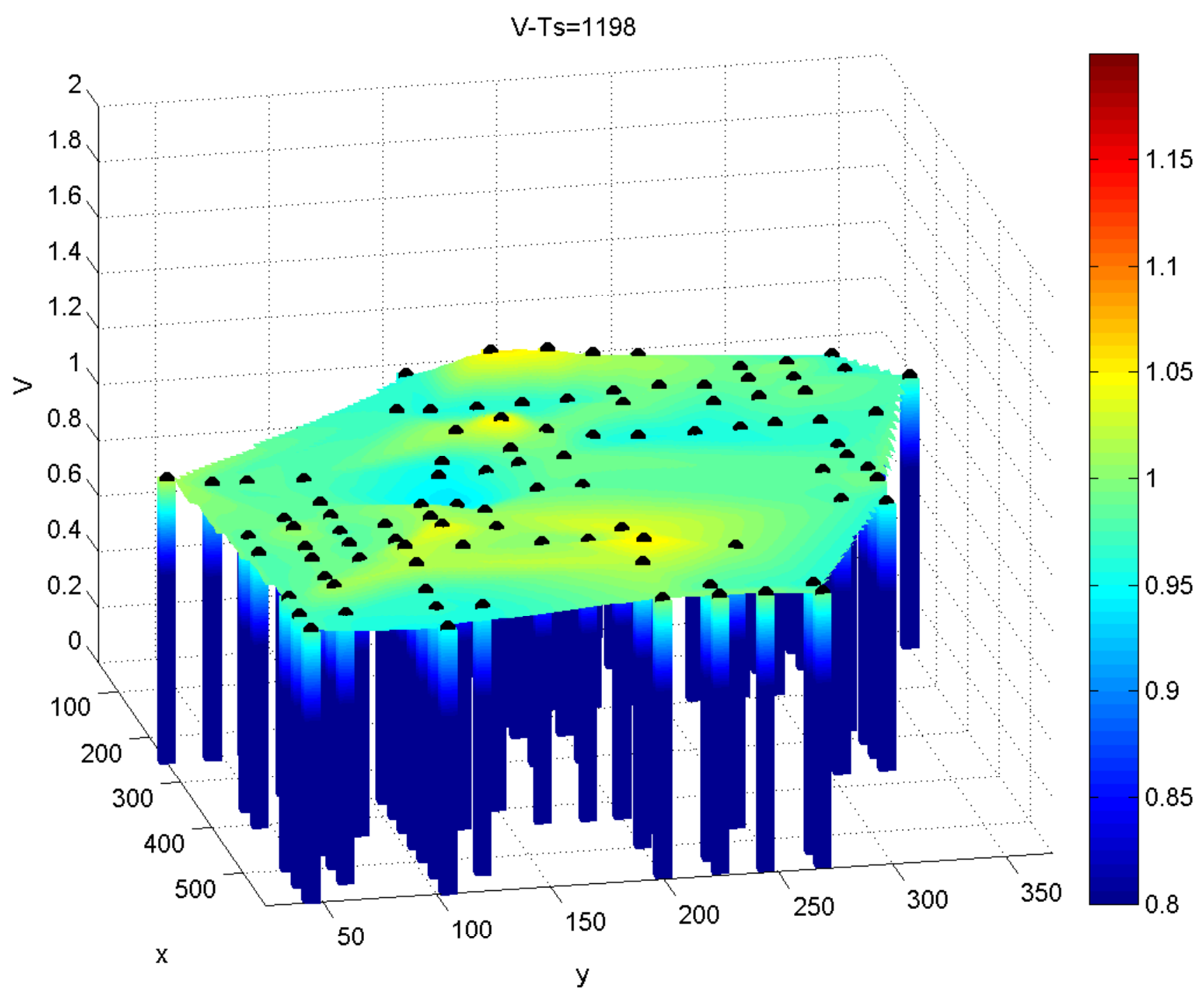}
}
\subfloat[\EtS{1250}{s}]{
\includegraphics[width=0.16\textwidth]{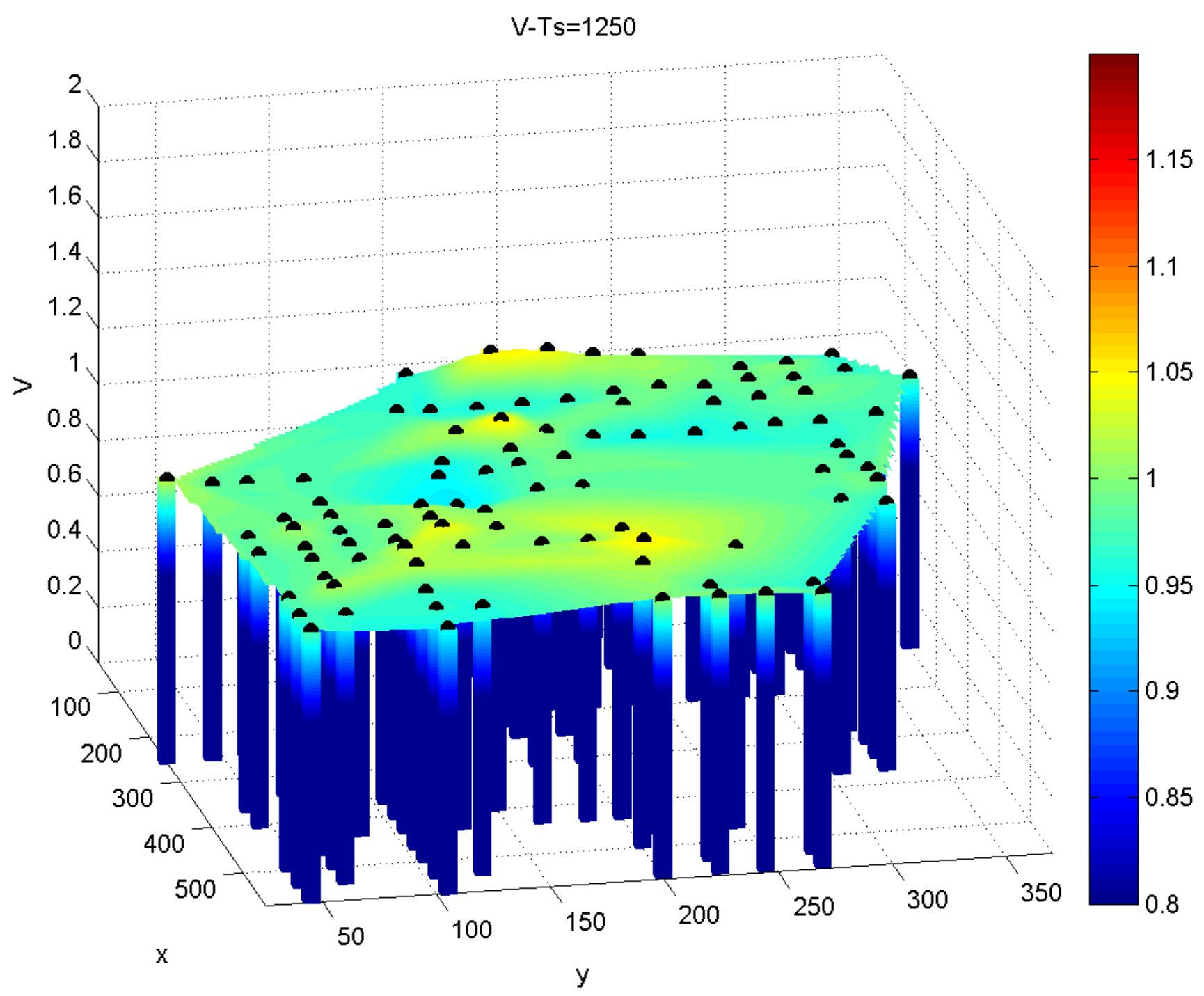}
}
\subfloat[\EtS{1350}{s}]{
\includegraphics[width=0.16\textwidth]{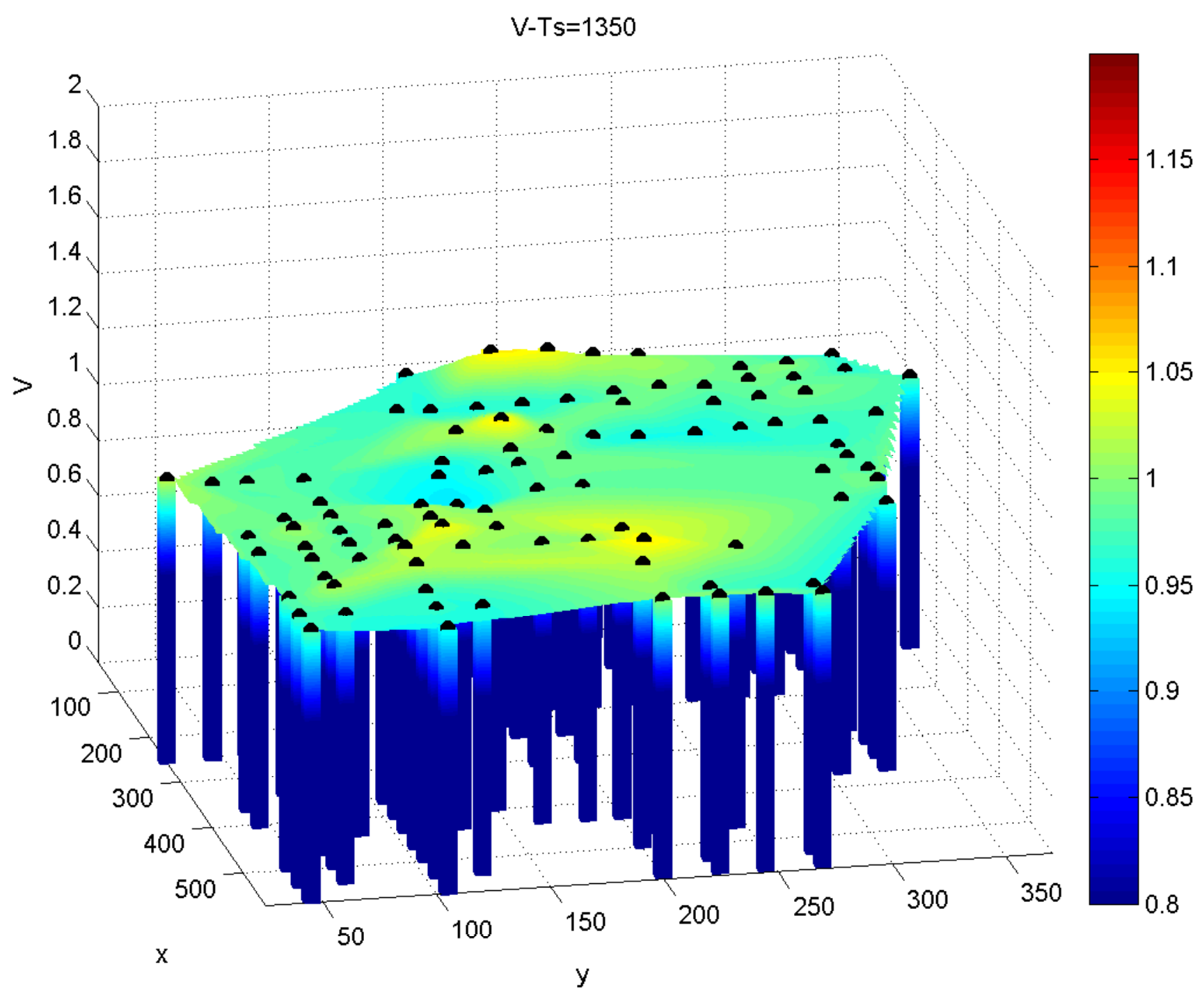}
}
\caption{Visualization of the Voltage \VV{} without Data Set of A3}
\label{fig:DataWithoutA3V}
\end{figure*}

We denote \Equs {\eta}{\VLES{LRF}/\STE{\VLES{LRF}}} as the high-dimensional indicator. For each time point, with an interpolation method \cite{weber2000voltage}, a 3D map is able to be plotted.
Figure \ref{fig:FullDataMSR}, \ref{fig:DataWithoutA3MSR} depict some key frames of the 3D power-map animation with $\eta$, whereas Figure \ref{fig:FullDataV}, \ref{fig:DataWithoutA3V} with the raw data  {\VV {}}:\\
a) For Figure \ref{fig:FullDataMSR}, at time \Et{601}{s}, {$\eta$} of the area around A3 changes relative dramatically; this trend last for the next \Equs{T-2}{238} sampling points (i.e.  \EtS{602\!:\!839}{s})  in the animation. Therefore, we conjecture that some event occurs in A3; even we can go further that the event is influential to A1, A2, A4, A5, and has little impact on A6. These conjectures, in a reasoning way, coincide with the common sense that there is a sudden change in A3 at \Et{601}{s}.\\
b) For Figure \ref{fig:FullDataMSR}, with sustainable growth of power demand at some bus (\VPbus{52} for this case), the whole system becomes more and more vulnerable. The vulnerability can be estimated, before the system has a breakdown due to voltage collapse, via the visualization of $\eta$.\\
c) Moreover, if the most related data (data of A3 for this case) are lost somehow, hardly any information can be gotten by {\VV {}} as Figure \ref{fig:DataWithoutA3V}, whereas the proper judgements can still be achieved by $\eta$ as Figure \ref{fig:DataWithoutA3MSR}.

In general, the combination of high-dimensional indicators (e.g. $\eta$) and 3D power-map is really a novel and feasible approach to EED in smart grids.

\subsection{Conduct EED Using Real Data}

\begin{figure}[H]
  \centering
  \includegraphics[width=0.5\textwidth]{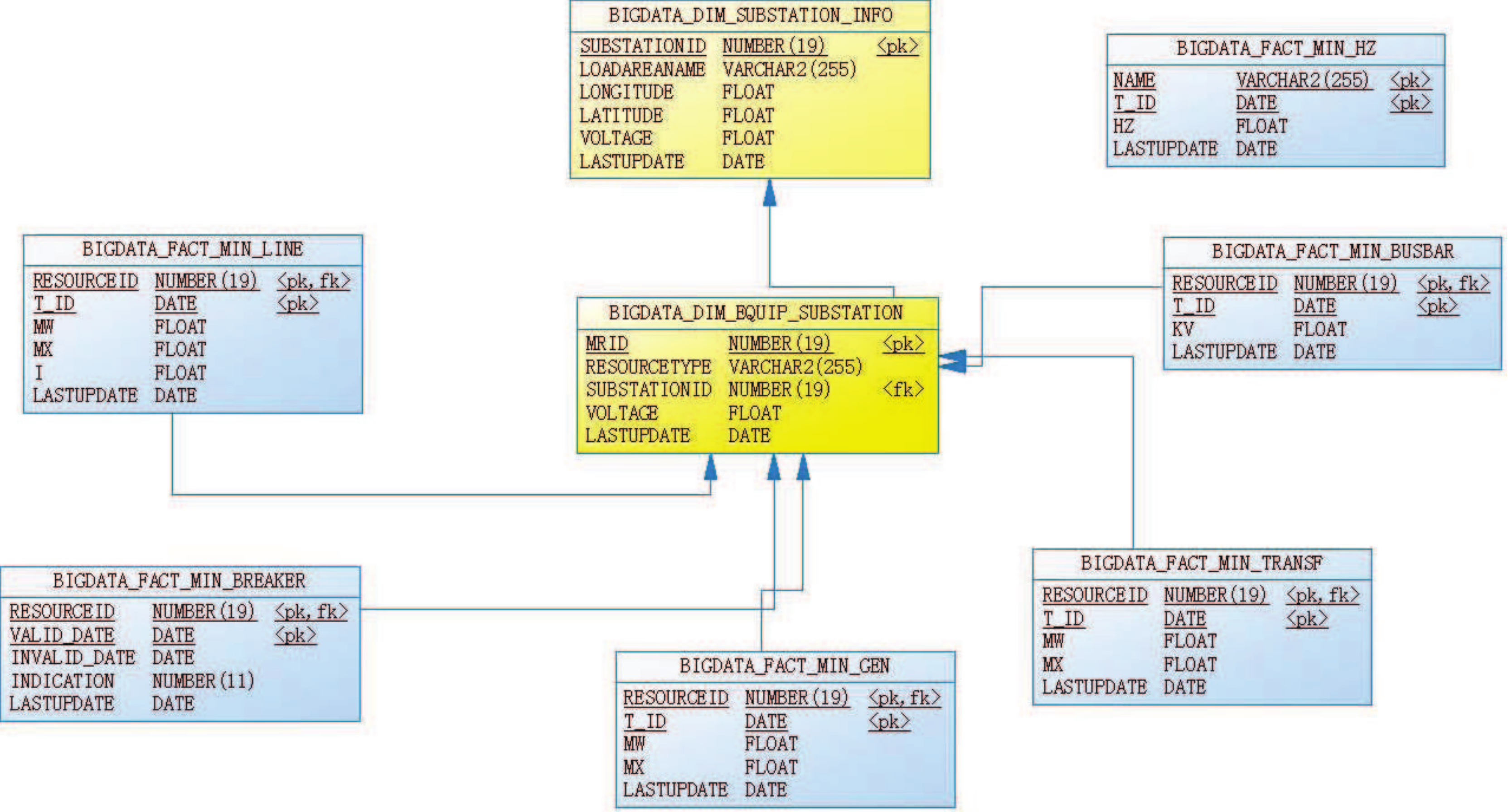}
  \caption{The structure of the data lists for a certain power grid in China}\label{fig:datalist}
\end{figure}

This case is based on the operation data of a certain interconnected power grid in China; the data lists are shown as Figure \ref{fig:datalist}. The data source includes data of substations, breakers, lines, buses, generators, frequencies, etc. These data are from 6 administrative regions---5 provincial ones and 1 directly affiliated one. The sampling frequency is once per minute and the sampling lasts for 3 days (4320 minutes).

\subsubsection{Model---Size \Data {42\! \times \!90}, voltage data, \VLES{MSR}}
\Text{\\}

\begin{figure}[H]
\centering

\includegraphics[width=0.46\textwidth]{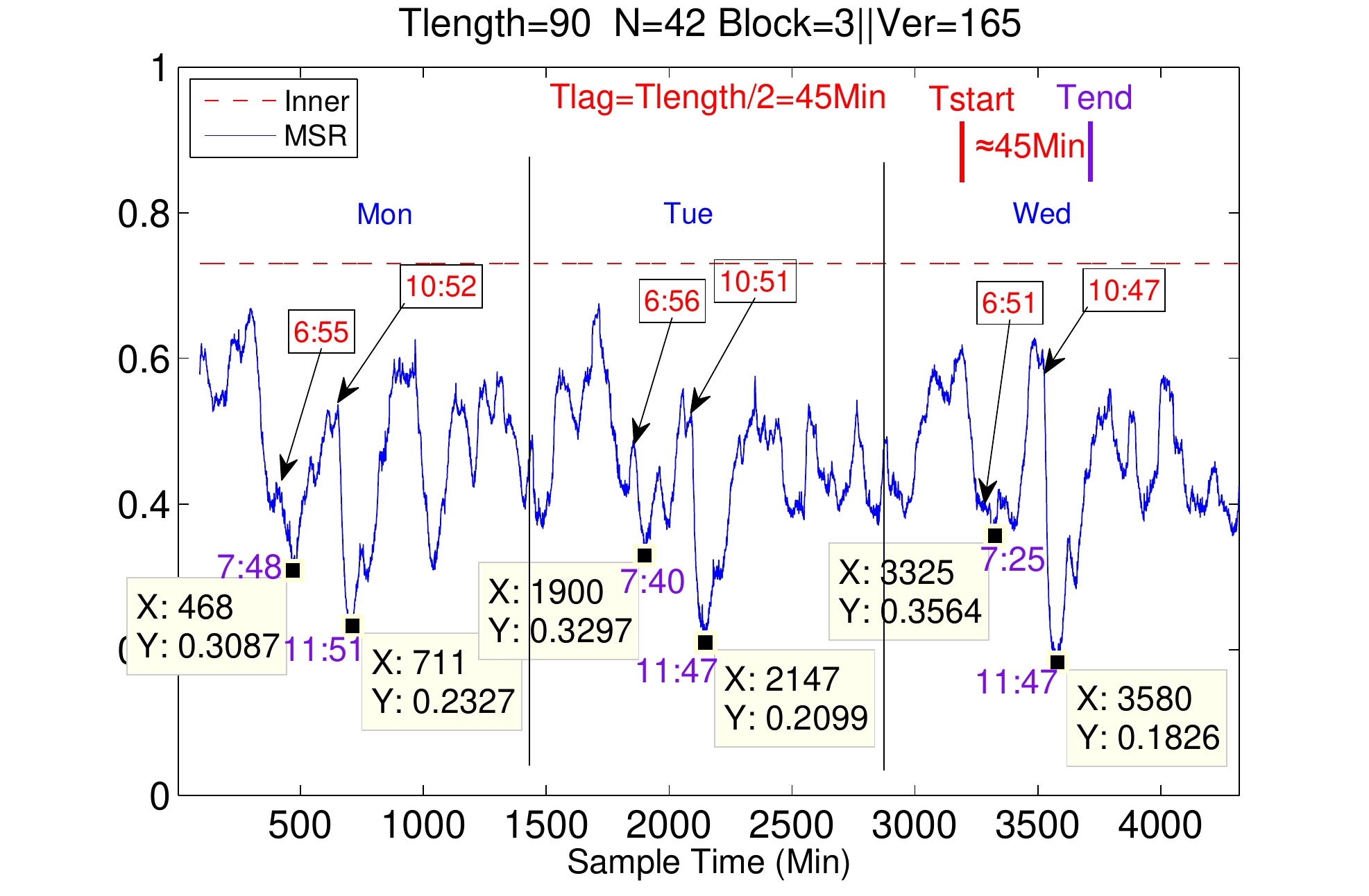}

\caption{\Cur {\VLES{MSR}}{t} Curve for Regional Bus Voltage Data}
\label{fig:RealMSR}
\end{figure}

\normalsize{}We only focus on a regional bus voltage data with 42 nodes; thus, the data source $\Omega$ is of \Equs {42\! \times \!4320}{181440} sampling voltage data.
Here, we choose \VLES{MSR} as the indicator. The  RMM \VRX{} at each sampling point is modeled in size of  \Data {42\! \times \!90}; the time length of the split-window is 90 (\Equs {T}{90} m, and \Equs {T/2}{45} m). Figure \ref{fig:RealMSR} shows the result.

In Figure \ref{fig:RealMSR}, we can extract some half "U"-shaped curves according to the legends (the red for start points and the purple for end ones). As our analyses above, the start points correspond to the right time when the event occurs. They are almost \Et {06:55}{}, maybe for the working start, \mbox{and  \Et {10:50}{},} maybe for the lunch break, respectively.

\subsubsection{Model--- \Equs {T}{90} m, power flow data, \VLES{LRF}}
\Text{\\}

\begin{figure}[H]
\centering

\includegraphics[width=0.46\textwidth]{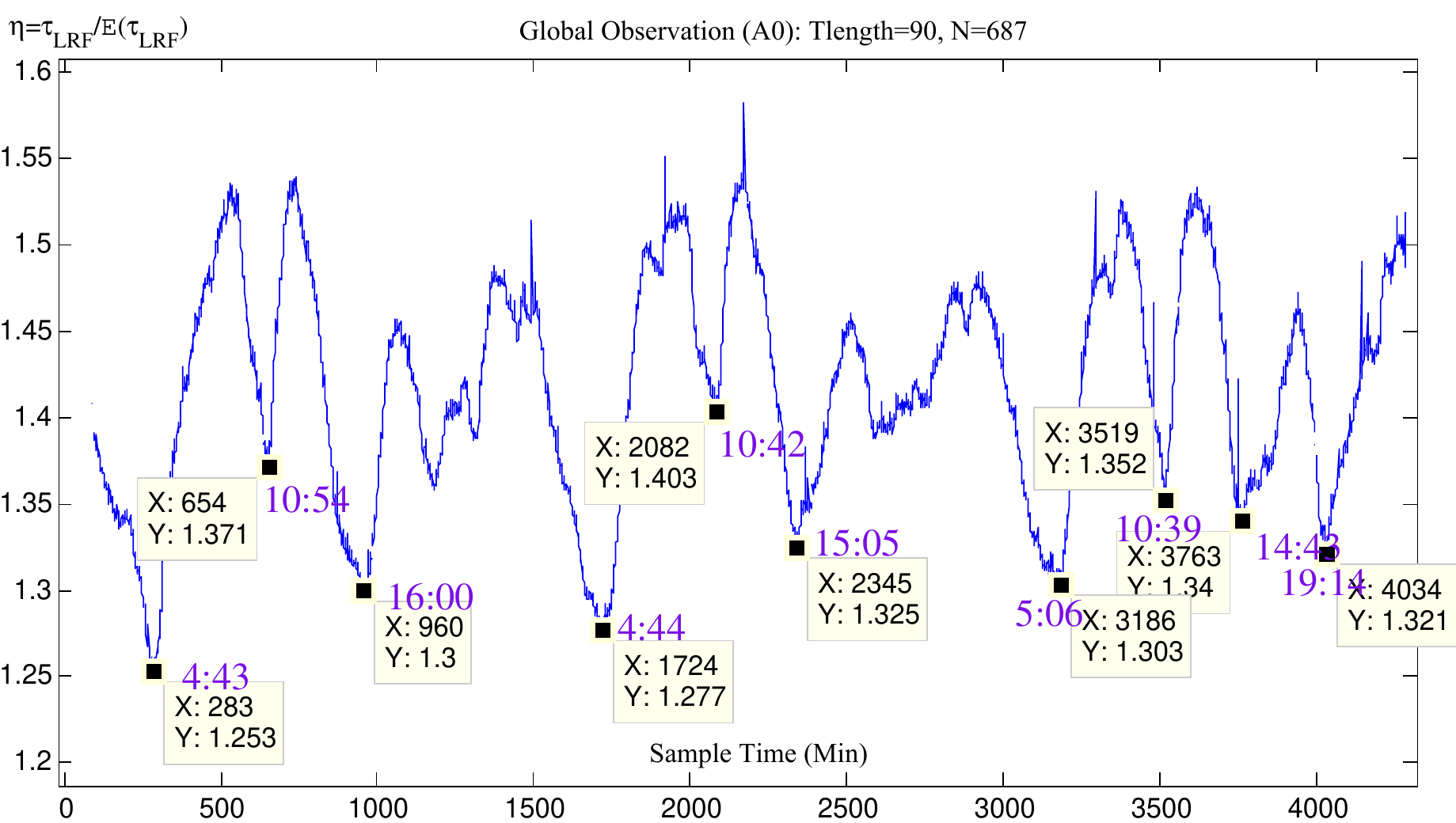}
\caption{\Cur {\eta}{t} Curve for Global Power Flow Data}
\label{fig:RealLRFA0}
\end{figure}

\begin{figure}[H]
\centering

\includegraphics[width=0.46\textwidth]{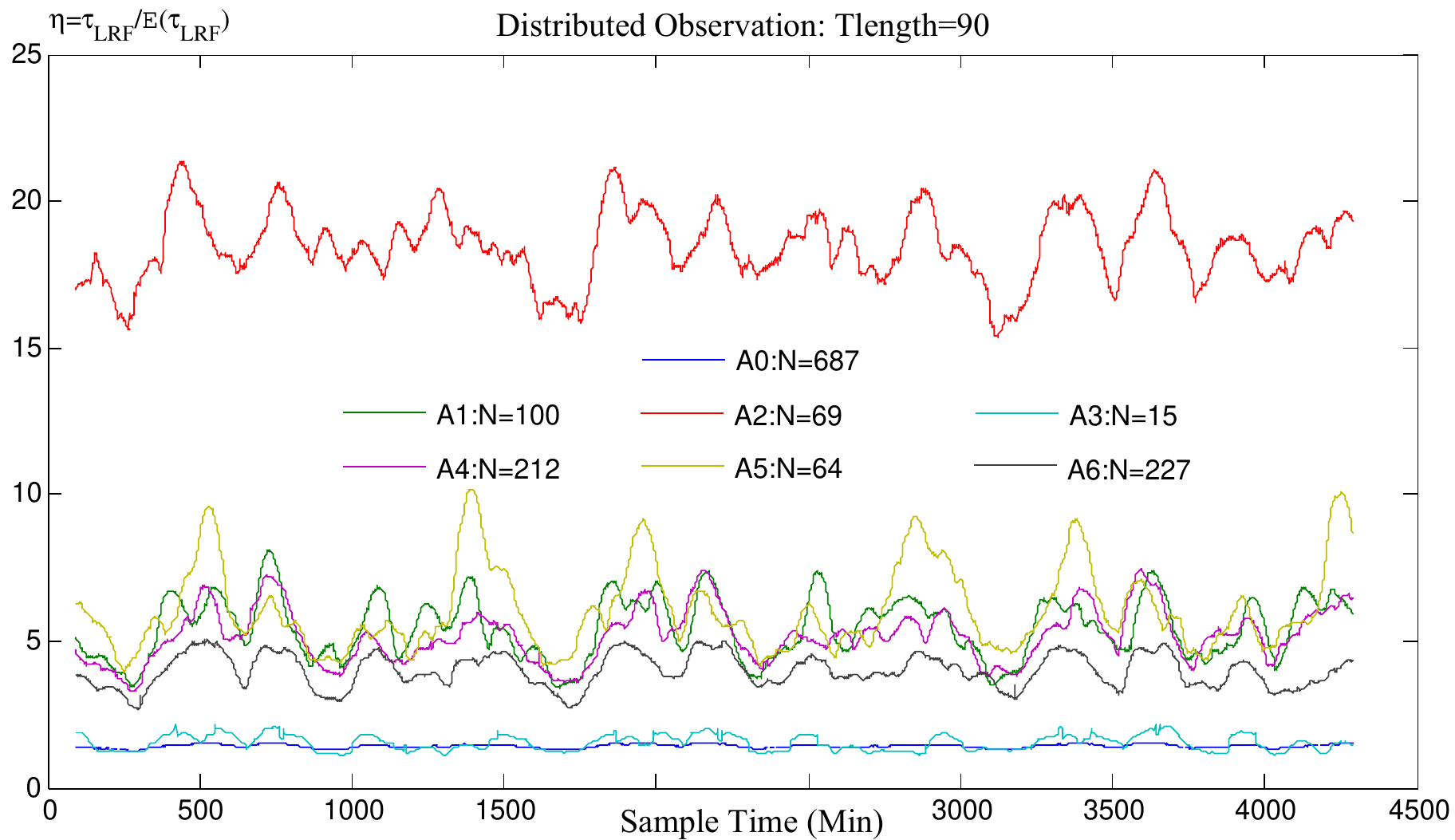}
\caption{\Cur {\eta}{t} Curves for Distributed Power Flow Data}
\label{fig:RealLRFALL}
\end{figure}

We focus on the power flow data (687 lines); we choose \VLES{LRF} and keep \Equs {T}{90} m.
Figure \ref{fig:RealLRFA0} shows the results; some special time points, such as \Et {4:44}{} and \Et {10:42}{}, are able to be observed.
For each administrative region, we conduct a similar procedure and  Figure \ref{fig:RealLRFALL} shows the results.

In Figure \ref{fig:RealLRFA0}, we regard the bottom part (the purple) but not the top as the start time. That is because the "U"-shaped curves for the \VLES{MSR} (Figure \ref{fig:Case 118}) and the \VLES{LRF} (Figure \ref{fig:Ples}) are reversed.

\subsubsection{Summary}
\Text{\\}

From both models above,  some key time points, as well as the daily periodicity,  are able to be observed. The bus voltage and the power flow have different engineering meanings essentially: the former is closely related to local loads, whereas the later is to power exchanges between connected regions. This may be the reason why their key time points are different. Moreover, we believe that the information loss during the processing proceeding in high dimensions should be much less than in low dimensions, even negligible with a proper designed LES (the LES \VLES{} is almost arbitrary; it is only related to the test function $\varphi({\lambda})$). In our opinions, the integration of statistic analyses in high dimensions and engineering explanations (see \textit{Procedure of Ring Law Analysis} in \textit {Sec \ROMAN 4}) are the two procedures to mine the potential values from the complex 4Vs data. More work (including more real data) are essential for further mining.

\section{Conclusion}
\normalsize{This paper proposes a data-driven unsupervised learning method for early event detection (EED). Based on  random matrix theories (RMTs) as our mathematical foundations, the linear eigenvalue statistics (LESs) are proposed as the key indicators. In addition, we compare the proposed method with a supervised one (\textit{a dimensionality reduction method based on PCA}). Case studies, with both simulated data and real ones, validate the effectiveness and the advantages of the proposed method. Especially, its robustness against bad data is highlighted---the combination of 3D power-map and high-dimensional indicators, even with data source loss in the core area, is still able to conduct EED effectively.}

This work is another attempt towards applying big data to power systems.
It mainly argues that random matrix models (RMMs) are a proper tool to reveal physical systems in high dimensional perspectives. This enables us to mine the potential values from the complex 4Vs data resources efficiently. Three main steps are essential as the procedure: 1) to build the RMM with raw data; 2) to conduct high-dimensional analyses with statistical transformations; 3) to interpret the results to human beings.
The proposed data-driven unsupervised approach is universal, as it mainly rely on raw data which are independent of empirical models or labelled parameters; it is also sensitive since it uses the high dimensional statistics as the indicators. Moreover, for some data processing ingredients, the theoretical values are predictable by the latest theorems.

However, our findings only make up a tiny fraction of high-dimensional analytics. More problems need to be further studied: 1) how to design the test function as a filter to detect potential anomalies in real systems (\textit{Sec \ROMAN 3, \ROMAN 5}); 2) how to choose data source or make data fusion to meet certain engineering requirements (\textit{Sec \ROMAN 3}); 3) how to explain the high-dimensional findings to human beings (\textit{Sec \ROMAN 4, \ROMAN 5}); 4) how to turn big data into tiny data (\textit{Sec \ROMAN 5}).
One wonders if this direction will be far-reaching in years to come toward the Big Data Age for smart grids.

\appendices
\section{Formulas for RMT}
\subsection{Convert \VRX{}\Belong {}{\VF CNT}  to \VTX{}\Belong {}{\VF CNT}:}

\FuncC {eq:StdMatrix}
{
\VTx{i}\!=\!\Div{\Sig{\VTx{i}}}{\Sig{\VRx{i}}}{(\VRx{i}\!-\!\Mu {\VRx{i}})}{\! + \! \Mu {\VTx{i}}}\,,\Data {1\!\LE{}\!i\LE{}\!N\!}
}
\small{where \Equs {\VRx{i}}{(\DatasNTt {\VRsx{}}{i,}{T})}  and \Equs {\Mu {\VTx{i}}}{0}, \Equs {\Sigg {\VTx{i}}}{1}.}\normalsize{}

\subsection{Convert \Belong {\VTX{}} {\VF CNT} to \Belong {\VX {u}} {\VF CNN}:}
\FuncC {eq:Xu}
{\Equ {\VX u}   {\Sqrt{\VTX{}{\Her{\VTX{}}}}\Vector U }}
\small{where \Belong {\Vector U} {\VF CNN} is a Haar unitary matrix; \Equu {\VX{u}\Her{\VX{u}}} {\VTX{}{\Her{\VTX{}}}}.}\normalsize{}

\subsection{Convert \VZ{}\Belong {}{\VF CNN} to \VTZ{}\Belong {}{\VF CNN}:}
\FuncC {eq:StdZ}{
{\Equ {\VTz{i}} {\Vz{i}/({\Sqrt N}\Sig{\Vz{i}})}}
\quad ,\Data {1\LE{}i\LE{}N}
}
\small{where \Equs {\Vz{i}}{(\DatasNTt {\Vsz{}}{i\!,}{N}) }, \Data {\VZ{}=\Prod{i=1}{L} \VX {u,i}}.}\normalsize{}

\section{Definition of the Gaussian orthogonal ensemble (GOE)}
\normalsize{This is a real symmetric \Muls n n random matrix:}
\FuncC {eq:GOE1}{
\Vector M=n^{-1/2}\Vector W, \ \Vector W=\left\{W_{j,k}\Belong{}{\Fdata R}, W_{j,k}=W_{k,j}\right\}^n_{j,k=1}
}
\small{defined by the probability law:}\normalsize{}
\FuncC {eq:GOE2}{
Z_{n1}^{-1}e^{-\text{Tr}{\,\Vector W^2}/4\omega^2}\prod_{1\LE{}j\LE{}k\LE{}n}\,dW_{j,k}
}
\small{where $Z_{n1}$ is the normalization constant. Since:}\normalsize{}
\[
\text{Tr}{\,\Vector W^2}=\sum_{1\LE{}j\LE{}n}{{W_{j,j}}^2+2\sum_{1\LE{}j<k\LE{}n}{W_{j,k}}^2}
\]
\normalsize{the above implies that $\left\{W_{j,k}\right\}_{1\LE{}j\LE{}k\LE{}n}$ are independent Gaussian random variables such that:}
\FuncC {eq:GOE3}{
\Fdata E({W_{j,k}})=0 \quad {\Fdata E({{W_{j,k}}^2})=\omega^2(1+\delta_{j,k})}
}

\section{Expectation of the Random Variable with Gaussian Distribution}
\normalsize{$X$ is a random variable with standard normal distribution; it is described by the probability density function (PDF):}
\[
f\left( x \right)=\frac{1}{\sqrt{2\pi }}{{e}^{-\frac{{{x}^{2}}}{2}}}
\]

\subsubsection{\STE{X},\STE{X^2},\STE{X^4}}
{\Text{\\}}

\begin{equation}
\label {eq:E124}
\begin{aligned}
\mathbb{E}\left( X \right)  &    =\int_{-\infty }^{\infty }{xf\left( x \right)dx}=0  \\
\mathbb{E}\left( {{X}^{2}} \right)    &     =\int_{-\infty }^{\infty }{{{x}^{2}}\frac{1}{\sqrt{2\pi }}{{e}^{-\frac{{{x}^{2}}}{2}}}dx}=\int_{-\infty }^{\infty }{{{x}^{2}}\frac{1}{\sqrt{2\pi }}\frac{1}{-x}d{{e}^{-\frac{{{x}^{2}}}{2}}}} \\
 &           =-\frac{1}{\sqrt{2\pi }}\left( \left[ x{{e}^{-{{x}^{2}}/2}} \right]_{-\infty }^{\infty }-\int_{-\infty }^{\infty }{{{e}^{-\frac{{{x}^{2}}}{2}}}dx} \right) \\
 &           =2\frac{\sqrt{2}}{\sqrt{2\pi }}\int_{0}^{\infty }{{{e}^{-\frac{{{x}^{2}}}{2}}}d\frac{x}{\sqrt{2}}} \\
 &            =\frac{2}{\sqrt{\pi }}\int_{0}^{\infty }{{{e}^{-{{t}^{2}}}}dt}\\
 &             =\frac{2}{\sqrt{\pi }}\frac{\sqrt{\pi }}{2}=1 \\
 \mathbb{E}\left( {{X}^{4}} \right)     &    =\int_{-\infty }^{\infty }{{{x}^{4}}\frac{1}{\sqrt{2\pi }}{{e}^{-\frac{{{x}^{2}}}{2}}}dx}=\int_{-\infty }^{\infty }{{{x}^{4}}\frac{1}{\sqrt{2\pi }}\frac{1}{-x}d{{e}^{-\frac{{{x}^{2}}}{2}}}} \\
 &             =-\frac{1}{\sqrt{2\pi }}\left( \left[ {{x}^{3}}{{e}^{-{{x}^{2}}/2}} \right]_{-\infty }^{\infty }-\int_{-\infty }^{\infty }{{{e}^{-\frac{{{x}^{2}}}{2}}}d{{x}^{3}}} \right) \\
 &            =2\frac{3}{\sqrt{2\pi }}2\sqrt{2}\int_{0}^{\infty }{{{e}^{-\frac{{{x}^{2}}}{2}}}\frac{{{x}^{2}}}{2}d\frac{x}{\sqrt{2}}}  \\
 &           =\frac{12}{\sqrt{\pi }}\int_{0}^{\infty }{{{e}^{-{{t}^{2}}}}{{t}^{2}}dt}\\
 &          =\frac{12}{\sqrt{\pi }}\frac{\sqrt{\pi }}{4}=3 \\
\end{aligned}
\end{equation}

\subsubsection{$\int_0^\infty  {{e^{ - {t^2}}}dt}$, $\int_0^\infty  {{t^2e^{ - {t^2}}}dt}$}
{\Text{\\}}
\\

Suppose:
\[
a=\int_{0}^{\infty }{{{e}^{-{{t}^{2}}}}dt}
\]
then:
\[
\begin{aligned}
{{a}^{2}}   &  =\int_{0}^{\infty }{{{e}^{-{{t}^{2}}}}dt}\int_{0}^{\infty }{{{e}^{-{{u}^{2}}}}du}\\
&      =\iint\limits_{t,u\ge 0}{{{e}^{-\left( {{t}^{2}}+{{u}^{2}} \right)}} dtdu}=\int_{0}^{\frac{\pi }{2}}{d\theta }\int_{0}^{\infty }{{{e}^{-{{r}^{2}}}}rdr} \\
 &      =\frac{\pi }{2}\frac{1}{2}\int_{0}^{\infty }{{{e}^{-t}}dt}=\frac{\pi }{2}\frac{1}{2}\left[ -{{e}^{-t}} \right]_{0}^{\infty }=\frac{\pi }{4} \\
\end{aligned}
\]
so:
\begin{equation}
\int_{0}^{\infty }{{{e}^{-{{t}^{2}}}}dt}=\frac{\sqrt{\pi }}{2}
\end{equation}

Because:
\[
\begin{aligned}
   a          &     = \int_{0}^{\infty }{{{e}^{-{{t}^{2}}}}dt}=\left[ t{{e}^{-{{t}^{2}}}} \right]_{0}^{\infty }-\int_{0}^{\infty }{td{{e}^{-{{t}^{2}}}}} \\
 &    =-\int_{0}^{\infty }{t\left( -2t \right){{e}^{-{{t}^{2}}}}d}t \\
 &    =2\int_{0}^{\infty }{{{t}^{2}}{{e}^{-{{t}^{2}}}}d}t \\
\end{aligned}
\]
so:
\begin{equation}
\int_{0}^{\infty }{{{t}^{2}}{{e}^{-{{t}^{2}}}}d}t=\frac{a}{2}=\frac{\sqrt{\pi }}{4} \\
\end{equation}

\section{}

\section{}
\begin{table}[H]
\caption {Series of Events}
\label{Tab: Event Series}
\centering

\begin{minipage}[htbp]{0.48\textwidth}
\centering

\begin{tabularx}{\textwidth} { >{\scshape}l !{\color{black}\vrule width1pt}    >{$}l<{$}    >{$}l<{$}   >{$}l<{$}  }  
\toprule[1.5pt]
\hline
\textbf {Par}$\backslash$\VAT{} & [001:600] & [601:1200] & [1201:1500]\\
\hline
\VPbus{52} (MW) & 0 & 300 & t-900\\

\toprule[1pt]
\end{tabularx}
\raggedright
\small {*\VPbus{52} is the power demand of bus-52}
\end{minipage}
\end{table}

\section{}

\small{}
\bibliographystyle{IEEEtran}
\bibliography{helx}

\end{document}